\documentclass[a4paper, 11pt]{article}
\usepackage[T1]{fontenc}
\usepackage{jheppub,graphicx,epsf,amssymb,amsbsy,amsfonts,amssymb,amsmath, empheq,mathrsfs,appendix,soul}
\usepackage{hyperref}

\def\be{\begin{equation}}
\def\ee{\end{equation}}
\def\bea{\begin{eqnarray}}
\def\eea{\end{eqnarray}}

\def\a{\alpha}
\def\b{\beta}
 
\def\d{\delta}
\def\e{\epsilon}
\def\f{\phi}

\def\l{\lambda}
\def\l{\lambda}

\def\bg{\bar{g}}
\def\beq{\begin{eqnarray}}\def\eeq{\end{eqnarray}}
\def\ba#1\ea{\begin{align}#1\end{align}}
\def\bg#1\eg{\begin{gather}#1\end{gather}}
\def\bm#1\em{\begin{multline}#1\end{multline}}
\def\bmd#1\emd{\begin{multlined}#1\end{multlined}}

\def\a{\alpha}
\def\b{\beta}

\def\d{\delta}
\def\D{\Delta}
\def\e{\epsilon}

\def\l{\lambda}

\def\({\left(}
\def\){\right)}
\def\[{\left[}
\def\]{\right]}

\def\a{\alpha}

\def\l{\lambda}
\def\b{\beta}

\def\d{\delta}
 
\def\om{\omega}

\def\f{\frac}

\def\D{\Delta}

\title{MHV Graviton Scattering Amplitudes and Current Algebra on the Celestial Sphere}

\author[1]{Shamik Banerjee}
\author[2]{Sudip Ghosh}
\author[3]{Partha Paul}
\affiliation[1]{Institute of Physics, Sachivalaya Marg, Bhubaneshwar, India-751005 \\ and Homi Bhabha National Institute, Anushakti Nagar, Mumbai, India-400085}
\affiliation[2]{Okinawa Institute of Science and Technology,1919-1 Tancha, Onna-son, Okinawa 904-0495,Japan}
\affiliation[3]{Chennai Mathematical Institute, SIPCOT IT Park, Siruseri 603103, India}
\emailAdd{banerjeeshamik.phy@gmail.com, sudip112phys@gmail.com, pl.partha13@gmail.com }

\abstract{The Cachazo-Strominger subleading soft graviton theorem for a positive helicity soft graviton is equivalent to the Ward identities for $\overline{SL(2,\mathbb C)}$ currents. This naturally gives rise to a $\overline{SL(2,\mathbb C)}$ current algebra living on the celestial sphere. The generators of the $\overline{SL(2,\mathbb C)}$ current algebra and the supertranslations, coming from a positive helicity leading soft graviton, form a closed algebra. We find that the OPE of two graviton primaries in the Celestial CFT, extracted from MHV amplitudes, is completely determined in terms of this algebra. To be more precise, 1) The subleading terms in the OPE are determined in terms of the leading OPE coefficient if we demand that both sides of the OPE transform in the same way under this local symmetry algebra. 2) Positive helicity gravitons have null states under this local algebra whose decoupling leads to differential equations for MHV amplitudes. An $n$ point MHV amplitude satisfies two systems of $(n-2)$ linear first order PDEs corresponding to $(n-2)$ positive helicity gravitons. We have checked, using Hodges' formula, that one system of differential equations is satisfied by any MHV amplitude, whereas the other system has been checked up to six graviton MHV amplitude. 3) One can determine the leading OPE coefficients from these differential equations. 

This points to the existence of an autonomous sector of the Celestial CFT which holographically computes the MHV graviton scattering amplitudes and is completely defined by this local symmetry algebra. The MHV-sector of the Celestial CFT is like a minimal model of $2$-D CFT. }

\begin{document}
\maketitle
\flushbottom

\section{Introduction}

The holographic dual of quantum theories of gravity in four dimensional ($4$-$\text{D}$) asymptotically flat spacetimes has been conjectured to be  two dimensional ($2$-$\text{D}$) conformal field theories (CFT) which live on the celestial sphere at null infinity \cite{Strominger:2017zoo,Kapec:2014opa,Kapec:2016jld,He:2017fsb,Ball:2019atb,Barnich:2009se,Barnich:2011ct,Pasterski:2016qvg,Pasterski:2017kqt, Cheung:2016iub, deBoer:2003vf,Banerjee:2018gce,Banerjee:2018fgd, Banerjee:2019prz,Pasterski:2017ylz, Schreiber:2017jsr, Cardona:2017keg,Lam:2017ofc,Banerjee:2017jeg, Stieberger:2018edy, Stieberger:2018onx,Himwich:2020rro,Bagchi:2016bcd}. These are aptly referred to as Celestial CFTs. The equivalence between the action of the $4$-D Lorentz group and global conformal transformations on the celestial sphere already hints at the possible existence of such a duality. In fact S-matrix elements, which are the primary observables of the bulk theory, can be expressed in a particular basis \cite{Pasterski:2016qvg,Pasterski:2017kqt} where they manifestly transform as correlation functions of primary operators in $2$-D CFT. For example in the case of massless particles this change of basis from the traditionally employed basis of plane waves is implemented via a Mellin transform \cite{Law:2019glh, Fotopoulos:2019vac, Pasterski:2016qvg, Pasterski:2017kqt, Banerjee:2018gce,Banerjee:2018fgd, Banerjee:2019prz,Pasterski:2017ylz, Schreiber:2017jsr, Cardona:2017keg,Lam:2017ofc,Banerjee:2017jeg, Stieberger:2018edy, Stieberger:2018onx}. This recasting of bulk scattering amplitudes, together with the remarkable connection between soft theorems and the infinite dimensional asymptotic symmetries \cite{Bondi:1962px,Sachs:1962zza,Barnich:2009se,Barnich:2011ct} of asymptotically Minkowski spacetimes \cite{Strominger:2013jfa,Strominger:2013lka,Kapec:2016jld,He,Strominger:2014pwa} reveals that Celestial CFTs are endowed with a much larger symmetry group in contrast with more conventional instances of CFTs. In this paper our objective is to explore the implications of these symmetries for the operator product expansion (OPE) in Celestial CFTs.

 
 \vskip 4pt
The celestial OPE between two primary operators can be obtained by considering the limit where the momenta associated with the corresponding particles in the $S$-matrix become collinear. At leading order in the collinear limit the scattering amplitude factorises into a lower point amplitude times a prefactor referred to as the splitting function. The OPE coefficient of the primary operator that contributes at leading order in the OPE can then be extracted from the Mellin transform of this splitting function \cite{Fan:2019emx, Pate:2019lpp}. Now, it is quite remarkable that one can actually calculate the leading OPE coefficients by demanding that both sides of the OPE transform in the same way under the (subleading) subsubleading soft symmetry \cite{Pate:2019lpp} for (gluons) gravitons. This is a holographic calculation of the leading OPE coeffiicients and is the primary motivation behind our work in this paper. 


 \vskip 4pt
Now depending on the helicities of the collinear particles it is possible to have different channels into which the scattering amplitude can factorise. As we elaborate further in Section \ref{mhvsector} of this paper, in the case of tree level graviton amplitudes in the MHV configuration in Einstein gravity, the only nontrivial factorisation channel is the one where the lower point amplitude is again MHV. Then the above mentioned correspondence between the collinear limit in the S-matrix and the OPE limit in the Celestial CFT tells us that the Mellin transformed MHV graviton amplitudes are effectively closed under taking the celestial OPE.  One of the central results in this paper is that this ``MHV-sector'' of the Celestial CFT is governed by an underlying infinite dimensional local symmetry algebra. This symmetry algebra comprises of a current algebra which is encoded in the Cachazo-Strominger subleading soft graviton theorem \cite{Cachazo:2014fwa} and supertranslations associated to the leading soft graviton theorem \cite{Weinberg:1965nx}, for a positive helicity soft graviton. The consequences of the existence of this symmetry, which we study in this paper, then suggest that the Celestial CFT dual to Einstein gravity admits an autonomous sector which provides a complete holographic description of tree level MHV graviton amplitudes.  In the rest of this introduction we present a summary of our main results and an outline of the paper. 

 \vskip 4pt
In Section \ref{curralgeb} we use the subleading soft graviton theorem  for a positive helicity soft graviton to define a set of three currents $J^{a}, a=1,0,-1$. The modes $J^{a}_{m}, \ m\in \mathbb{Z}$, of these currents are shown to generate a level zero current algebra on the celestial sphere. The zero modes of these currents are the global $\overline{SL(2,\mathbb{C})}$, i.e., antiholomorphic Lorentz transformation generators. Throughout this paper, we refer to the algebra of the modes $J^{a}_{m}$ as the $\overline{SL(2,\mathbb{C})}$ current algebra. In Section \ref{shope} we derive the OPE between the subleading soft graviton operator and a generic conformal matter primary. This OPE allows us to define descendants created by the current algebra generators and reveals that conformal primaries are also primaries under the $\overline{SL(2,\mathbb{C})}$ current algebra. This also serves to provide an important consistency check that we use in later sections of the paper.  In Sections \ref{leadsoft} and \ref{opestcp} we use the leading soft graviton theorem to define a set of two supertranslation currents $P_{0}, P_{-1}$. Following that we compute the OPE between the leading soft graviton operator and a ``hard'' conformal primary. Form this result we identify the supertranslation descendants created by the modes $\{P_{n,0}, P_{n,-1}\}$ of $P_{0} $ and $P_{-1}$ respectively.  As in the case of the subleading soft graviton, this OPE also provides a consistency check for the OPEs of graviton primaries that we extract from Mellin transform of MHV graviton amplitudes, subsequently in this paper. Section \ref{mxcom} is devoted to showing that the $\overline{SL(2,\mathbb{C})}$ current algebra and above mentioned supertranslation generators form a closed algebra.  Section \ref{extalgebsum} contains a summary of all the relevant commutation relations, including those with the global $SL(2,\mathbb{C})$, i.e., holomorphic Lorentz generators $\{L_{0}, L_{\pm 1} \}$,  and also the definition of a primary operator under the extended symmetry algebra generated by $\{J^{a}_{n}, P_{n,0}, P_{n,-1}, L_{0}, L_{\pm 1} \}, \ n\in \mathbb{Z}$.  

 \vskip 4pt
In Section \ref{ope4pt} we show that consistency of the first set of subleading terms in the celestial OPE derived from the Mellin transform of the $4$-point MHV graviton amplitude \cite{Banerjee:2020kaa} requires a particular combination of descendants to decouple from the $3$-point Mellin amplitude. This is analogous to null state relations familiar from $2$-D CFTs and leads to a first order linear partial differential equation that must be satisfied by the $3$-point amplitude. But it turns out that the due to kinematic constraints some modes of the local symmetry algebra generators trivially annihilate the $3$-point function. However this does not happen for higher point amplitudes. Therefore in Section \ref{6mhvu} we consider the Mellin transform of the $6$-point MHV graviton amplitude and extract the celestial OPE for positive helicity outgoing gravitons upto first few subleading orders.  We find that this OPE can indeed be organised according to representations of the extended symmetry algebra that we referred to above. In this case we also encounter null state relations which give rise to differential equations for the $5$-point Mellin amplitude. These differential equations play a fundamental role in the analysis which we undertake in the following sections of the paper. Our findings from this section are summarised in Section \ref{csope6ptfinal}.

 \vskip 4pt
In Section \ref{NULL} we show that the local symmetry algebra under consideration admits null states. We consider in particular two such null states which were independently found by studying the OPE decomposition of the $6$-point MHV amplitude in the previous section. We then impose that these null states must also decouple from the Mellin transform of $n$-point MHV amplitudes. In turn this gives rise to partial differential equations for these $n$-point amplitudes. These differential equations can also be translated directly to the Fock space MHV graviton amplitudes as we describe in Section \ref{fspde}. We then turn to explore some of the remarkable implications of the above mentioned differential equations and the  local symmetry algebra for the celestial OPE of graviton primary operators in the ``MHV-sector''. In Sections \ref{genope},  \ref{leadopede} and \ref{subleadopede}, we show that these equations can be used to completely determine the structure the of leading OPE.  We also illustrate via particular examples how the subleading OPE coefficients can be systematically obtained by demanding both sides of the OPE under the action of the extended local symmetry algebra generators. 

 \vskip 4pt
We conclude the paper with several Appendices in Section \ref{appendix} which collect some of the results that have been used in various sections of this paper. Amongst these we would like to highlight in particular Section \ref{5direct} and Section \ref{hodge} where we present a direct check of one of the null state decoupling relations from Section \ref{NULL} for the Mellin transform of the $5$-point MHV graviton amplitude and prove the other null state decoupling relation for the Mellin transform of any $n$-point MHV graviton amplitude. 


\section{MHV-sector of the Celestial CFT}\label{mhvsector}

In this paper we consider the mostly plus MHV graviton scattering amplitudes \\
$\langle{1^{-}2^- 3^+ 4^+\cdots n^+}\rangle$. In GR, at tree-level, the scattering amplitudes with only one or no negative helicity gravitons vanish, i.e, 
\be\label{GR}
\langle{1^{-}2^+ 3^+ 4^+\cdots n^+}\rangle = 0, \quad \langle{1^+2^+ 3^+ 4^+\cdots n^+}\rangle = 0
\ee
This has the following consequences : 
\begin{enumerate}
\item  Consider the collinear limit of two gravitons. We write the factorisation channels schematically as \cite{Bern:1998sv},
\be\label{coll}
+ + \rightarrow + , \quad + - \rightarrow (-) + (+) , \quad - - \rightarrow - 
\ee
This, together with \eqref{GR}, imply that \textit{the set of MHV amplitudes is \ul{closed under the collinear limit}.} Now since the collinear limit corresponds to the OPE limit in the Celestial CFT, the corresponding statement in the celestial CFT is that the \textit{set of MHV amplitudes is \underline{closed under OPE}.} 

\item Now one can use the correspondence between soft theorems and Ward identities \cite{Strominger:2013jfa,Strominger:2013lka,Kapec:2016jld,He,Strominger:2014pwa} to determine the symmetries of the \textit{set of all MHV amplitudes.} So we can make a positive helicity graviton soft and the lower point amplitude that we get due to soft factorisation is again a MHV amplitude. On the contrary, if we make a negative helicity graviton soft then, due to \eqref{GR}, the amplitude vanishes. This means that \textit{there is \underline{no} negative helicity soft graviton in the MHV sector.} This simplifies the structure of the symmetry algebra significantly. As we will see, the subleading soft graviton theorem for a positive helicity soft graviton can be cast as the Ward identity for $\overline{SL(2,\mathbb C)}$ current algebra and we also have the supertranslations coming from the positive helicity soft graviton. They together form a closed algebra and can be thought of as the symmetry algebra in the MHV sector of the celestial CFT. 

\end{enumerate}

Now the graviton-graviton OPE extracted from MHV amplitudes, holds as operator statements \underline{only} in the MHV sector of the Celestial CFT. This is of course true by construction. But, what is most surprising is that the OPE \textit{is completely determined by the $\overline{SL(2,\mathbb C)}$ current algebra together with the supertranslations coming from the positive helicity soft graviton.} By completely we mean including the leading OPE coefficients. This happens because this infinite dimensional symmetry algebra has null states whose decoupling forces the MHV amplitudes to satisfy certain linear first order partial differential equations. One can get the leading OPE coefficients from these equations. 

\vskip 4pt
We believe that these facts should be interpreted in the following way : 
\vskip 5pt
At tree-level, the Celestial CFT has \textit{an \underline{autonomous} sector which holographically computes the MHV graviton scattering amplitudes and is governed by the $\overline{SL(2,\mathbb C)}$ current algebra and supertranslation symmetry coming from positive helicity soft gravitons.} This sector is neither local nor unitary. But, it is very likely that it is exactly solvable in the same way as minimal models of $2$-D CFT are exactly solvable. 

\vskip 4pt
Before we end this section let us point out that the OPE is particularly simple in the MHV sector. Because of \eqref{GR} the OPE in this sector is schematically given by, 
\be\label{mhvope}
+ + \sim + , \quad + - \sim - , \quad - - \sim 0
\ee

We will see that \eqref{mhvope} also independently follows from the invariance of the OPE under the infinite dimensional local symmetry algebra. 

\section{$\overline{SL(2,\mathbb C)}$ current algebra from the subleading soft graviton theorem}
\label{curralgeb}

Let us start with the subleading soft graviton theorem \cite{Cachazo:2014fwa} for a positive helicity soft graviton in Mellin space \footnote{For a brief review of Mellin amplitudes for massless particles and notation used in the paper, please see the Appendix \eqref{mellin}. We are also setting $\kappa = \sqrt{32\pi G_N}=2$.},
\be\label{ss}
\big\langle{S^+_1(z,\bar z)} \prod_{i=1}^n \phi_{h_i,\bar h_i}(z_i,\bar z_i)\big\rangle =  \sum_{k=1}^n \frac{(\bar z- \bar z_k)^2}{z-z_k}  \[\frac{2\bar h_k}{\bar z- \bar z_k} - \bar\partial_k\] \langle{\prod_{i=1}^n \phi_{h_i, \bar h_i}(z_i,\bar z_i)}\rangle
\ee
where
\be
S^+_1(z,\bar z) = \lim_{\D\rightarrow 0} \D G^+_{\D}(z,\bar z)
\ee
is the subleading conformally soft \cite{Donnay:2018neh,Pate:2019mfs,Fan:2019emx,Nandan:2019jas,Adamo:2019ipt,Puhm:2019zbl,Guevara:2019ypd} graviton and
\be
\bar h_k = \frac{\D_k -\sigma_k}{2} 
\ee

We treat $z$ and $\bar z$ as independent complex variables and follow a procedure similar to what was applied to subsubleading soft graviton theorem in \cite{Pate:2019lpp}. The only difference is that we are also allowing the (singular) local transformations. The basic idea is the following. The structure of the soft factor allows us to think of the soft operator as a generating function of currents.  For example, we can see that the R.H.S of \eqref{ss} is a polynomial in the coordinate $\bar z$ of the subleading soft operator $S^+_1(z,\bar z)$. So it makes sense to expand the soft operator $S^+_1(z,\bar z)$ in powers of $\bar z$ and then from \eqref{ss} it follows that the expansion coefficients are conserved currents whose correlation functions with a collection of primary operators are already contained in the soft theorem. We will also apply the same procedure to the leading soft operator. 

 \vskip 4pt
So we start by expanding the R.H.S of \eqref{ss} in powers of $\bar z$
\be
\begin{gathered}
\big\langle{S^+_1(z,\bar z)} \prod_{i=1}^n \phi_{h_i,\bar h_i}(z_i,\bar z_i)\big\rangle \\ = - \sum_{k=1}^n \frac{\bar z_k^2 \bar\partial_k + 2\bar h_k \bar z_k}{z- z_k} \ \langle{\prod_{i=1}^n \phi_{h_i,\bar h_i}(z_i,\bar z_i)}\rangle + 2 \bar z \sum_{k=1}^n \frac{\bar h_k + \bar z_k \bar\partial_k}{z- z_k} \ \langle{\prod_{i=1}^n \phi_{h_i,\bar h_i}(z_i,\bar z_i)}\rangle \\ - \bar z^2 \sum_{k=1}^n \frac{1}{z- z_k} \frac{\partial}{\partial \bar z_k} \ \langle{\prod_{i=1}^n \phi_{h_i,\bar h_i}(z_i,\bar z_i)}\rangle 
\end{gathered}
\ee

Using this we define three currents $J^a(z)$ where $a= 0, \pm 1$ whose correlation functions are given by, 
\be
\big\langle{J^1(z)} \prod_{i=1}^n \phi_{h_i,\bar h_i}(z_i,\bar z_i)\big\rangle = \sum_{k=1}^n \frac{\bar z_k^2 \bar\partial_k + 2\bar h_k \bar z_k}{z- z_k} \ \langle{\prod_{i=1}^n \phi_{h_i,\bar h_i}(z_i,\bar z_i)}\rangle = \sum_{k=1}^n \frac{\bar L_1(k)}{z- z_k} \ \langle{\prod_{i=1}^n \phi_{h_i,\bar h_i}(z_i,\bar z_i)}\rangle
\ee

\be
\big\langle{J^0(z)} \prod_{i=1}^n \phi_{h_i,\bar h_i}(z_i,\bar z_i)\big\rangle = \sum_{k=1}^n \frac{\bar h_k + \bar z_k \bar\partial_k}{z- z_k} \ \langle{\prod_{i=1}^n \phi_{h_i,\bar h_i}(z_i,\bar z_i)}\rangle = \sum_{k=1}^n \frac{\bar L_0(k)}{z- z_k} \ \langle{\prod_{i=1}^n \phi_{h_i,\bar h_i}(z_i,\bar z_i)}\rangle
\ee

\be
\big\langle{J^{-1}(z)} \prod_{i=1}^n \phi_{h_i,\bar h_i}(z_i,\bar z_i)\big\rangle = \sum_{k=1}^n \frac{1}{z- z_k} \frac{\partial}{\partial \bar z_k} \ \langle{\prod_{i=1}^n \phi_{h_i,\bar h_i}(z_i,\bar z_i)}\rangle = \sum_{k=1}^n \frac{\bar L_{-1}(k)}{z- z_k} \ \langle{\prod_{i=1}^n \phi_{h_i,\bar h_i}(z_i,\bar z_i)}\rangle
\ee

where we have defined the differential operators, 
\be
\bar L_1(k) = \bar z_k^2 \bar\partial_k + 2\bar h_k \bar z_k
\ee

\be
\bar L_0(k) = \bar h_k + \bar z_k \bar\partial_k
\ee

and 
\be
\bar L_{-1}(k) = \frac{\partial}{\partial \bar z_k}
\ee

which are the generators of $\overline{SL(2,\mathbb C)}$ or the antiholomorphic Lorentz transformations and $J^a(z)$ are the corresponding currents. In terms of the currents we can write the soft graviton operator $S^+_1(z,\bar z)$ as, 
\be
S^+_1(z,\bar z) = - J^{1}(z) + 2\bar z J^{0}(z) - \bar z^2 J^{-1}(z) 
\ee

Therefore the soft operator is a generating function for the $\overline{SL(2,\mathbb C)}$ currents $J^a(z), \ a=0,\pm 1$. 

 \vskip 4pt
The modes of the currents $J^a(z)$, which generate local $\overline{SL(2,\mathbb C)}$ transformations, are defined in the standard way and their (classical) commutator is given by, 
\be\label{g0}
\[J^a_m, J^b_n\] = (a-b) J^{a+b}_{m+n}
\ee

The zero-modes or the global symmetry generators are given by $J^1_0 = \bar L_1, J^0_0 = \bar L_0$ and $J^{-1}_0= \bar L_{-1}$ which are the generators of the antiholomorphic Lorentz transformations. We will be working with this level zero $\overline{SL(2,\mathbb C)}$ current algebra.  

 \vskip 4pt
The commutator of a primary and the current is given by, 
\be\label{gcm}
\begin{gathered}
\[J^1_n, \phi_{h,\bar h}(z,\bar z)\] = z^n \(\bar z^2 \bar\partial + 2\bar h \bar z\)\phi_{h,\bar h}(z,\bar z) \\
\[J^0_n, \phi_{h,\bar h}(z,\bar z)\] = z^n \(\bar z \bar\partial + \bar h\)\phi_{h,\bar h}(z,\bar z) \\
\[J^{-1}_n, \phi_{h,\bar h}(z,\bar z)\] = z^n \bar\partial\phi_{h,\bar h}(z,\bar z)
\end{gathered}
\ee

Now note that here for simplicity we have expanded the soft operator around $\bar z=0$ but we could have expanded around an arbitrary point in the $\bar z$ plane. The effect of this is nothing but to translate the current $J^a(z)$ from the point $\bar z=0$ to the point in the $\bar z$ plane around which we are expanding the soft operator. In fact for the purpose of operator product expansion we will expand it around points other than $\bar z=0$. In the general case the expansion around a fixed point $\bar z = \bar z'$ gives,
\be
S^+_1(z,\bar z) = - J^{1}(z, \bar z') + 2(\bar z - \bar z') J^{0}(z, \bar z') - (\bar z - \bar z')^2 J^{-1}(z,\bar z') 
\ee

The new and the old currents are related by conjugation by the translation operator $exp(- \bar z' \bar L_{-1})$ and so none of the operator relations change. Similarly, the modes of the current $J^a(z,\bar z')$ can be defined with respect to some fixed point $z=z'$ other than $z=0$ and so in general we should denote the modes by $J^a_n(z',\bar z')$. They satisfy the same commutation relation 
\be
[J^a_m(z',\bar z'), J^b_n(z',\bar z')] = (a-b) J^{a+b}_{m+n}(z',\bar z')
\ee

because commutation relations do not change under conjugation. The generators $J^a_n$ defined in \eqref{g0} or \eqref{gcm} is given by $J^a_n = J^a_n(z'=0, \bar z'=0)$.

\subsection{Interpretation as diffeomorphism}

Here we want to give a geometric interpretation to this algebra because it may be useful in order to interpret this as an asymptotic symmetry \cite{Campiglia:2014yka,Donnay:2020guq,Campiglia:2020qvc,Compere:2018ylh}. So let us consider infinitesimal diffeomorphisms of the form,
\be
z \rightarrow z, \quad \bar z \rightarrow \bar z + A(z) + B(z) \bar z + C(z)\bar z^2
\ee

where $A,B,C$ are meromorphic functions. We can do a mode expansion of the functions and define the following vector fields which form a basis,
\be
J^{-1}_n = - z^{n} \frac{d}{d\bar z}, \quad J^{0}_n = - z^{n} \bar z \frac{d}{d\bar z}, \quad J^{1}_n = - z^{n} \bar z^2 \frac{d}{d\bar z}
\ee

The commutator of these vector fields is given by,
\be
[J^a_m, J^b_n] = (a-b) J^{a+b}_{m+n}, \quad a,b = -1,0,1
\ee

which is again the level zero $\overline{SL(2,\mathbb C)}$ current algebra. So this is a subalgebra of the algebra of diffeomorphisms on the plane, with analytic singularities.

\section{OPE between the subleading soft graviton and a conformal primary}\label{shope}

In order to construct the OPE between two hard gravitons it is very useful to know the OPE between a soft graviton and a hard graviton. Since a hard graviton can be converted into a soft graviton by taking conformal soft limit \cite{Donnay:2018neh,Pate:2019mfs,Fan:2019emx,Nandan:2019jas,Adamo:2019ipt,Puhm:2019zbl,Guevara:2019ypd}, the OPE between soft and hard acts as a boundary condition which always needs to be satisfied \cite{Banerjee:2020kaa}. Secondly, the OPE between the soft operator and a hard operator also clarifies the definition of the primary state with respect to the extended symmetry algebra. 

 \vskip 4pt
This OPE can be constructed from the subleading soft theorem by bringing the soft graviton close to the hard operator. So we start from the subleading soft theorem \eqref{ss} 
\be
\big\langle{S^+_1(z,\bar z)} \prod_{i=1}^n \phi_{h_i,\bar h_i}(z_i,\bar z_i)\big\rangle = - \sum_{k=1}^n \frac{\(\bar z_k - \bar z\)^2 \bar\partial_k + 2\bar h_k\( \bar z_k - \bar z\)}{z- z_k} \ \langle{\prod_{i=1}^n \phi_{h_i,\bar h_i}(z_i,\bar z_i)}\rangle
\ee

and suppose we want to know the OPE between $S^+_1(z,\bar z)$ and $\phi_{h_1,\bar h_1}(z_1,\bar z_1)$. So we simply have to expand the subleading soft factor in powers of $(z-z_1)$ and $(\bar z - \bar z_1)$ and there will be both singular and non-singular terms. Doing this expansion we get, 
\be\label{se}
\begin{gathered}
\big\langle{S^+_1(z,\bar z)} \prod_{i=1}^n \phi_{h_i,\bar h_i}(z_i,\bar z_i)\big\rangle \\ 
= - \( \frac{0}{z - z_1} + \sum_{p=1}^{\infty} (z - z_1)^{p-1} \mathcal J^1_{-p}(z_1,\bar z_1)\) \ \langle{\prod_{i=1}^n \phi_{h_i,\bar h_i}(z_i,\bar z_i)}\rangle \\ 
+ 2(\bar z - \bar z_1) \( \frac{\bar h_1}{z-z_1} + \sum_{p=1}^{\infty} (z-z_1)^{p-1} \mathcal J^0_{-p}(z_1,\bar z_1) \) \big\langle{\prod_{i=1}^n \phi_{h_i,\bar h_i}(z_i,\bar z_i)}\big\rangle \\
-(\bar z - \bar z_1)^2  \( \frac{1}{z- z_1} \frac{\partial}{\partial \bar z_1}+ \sum_{p=1}^{\infty} (z - z_1)^{p-1} \mathcal J^{-1}_{-p}(z_1,\bar z_1)\) \big\langle{ \prod_{i=1}^n \phi_{h_i,\bar h_i}(z_i,\bar z_i)}\big\rangle
\end{gathered}
\ee

where the differential operators $\mathcal J^a_{-p}(z_1,\bar z_1)$ are defined as, 
\be\label{nu1}
\begin{gathered}
\mathcal J^{1}_{-p}(z_1,\bar z_1) \langle{\prod_{i=1}^n \phi_{h_i,\bar h_i}(z_i,\bar z_i)}\rangle = \langle{ \(J^{1}_{-p}\phi_{h_1,\bar h_1}\)(z_1, \bar z_1)\prod_{i=2}^n \phi_{h_i,\bar h_i}(z_i,\bar z_i)}\rangle \\ =  -  \sum_{k\ne 1} \frac{\(\bar z_k - \bar z_1\)^2 \bar\partial_k + 2\bar h_k\( \bar z_k - \bar z_1\)}{\(z_k- z_1\)^{p}} \ \langle{\prod_{i=1}^n \phi_{h_i,\bar h_i}(z_i,\bar z_i)}\rangle
\end{gathered}
\ee

\be\label{nu2}
\begin{gathered}
\mathcal J^{0}_{-p}(z_1,\bar z_1) \langle{\prod_{i=1}^n \phi_{h_i,\bar h_i}(z_i,\bar z_i)}\rangle = \langle{ \(J^{0}_{-p}\phi_{h_1,\bar h_1}\)(z_1, \bar z_1)\prod_{i=2}^n \phi_{h_i,\bar h_i}(z_i,\bar z_i)}\rangle \\ 
= -\sum_{k\ne 1} \frac{\bar h_k + \(\bar z_k - \bar z_1\) \bar\partial_k}{(z_k - z_1)^{p}} \big\langle{\prod_{i=1}^n \phi_{h_i,\bar h_i}(z_i,\bar z_i)}\big\rangle
\end{gathered}
\ee

and 
\be\label{nu3}
\begin{gathered}
\mathcal J^{-1}_{-p}(z_1,\bar z_1) \langle{\prod_{i=1}^n \phi_{h_i,\bar h_i}(z_i,\bar z_i)}\rangle = \langle{ \(J^{-1}_{-p}\phi_{h_1,\bar h_1}\)(z_1, \bar z_1)\prod_{i=2}^n \phi_{h_i,\bar h_i}(z_i,\bar z_i)}\rangle \\
= - \( \sum_{k\ne 1} \frac{1}{(z_k - z_1)^{p}}\frac{\partial}{\partial \bar z_k}\) \big\langle{ \prod_{i=1}^n \phi_{h_i,\bar h_i}(z_i,\bar z_i)}\big\rangle
\end{gathered}
\ee

Here the notation is standard. We have defined the current algebra descendant \\ $\(J^a_{-p}\phi_{h_1,\bar h_1}\)(z_1,\bar z_1)$, $ p> 0$,  whose correlation function with a collection of primary operators is given by the above equations. The residues of the single pole terms in $(z-z_1)$, appearing in \eqref{se}, has the following explanation. The residues are just the operators $J^a_0\phi_{h_1,\bar h_1}(z_1,\bar z_1)$. Now
\be
J^1_0(z_1,\bar z_1) = \bar L_1(\bar z_1), \quad J^0_0(z_1,\bar z_1) = \bar L_0(\bar z_1), \quad J^{-1}_0(z_1,\bar z_1) = \bar L_{-1}(\bar z_1)
\ee

and so
\be
J^1_0 \phi_{h_1,\bar h_1}(z_1,\bar z_1) = 0, \hspace{0.2cm} J^0_0 \phi_{h_1,\bar h_1}(z_1,\bar z_1) = \bar h_1 \phi_{h_1,\bar h_1}(z_1,\bar z_1), \hspace{0.2cm}  J^{-1}_0 \phi_{h_1,\bar h_1}(z_1,\bar z_1) = \frac{\partial}{\partial \bar z_1}\phi_{h_1,\bar h_1}(z_1,\bar z_1)
\ee

This is exactly what we have obtained as the residue of the pole term in $(z-z_1)$ in \eqref{se}. Now \eqref{se} can be written in the form of OPE as, 
\be\label{se1}
\begin{gathered}
S^+_1(z,\bar z) \phi_{h_1,\bar h_1}(z_1,\bar z_1) \\ 
= - \sum_{p=1}^{\infty} (z - z_1)^{p-1} \(J^1_{-p}\phi_{h_1,\bar h_1}\)(z_1,\bar z_1) \\ 
+ 2(\bar z - \bar z_1) \( \frac{\bar h_1}{z-z_1} \phi_{h_1,\bar h_1}(z_1,\bar z_1)+ \sum_{p=1}^{\infty} (z-z_1)^{p-1} \(J^0_{-p}\phi_{h_1,\bar h_1}\)(z_1,\bar z_1) \) \\
-(\bar z - \bar z_1)^2  \( \frac{1}{z- z_1} \frac{\partial}{\partial \bar z_1} \phi_{h_1,\bar h_1}(z_1,\bar z_1)+ \sum_{p=1}^{\infty} (z - z_1)^{p-1} \(J^{-1}_{-p}\phi_{h_1,\bar h_1}\)(z_1,\bar z_1)\)
\end{gathered}
\ee

If we compute the OPE between a positive helicity graviton $G^+_{\Delta}(z,\bar z)$ and a matter field $\phi_{h_1,\bar h_1}(z_1,\bar z_1)$, which can be another graviton, then in the subleading conformal soft limit we must have,
\be
\lim_{\Delta\rightarrow 0} \Delta G^+_{\Delta}(z,\bar z) \phi_{h_1,\bar h_1}(z_1,\bar z_1) = \eqref{se1}
\ee 

This is an important constraint which needs to be satisfied by any (tree-level) OPE involving a positive helicity graviton. 

Another thing we want to point out is that the OPE \eqref{se1} implies that for any conformal primary $\phi_{h,\bar h}(z,\bar z)$ we have, 
\be
\boxed{
\(J^a_n\phi_{h,\bar h}\)(z,\bar z) =0}, \quad \forall n>0, \quad a = 0, \pm 1
\ee

For $a=1$ we also have the additional condition 
\be
\(J^1_0 \phi_{h,\bar h}\)(z,\bar z) = 0 
\ee

which is nothing but $\bar L_1(\bar z)\phi_{h,\bar h}(z,\bar z)=0$.

 \vskip 4pt
This means that the \textit{conformal primaries are also primaries of the $\overline{SL(2,\mathbb C)}$ current algebra} in the standard sense. We want to emphasise the word "standard sense" because the generators $J^a_n$ are not quite standard because they are not purely holomorphic. The antiholomorphic scaling dimension of $J^1_n$ is $-1$ and that of $J^{-1}_n$ is $+1$. The generator $J^0_n$ has antiholomorphic scaling dimension 0. 

 \vskip 4pt
Before we end this section let us mention that in the intermediate stages of our calculation we take the conformal primaries to be functions of time \cite{Banerjee:2018gce,Banerjee:2018fgd,Banerjee:2019prz} also and denote them by $\phi_{h,\bar h}(u,z,\bar z)$. With the introduction of time, the new formulas for the differential operators \eqref{nu1} and \eqref{nu2} are obtained by making the following replacement
\be\label{replace}
\bar h_i \rightarrow  \bar h_i + \frac{1}{2} (u_i - u_1) \frac{\partial}{\partial u_i}
\ee

This can be obtained from the subleading soft graviton theorem written in Mellin space in the presence of the time coordinate and is discussed in the Appendix \eqref{sstt}.  

\subsection{Commutator with $SL(2,\mathbb C)$ generators}

The generators of $SL(2,\mathbb C)$ are given by $\{L_0, L_{\pm1}\}$ with the commutator algebra,
\be
\[L_m,L_n\] = (m-n) L_{m+n}, \qquad m,n = 0, \pm1
\ee

Note that $\{L_0, L_{\pm 1}\}$ which generate Lorentz transformations, do not belong to the $\overline{SL(2,\mathbb{C})}$ current algebra and so we need to specify their action and commutator separately.  These generators act on a primary field $\phi_{h_i,\bar h_i}(z_i,\bar z_i)$ according to, 
\be
\[L_n, \phi_{h_i,\bar h_i}(z_i,\bar z_i)\] = \( z_i^{n+1} \partial_i + (n+1) h_i z_i^n \) \phi_{h_i,\bar h_i}(z_i,\bar z_i)
\ee
where 
\be
2 h_i = \D_i + \sigma_i 
\ee

The commutator between the $SL(2,\mathbb C)$ generators $\{L_0, L_{\pm1}\}$ and the generators $J^a_n$ of the $\overline{SL(2,\mathbb C)}$ current algebra is given by, 
\be
[ L_m, J^a_n] = -n J^a_{m+n}, \quad m = 0, \pm 1
\ee

\section{Leading soft theorem and supertranslation generators} \label{leadsoft}

The leading soft theorem \cite{Weinberg:1965nx} can be written as \cite{Banerjee:2018fgd},\footnote{In \eqref{ws} we keep the time coordinate because it is more convenient. If we do not want to keep it then we just have to replace the $i \partial / \partial u_k$ by $\epsilon_k \mathcal P_k$ where $\mathcal P_k \phi_{h_i,\bar h_i}(z_i,\bar z_i) = \delta_{ki}\phi_{h_i+ 1/2, \bar h_i + 1/2}(z_i,\bar z_i)$ and $\epsilon_k = \pm 1$ for an outgoing (incoming) particle.} 
\be\label{ws}
\big\langle{S^+_0(z,\bar z)} \prod_{i=1}^n \phi_{h_i,\bar h_i}(u_i,z_i,\bar z_i)\big\rangle = - \(\sum_{k=1}^n \frac{\bar z - \bar z_k}{z - z_k} i\frac{\partial}{\partial u_k}\) \big\langle{\prod_{i=1}^n \phi_{h_i,\bar h_i}(u_i,z_i,\bar z_i)}\big\rangle
\ee

where $S^+_0(z,\bar z)$ is the leading soft graviton which we take to be outgoing. The following discussion is essentially identical to that in the case of subleading soft graviton and so we discuss only the essential formulas which will be used in the paper. 

 \vskip 4pt
We start by expanding the R.H.S of \eqref{ws} in powers of $\bar z$,
\be
\begin{gathered}
\big\langle{S^+_0(z,\bar z)} \prod_{i=1}^n \phi_{h_i,\bar h_i}(u_i,z_i,\bar z_i)\big\rangle \\
= \(\sum_{k=1}^n \frac{\bar z_k}{z - z_k} i\frac{\partial}{\partial u_k}\) \big\langle{\prod_{i=1}^n \phi_{h_i,\bar h_i}(u_i,z_i,\bar z_i)}\big\rangle - \bar z \(\sum_{k=1}^n \frac{1}{z - z_k} i\frac{\partial}{\partial u_k}\) \big\langle{\prod_{i=1}^n \phi_{h_i,\bar h_i}(u_i,z_i,\bar z_i)}\big\rangle
\end{gathered}
\ee

This leads us to define two currents $P_0(z)$ and $P_{-1}(z)$, i.e,
\be
S^+_0(z,\bar z) = P_0(z) - \bar z P_{-1}(z)
\ee

whose correlation functions or Ward identities are given by, 
\be\label{c1}
\big\langle{P_0(z) \prod_{i=1}^n \phi_{h_i,\bar h_i}(u_i,z_i,\bar z_i)}\big\rangle = \(\sum_{k=1}^n \frac{\bar z_k}{z - z_k} i\frac{\partial}{\partial u_k}\) \big\langle{\prod_{i=1}^n \phi_{h_i,\bar h_i}(u_i,z_i,\bar z_i)}\big\rangle
\ee

and 
\be\label{c2}
\big\langle{P_{-1}(z) \prod_{i=1}^n \phi_{h_i,\bar h_i}(u_i,z_i,\bar z_i)}\big\rangle = \(\sum_{k=1}^n \frac{1}{z - z_k} i\frac{\partial}{\partial u_k}\) \big\langle{\prod_{i=1}^n \phi_{h_i,\bar h_i}(u_i,z_i,\bar z_i)}\big\rangle
\ee

We can see from the Ward identities \eqref{c1} and \eqref{c2} that $P_0(z)$ and $P_{-1}(z)$ generate the infinitesimal global symmetries
\be
\delta_{\xi}\phi_{h,\bar h}(u,z,\bar z) = \xi \bar z  i\frac{\partial}{\partial u}\phi_{h,\bar h}(u,z,\bar z)
\ee

and 
\be
\delta_{\xi}\phi_{h,\bar h}(u,z,\bar z) = \xi  i\frac{\partial}{\partial u}\phi_{h,\bar h}(u,z,\bar z)
\ee

These are the transformation laws of the fields under the global space-time translations given by $u\rightarrow u + \xi \bar z$ and $u\rightarrow u + \xi$, respectively. 

 \vskip 4pt
Now we can define the modes of the currents as $P_{n,0}$ and $P_{n,-1}$ which are the generators of the supertranslation and they commute among themselves. Their actions on a primary operator are given by \cite{Banerjee:2018gce,Banerjee:2018fgd},
\be
\[P_{n,0}, \phi_{h,\bar h}(u,z,\bar z)\] = z^{n+1}\bar z i\frac{\partial}{\partial u} \phi_{h,\bar h}(u,z,\bar z)
\ee
and 
\be
\[P_{n,-1}, \phi_{h,\bar h}(u,z,\bar z)\] = z^{n+1} i\frac{\partial}{\partial u} \phi_{h,\bar h}(u,z,\bar z)
\ee

If we take the conformal primaries to be time-independent then the action of the time derivative is given by,
\be\label{sub}
i \frac{\partial}{\partial u} \phi^{\epsilon}_{h,\bar h}(u,z,\bar z) \rightarrow \epsilon \phi^{\epsilon}_{h+\frac{1}{2},\bar h +\frac{1}{2}}(z,\bar z) = \epsilon \mathcal P \phi^{\epsilon}_{h,\bar h}(z,\bar z)
\ee

where $\epsilon= \pm 1$ for an outgoing (incoming) particle and $\mathcal P \phi^{\epsilon}_{h,\bar h}(z,\bar z)= \phi^{\epsilon}_{h+1/2, \bar h + 1/2}(z,\bar z)$. In other words, the above commutators should be written as,
\be
\[P_{n,0}, \phi^{\epsilon}_{h,\bar h}(z,\bar z)\] = \epsilon z^{n+1}\bar z \phi^{\epsilon}_{h+\frac{1}{2},\bar h+\frac{1}{2}}(z,\bar z) = \epsilon z^{n+1}\bar z \mathcal P\phi^{\epsilon}_{h,\bar h}(z,\bar z)
\ee

and
\be
\[P_{n,-1}, \phi^{\epsilon}_{h,\bar h}(z,\bar z)\] = \epsilon z^{n+1} \phi^{\epsilon}_{h+\frac{1}{2},\bar h+\frac{1}{2}}(z,\bar z) = \epsilon z^{n+1} \mathcal P \phi^{\epsilon}_{h,\bar h}(z,\bar z)
\ee 

Note that every primary carries the additional index $\epsilon$ whose value tells us whether it is incoming or outgoing. In this paper we have mostly kept it implicit unless its specification is absolutely necessary for our purpose. 


\section{OPE between the leading soft graviton and a conformal primary }\label{opestcp}

Proceeding in the same way as in the case of subleading soft theorem we get the following OPE between the leading soft operator and a conformal primary field,
\be\label{wsope}
\begin{gathered}
S^+_0(z,\bar z) \phi_{h_1,\bar h_1}(u_1,z_1,\bar z_1) \\ 
=  \sum_{a=2}^{\infty} (z- z_1)^{a-2} \(P_{-a, 0}\phi_{h_1,\bar h_1}\)(u_1,z_1,\bar z_1) \\
 - (\bar z - \bar z_1) \(\frac{1}{z - z_1} i \frac{\partial}{\partial u_1}\phi_{h_1,\bar h_1}(u_1, z_1,\bar z_1) + \sum_{a=2}^{\infty} (z- z_1)^{a-2} \(P_{-a, -1}\phi_{h_1,\bar h_1}\)(u_1,z_1,\bar z_1)\)
 \end{gathered}
\ee

where the correlation functions of the supertranslation descendants are given by,
\be\label{sd1}
\begin{gathered}
 \big\langle{\(P_{-a,0}\phi_{h_1,\bar h_1}\)(u_1,z_1,\bar z_1)\prod_{i=2}^n \phi_{h_i,\bar h_i}(u_i,z_i,\bar z_i)}\big\rangle = \mathcal P_{-a,0}(u_1,z_1,\bar z_1)\big\langle{\prod_{i=1}^n \phi_{h_i,\bar h_i}(u_i,z_i,\bar z_i)}\big\rangle \\ = - \sum_{k\ne 1} \frac{\bar z_k - \bar z_1}{(z_k - z_1)^{a-1}} i \frac{\partial}{\partial u_k}\big\langle{\prod_{i=1}^n \phi_{h_i,\bar h_i}(u_i,z_i,\bar z_i)}\big\rangle
\end{gathered}
\ee

and \cite{Banerjee:2020kaa}
\be\label{p}
\begin{gathered}
\big\langle{ \(P_{-a,-1}\phi_{h_1,\bar h_1}\)(u_1,z_1,\bar z_1) \prod_{i=2}^n \phi_{h_i,\bar h_i}(u_i,z_i,\bar z_i)}\big\rangle  = \mathcal P_{-a,-1}(u_1,z_1,\bar z_1)\big\langle{\prod_{i=1}^n \phi_{h_i,\bar h_i}(u_i,z_i,\bar z_i)}\big\rangle \\ = -\sum_{k\ne 1}\frac{1}{(z_k - z_1)^{a-1}} i \frac{\partial}{\partial u_k}\big\langle{\prod_{i=1}^n \phi_{h_i,\bar h_i}(u_i,z_i,\bar z_i)}\big\rangle
\end{gathered}
\ee

In the absence of the time coordinate the time derivatives are replaced by the substitution \eqref{sub}. 

 \vskip 4pt
Now if we compute the OPE between a positive helicity outgoing graviton $G^+_{\D}(z,\bar z)$ and a conformal primary then the following constraint serves as a boundary condition which always needs to be satisfied, 
\be
\label{lsoftopecond}
\lim_{\D\rightarrow 1} (\D -1) G^+_{\D}(z_,\bar z) \phi_{h_1,\bar h_1}(z_1,\bar z_1) = \eqref{wsope}
\ee

Here we have taken the leading conformal soft limit \cite{Donnay:2018neh,Pate:2019mfs,Fan:2019emx,Nandan:2019jas,Adamo:2019ipt,Puhm:2019zbl,Guevara:2019ypd} on the graviton operator.  

 \vskip 4pt
Now the OPE \eqref{wsope} unambiguously implies that the following relations hold, 
\be
\boxed{
\(P_{n,0}\phi_{h,\bar h}\)(z,\bar z) = 0}, \quad \forall n \ge -1
\ee
and 
\be
\boxed{
\(P_{n,-1}\phi_{h,\bar h}\)(z,\bar z) = 0}, \quad \forall n > -1
\ee

We can see that although the supertranslation generator $P_{n,0}$ has \textit{negative} antiholomorphic scaling dimension given by $-\frac{1}{2}$, $P_{n,0}$, for $n\le -2$, does \textit{not} annihilate a primary operator. Rather, acting on a primary operator, $P_{n,0}$ creates a descendant whose correlation function with a collection of other primaries is given by \eqref{sd1}. We will see that the generators $P_{-2,0}$ and $J^1_{-1}$ play a central role in the later part of the paper. 


\section{Commutator between supertranslation and $\overline{SL(2,\mathbb C)}$ current algebra generators}\label{mxcom}
Let us start with the supertranslation generators $P_{m,n}$ which acts on a conformal primary field as \cite{Banerjee:2018gce,Banerjee:2018fgd}, 
\be
[P_{m,n}, \phi_{h,\bar h}(u,z,\bar z)] = i z^{m+1} {\bar z}^{n+1} \partial_u \phi_{h,\bar h}(u,z,\bar z)
\ee
The supertranslation generators commute
\be
[P_{m,n}, P_{m',n'}] =0
\ee

The subset of generators given by $P_{n,-1}$ and $P_{n,0}$, coming from the positive helicity soft graviton, form a closed algebra with the generators of the $\overline{SL(2,\mathbb C)}$ current algebra,
\be
\quad [J^1_m, P_{n,-1}] =  P_{m+n,0}, \quad  [J^0_m, P_{n,-1}] = \frac{1}{2} P_{m+n,-1}, \quad [J^{-1}_m, P_{n,-1}] = 0 
\ee

\be
\quad \[ J^1_m, P_{n,0}\] =0, \quad \[J^0_m, P_{n,0}\] = - \frac{1}{2}P_{m+n,0} , \quad [J^{-1}_m, P_{n,0}] = - P_{m+n,-1} 
\ee
 
Note that the global space-time translation generators, given by $\{P_{-1,-1}, P_{0,-1}, P_{-1,0}, P_{0,0}\}$, are part of the algebra generated by $\{P_{n,-1}, P_{n,0}, J^a_{n}\}$.  

 \vskip 4pt
Now the commutator of the supertranslations $\{P_{n,-1}, P_{n,0}\}$ with the $SL(2,\mathbb{C})$ generators, which are not part of the algebra generated by the positive helicity soft graviton, is given by
\be
\[ L_m, P_{n,a}\] = \( \frac{m-1}{2} - n\) P_{n+m,a}, \quad a = 0,-1
\ee

where $m= 0, \pm 1$.

\section{Summary : Extended symmetry algebra}
\label{extalgebsum}
For the sake of convenience of the reader, in this section we summarise the commutation relations and the definition of primary under the extended algebra coming from the positive helicity soft graviton. 
\subsection{$\overline{SL(2,\mathbb C)}$ current algebra}
\be
\[ J^a_m, J^b_n\] = (a-b) J^{a+b}_{m+n}, \quad a,b = 0, \pm 1, \quad m,n \in \mathbb{Z}
\ee

\be
J^1_0 = \bar L_1, \quad J^0_0 = \bar L_0, \quad J^{-1}_0 = \bar L_{-1}
\ee

\subsection{Global $SL(2,\mathbb C)$}
\be
\[ L_m, L_n\] = (m-n) L_{m+n}, \quad m,n = 0, \pm 1
\ee

\subsection{Supertranslation}
\be
\[ P_{m,n}, P_{m',n'}\] = 0, \quad n, n' = 0, -1
\ee

\subsection{Mixed commutators}
\be
\quad [J^1_m, P_{n,-1}] =  P_{m+n,0}, \quad  [J^0_m, P_{n,-1}] = \frac{1}{2} P_{m+n,-1}, \quad [J^{-1}_m, P_{n,-1}] = 0 
\ee 

\be
\quad \[ J^1_m, P_{n,0}\] =0, \quad \[J^0_m, P_{n,0}\] = - \frac{1}{2}P_{m+n,0} , \quad [J^{-1}_m, P_{n,0}] = - P_{m+n,-1} 
\ee

\be
[L_m, J^a_n] = -n J^a_{m+n}, \quad m = 0, \pm 1, \quad n\in \mathbb{Z}
\ee

\be
\[ L_n, P_{a,b}\] = \( \frac{n-1}{2} - a\) P_{a+n,b}, \quad n = 0, \pm 1, \quad b =0, -1
\ee

\subsection{Definition of Primary (under extended algebra)}
\be
\(J^a_n \phi_{h,\bar h}\)(z,\bar z) =0, \quad \forall n > 0, \quad a= 0, \pm 1
\ee

\be
\(J^1_0 \phi_{h,\bar h}\)(z,\bar z) = 0 = \bar L_1(\bar z) \phi_{h,\bar h}(z,\bar z) 
\ee

\be
L_1(z)\phi_{h,\bar h}(z,\bar z) = 0 
\ee

\be
\(P_{n,0}\phi_{h,\bar h}\)(z,\bar z) = 0 , \quad n \ge -1
\ee

\be
\(P_{n,-1}\phi_{h,\bar h}\)(z,\bar z) = 0 , \quad n \ge 0
\ee

\be
L_0(z) \phi_{h,\bar h}(z,\bar z) =  h \phi_{h,\bar h}(z,\bar z), \quad \bar L_0(\bar z) \phi_{h,\bar h}(z,\bar z) =  \bar h \phi_{h,\bar h}(z,\bar z)
\ee

\subsection{Scaling Dimensions of Generators}

\be
J^a_n \longrightarrow \( -n, -a\), \quad a = 0, \pm 1 
\ee

\be
P_{n,-1} \longrightarrow \(-n - \frac{1}{2}, \frac{1}{2}\)
\ee

\be
P_{n,0} \longrightarrow \(-n - \frac{1}{2}, - \frac{1}{2}\)
\ee

\be
L_n \longrightarrow (-n,0), \quad n = 0, \pm 1
\ee

\section{Differential equation for three graviton scattering amplitude}\label{ope4pt}
In this section we obtain a differential equation for the three graviton scattering amplitude in Mellin space \cite{Pasterski:2016qvg,Pasterski:2017kqt,Banerjee:2018gce,Banerjee:2019prz}, denoted by $\langle{G^-_{\D_1}(1)G^-_{\D_2}(2)G^+_{\D_4 }(4)}\rangle$ \footnote{Here $G^{\pm}_{\D}(i)$ denotes either $G^{\pm}_{\D}(z_i,\bar z_i)$ or $G^{\pm}_{\D}(u_i,z_i,\bar z_i)$ depending on how we choose to regulate the graviton Mellin amplitude in GR. But the following arguments do not depend on the choice of regulator as long as it respects all the symmetries of the problem. As we will see, these differential equations are ultimately determined by the underlying symmetry algebra. The only thing that depends on the choice of regulator is the explicit form of the differential operators which we have already explained in \eqref{replace}.}. This is obtained in the following way. We start with the four graviton scattering amplitude denoted by $\langle{G^-_{\D_1}(1)G^-_{\D_2}(2)G^+_{\D_3}(3)G^+_{\D_4}(4)}\rangle$ and perform the following operations :

\begin{enumerate}
\item We first take the OPE limit $3^+\rightarrow 4^+$ and then make the graviton $3^+$ conformally soft \cite{Donnay:2018neh,Pate:2019mfs,Fan:2019emx,Nandan:2019jas,Adamo:2019ipt,Puhm:2019zbl,Guevara:2019ypd}. 

\item Reverse the sequence, i.e, we first make the graviton $3^+$ conformally soft and then take the OPE limit $3^+\rightarrow 4^+$. The result of this is described by the soft theorems as discussed in sections \eqref{shope} and \eqref{opestcp}. 

Finally, we demand that these two operations, performed on the four graviton scattering amplitude, should yield the same result order by order in $z_{34}$ and $\bar z_{34}$. As we will see now this gives rise to non-trivial differential equation for the three graviton scattering amplitude.
\end{enumerate}

So let us start with the four graviton scattering amplitude in Mellin space
\be
\mathcal{M}_4(1^-2^-3^+4^+) = \langle{G^-_{\D_1}(1)G^-_{\D_2}(2)G^+_{\D_3}(3)G^+_{\D_4}(4)}\rangle
\ee 

where both the gravitons $3^+$ and $4^+$ are outgoing. In the OPE limit $3^+ \rightarrow 4^+$, the amplitude $\mathcal M_4(1^-2^-3^+4^+)$ factories as \cite{Banerjee:2020kaa},
\be\label{sope}
\begin{gathered}
\mathcal M_4(1^-2^-3^+4^+) \\ =  - B(i\l_3,i\l_4)\frac{\bar z_{34}}{z_{34}} \mathcal P_{-1,-1}(4) \mathcal M_3(1^-2^-4^+) \\ - B(i\l_3,i\l_4) \bar z_{34}\( \frac{i\l_4 - i\l_3}{i\l_4 + i\l_3} \ \mathcal P_{-2,-1}(4) + \frac{i\l_3}{i\l_3 + i\l_4} \  \mathcal L_{-1}(4) \mathcal P_{-1,-1}(4)\) \mathcal M_3(1^-2^-4^+) + \cdots
\end{gathered}
\ee

where we have defined
\be
\mathcal M_3(1^-2^-4^+) =  \langle{G^-_{\D_1}(1)G^-_{\D_2}(2)G^+_{\D_3 + \D_4 -1}(4)}\rangle
\ee

Now we take the subleading conformal soft limit on the graviton $3^+$ and using \eqref{sope}, we get
\be\label{ope4}
\begin{gathered}
\lim_{\D_3\rightarrow 0} \D_3 \mathcal{M}_4(1^-2^-3^+4^+)\\
= \bar z_{34}\[ \frac{2 \bar h_4}{z_{34}} \mathcal P_{-1,-1}(4) + \{(1+ i\lambda_4) \mathcal P_{-2,-1}(4) -  \mathcal L_{-1}(4) \mathcal P_{-1,-1}(4)\}\] \mathcal M'_3(1^-2^-4^+) + \cdots
\end{gathered}   
\ee

where 
\be
\mathcal M'_3(1^-2^-4^+) = \langle{G^-_{\D_1}(1)G^-_{\D_2}(2)G^+_{\D_4 -1}(4)}\rangle
\ee

But we also know from the discussion of the subleading soft graviton theorem in section \eqref{shope} that \eqref{ope4} must be equal to \eqref{se}, i.e, 
\be\label{ope4'}
\lim_{\D_3\rightarrow 0} \D_3 \mathcal{M}_4(1^-2^-3^+4^+) = \bar z_{34}\[ \frac{2 \bar h_4}{z_{34}} + 2 \mathcal J^0_{-1}(4) \]\langle{G^-_{\D_1}(1)G^-_{\D_2}(2)G^+_{\D_4}(4)}\rangle  + \cdots
\ee

So by equating \eqref{ope4} to \eqref{ope4'} we get,
\be\label{fde1}
\begin{gathered}
\( \D_4 \mathcal P_{-2,-1}(4) -  \mathcal L_{-1}(4) \mathcal P_{-1,-1}(4) \)\mathcal M'_3  = 2 \mathcal J^0_{-1}(4)\langle{G^-_{\D_1}(1)G^-_{\D_2}(2)G^+_{\D_4}(4)}\rangle  \\ = 2 \mathcal J^0_{-1}(4) \mathcal P_{-1,-1}(4) \langle{G^-_{\D_1}(1)G^-_{\D_2}(2)G^+_{\D_4-1}(4)}\rangle
\end{gathered}
\ee  

Now by shifting the dimension $\D_4\rightarrow \D_4 + 1$ we can write \eqref{fde1} as
\be\label{fde}
\boxed{
\bigg[ \mathcal L_{-1}(4) \mathcal P_{-1,-1}(4) + 2 \mathcal J^0_{-1}(4) \mathcal P_{-1,-1}(4) - (\D_4 +1) \mathcal P_{-2,-1}(4) \bigg] \langle{G^-_{\D_1}(1)G^-_{\D_2}(2)G^+_{\D_4}(4)}\rangle = 0}
\ee 

This is a first-order linear partial differential equation for the three graviton scattering amplitude where the positive helicity graviton is outgoing. One can check using the explicit expression for the differential operators, given in sections \eqref{shope} and \eqref{opestcp} and the three graviton scattering amplitude in Mellin space, that this equation is indeed satisfied. Note that the leading conformal soft limit does not produce anything new because the supertranslation generator $P_{-2,-1}$ already appears in the OPE \eqref{sope} and so it trivially satisfies the condition in equation \eqref{lsoftopecond}. 

 \vskip 4pt
Now \eqref{fde} is a decoupling equation which tells us that a certain linear combination of descendants of the graviton $4^+$ vanish. The vanishing condition must be a tensor equation so that it does not violate any of the symmetries of the theory. This means that the linear combination, which vanishes, must also be a primary of the symmetry algebra. Since the descendants generated by the (singular) local transformations appear in \eqref{fde}, we should \textit{look for null states or primary descendants of the extended symmetry algebra}. 

 \vskip 4pt
The relevance of null states of (conformally) \textit{soft} operators in the context of Celestial CFT were studied in \cite{Banerjee:2019aoy,Banerjee:2019tam}. The null states studied there were null states of the \textit{global} conformal group. The null states that we will study in this paper are null states of \textit{hard} operators under the \textit{local} infinite dimensional symmetry algebra. So they are supposed to be much more powerful, as we will see. 

\subsection{Limitations of the three point function} 

Let us now point out some of the limitations of the three point function which follows from energy-momentum conservation. We know that 
\be
\langle{G^-_{\D_1}(1)G^-_{\D_2}(2)G^+_{\D_4}(4)}\rangle \propto  \delta(\bar z_{14}) \delta(\bar z_{24})
\ee

Now let us consider a subset of supertranslation generators of the form $P_{-a,0}, a\ge 2$ or the $\overline{SL(2,\mathbb C)}$ current algebra generators $J^1_{-n}, n\ge 1$. Their actions on the three point function, following from \eqref{sd1} and \eqref{nu1}, are given by
\be\label{ls}
\begin{gathered}
\mathcal P_{-a,0}(4) \langle{G^-_{\D_1}(1)G^-_{\D_2}(2)G^+_{\D_4}(4)}\rangle \\
= - \sum_{k\ne 4} \frac{\bar z_k - \bar z_4}{(z_k - z_4)^{a-1}} i \frac{\partial}{\partial u_k} \langle{G^-_{\D_1}(1)G^-_{\D_2}(2)G^+_{\D_4}(4)}\rangle = 0
\end{gathered} 
\ee

\be
\begin{gathered}
\mathcal J^1_{-n}(4) \langle{G^-_{\D_1}(1)G^-_{\D_2}(2)G^+_{\D_4}(4)}\rangle \\
= -  \sum_{k\ne 4} \frac{\(\bar z_k - \bar z_4\)^2 \bar\partial_k + 2\bar h_k\( \bar z_k - \bar z_4\)}{\(z_k- z_4\)^{n}} \langle{G^-_{\D_1}(1)G^-_{\D_2}(2)G^+_{\D_4}(4)}\rangle = 0
\end{gathered}
\ee

So they annihilate the three point function because of the delta function coming from the energy-momentum conservation. This means that one cannot get all possible terms in the OPE, consistent with the extended symmetry algebra, just by studying the four graviton scattering amplitude. In higher point amplitudes this does not happen and all the symmetry generators are nontrivial. In this paper we will study the OPE and null states both from the (extended) symmetry point of view and also explicitly by starting from the six graviton MHV amplitude. 

\section{ OPE from $6$-point MHV Amplitude}\label{6mhvu} 

In this section we analyse the modified Mellin transform of the $6$-point tree level MHV graviton amplitude in Einstein gravity with the goal of extracting the leading as well first few subleading order terms in the celestial OPE of outgoing positive helicity gravitons. The subleading terms in the OPE that we obtain can potentially be related to the subleading terms in the collinear expansion of the momentum space amplitude which has been investigated in \cite{Nandan:2016ohb}. But we will not pursue that direction in this paper. 

\subsection{$6$-point MHV Graviton Amplitude}

The elegant representation of tree level MHV graviton amplitudes due to Hodges \cite{Hodges:2011wm, Hodges:2012ym} turns out to be the most useful for our purposes of studying the celestial OPE. According to Hodges' formula, the tree level MHV $n$-point stripped amplitude is given by
\begin{equation}
\label{hodgesnpt}
\begin{split}
& \mathcal{A}_{n} (1^{-},2^{-},3^{+},\ldots,n^{+})=  \langle 1 2 \rangle^{8} \ \frac{ \text{det}( \Phi^{i j k}_{p q r })}{\langle i j \rangle \langle i k \rangle\langle j k \rangle\langle p q \rangle\langle p r \rangle\langle q r \rangle }
\end{split}
\end{equation}

where $(1,2)$ label the negative helicity gravitons. $\Phi^{ijk}_{pqr}$ is a $(n-3)\times (n-3)$ matrix obtained by deleting the set of rows $\{i,j,k\}$ and columns $\{p,q,r\}$ from a $n\times n$ matrix  $\Phi$ whose elements are defined as follows
  \begin{equation}
  \label{Phimatrix}
    \Phi_{ij}= 
\begin{cases}
    \frac{[i j]}{\langle i j \rangle},&  i\ne j\\
    - \sum\limits_{k\ne i} \frac{[i k]\langle x k \rangle \langle y k \rangle }{\langle i k \rangle \langle x i \rangle \langle y i \rangle},              & i=j
\end{cases}
\end{equation}  

where $x, y$ denote reference spinors. As shown in \cite{Hodges:2011wm}, the representation \eqref{hodgesnpt} makes the $S_{n}$ permutation symmetry of the amplitude manifest. Consequently there is no a priori preferred choice for the set of rows and columns to be removed in order to obtain $\Phi^{ijk}_{pqr}$.  Another salient feature of Hodges' formula is that it elucidates the behaviour of the amplitude under soft limits. 

 \vskip 4pt
Now let us consider the $n=6$-point amplitude. In this case a convenient choice for the rows and columns to be removed from $\Phi$ is $\{i,j,k\}=\{1,2,3\}$ and $\{p,q,r\}=\{4,5,6\}$ respectively. Then the $6$-point MHV graviton amplitude takes the form
\begin{equation}
\label{hodges6pt}
\begin{split}
 \mathcal{A}_{6} (1^{-},2^{-},3^{+},4^{+}, 5^{+}, 6^{+}) & =  \frac{\langle 1 2 \rangle^{8} \ \text{det}\left( \Phi^{123}_{456}\right)}{\langle 1 2 \rangle \langle 1 3 \rangle\langle 2 3 \rangle\langle 45 \rangle\langle 46 \rangle\langle 56 \rangle } 
 \end{split}
\end{equation}

where $\Phi^{123}_{456}$ is a $3\times 3$ matrix and it's determinant is given by
 \begin{equation}
\label{6ptPhidet}
\begin{split}
 \text{det}\left( \Phi^{123}_{456}\right) & =  \frac{[14]}{\langle 1 4 \rangle} \left( \frac{[25]}{\langle 25 \rangle} \frac{[36]}{\langle 36 \rangle} - \frac{[26]}{\langle 26 \rangle} \frac{[35]}{\langle 35 \rangle}\right)  -  \frac{[24]}{\langle 2 4 \rangle} \left( \frac{[15]}{\langle 15 \rangle} \frac{[36]}{\langle 36 \rangle} - \frac{[16]}{\langle 16 \rangle} \frac{[35]}{\langle 35 \rangle}\right) \\
  & +  \frac{[34]}{\langle 3 4 \rangle} \left( \frac{[26]}{\langle 26 \rangle} \frac{[15]}{\langle 15 \rangle} - \frac{[25]}{\langle 25 \rangle} \frac{[16]}{\langle 16 \rangle}\right)
\end{split}
\end{equation}

Now using the following parametrisation of null momenta
\begin{equation}
\label{nullmom}
\begin{split}
 p^{\mu} &= \epsilon \hspace{0.03cm}\omega \hspace{0.03cm} q^{\mu}(z,\bar{z}), \quad q^{\mu}(z,\bar{z}) = (1+z \bar{z}, z+\bar{z}, -i(z-\bar{z}), 1- z\bar{z})\\
\end{split}
\end{equation}

the spinor helicity brackets can be written as
\begin{equation}
\label{angsqbr}
\begin{split}
\langle i j \rangle = -2 \hspace{0.04cm} \epsilon_{i}\epsilon_{j} \sqrt{\omega_{i}\omega_{j}} \hspace{0.04cm}z_{ij}, \quad [ i j ] = 2 \sqrt{\omega_{i}\omega_{j}} \hspace{0.04cm} \bar{z}_{ij}
\end{split}
\end{equation}

where $\varepsilon_{i}=\pm 1$ for outgoing (incoming) particles. Then the amplitude in \eqref{hodges6pt} becomes
\begin{equation}
\label{hodges6pt1}
\begin{gathered}
 \mathcal{A}_{6} (1^{-},2^{-},3^{+},4^{+}, 5^{+}, 6^{+})  \\
 = - 4 \left( \frac{\omega_{1}^{3}\omega_{2}^{3}}{\omega_{3}\omega_{4}\omega_{5}\omega_{6}} \right)   \bigg[  \frac{\bar{z}_{14}}{z_{14}} \left( \frac{\bar{z}_{25}\bar{z}_{36}}{z_{25}z_{36}} - \frac{\bar{z}_{26}\bar{z}_{35}}{z_{26}z_{35}}\right) - \frac{\bar{z}_{24}}{z_{24}} \left( \frac{\bar{z}_{15}\bar{z}_{36}}{z_{15}z_{36}} - \frac{\bar{z}_{16}\bar{z}_{35}}{z_{16}z_{35}}\right) \\
  + \frac{\bar{z}_{34}}{z_{34}} \left( \frac{\bar{z}_{15}\bar{z}_{26}}{z_{15}z_{26}} - \frac{\bar{z}_{16}\bar{z}_{25}}{z_{16}z_{25}}\right) \bigg]  \frac{z_{12}^{8}}{z_{12}z_{13}z_{23}z_{45}z_{46}z_{56}} \prod_{i=1}^{6}  \epsilon_{i}
\end{gathered}
\end{equation}


\subsection{$5$-point MHV Graviton Amplitude}

In order to obtain the celestial OPE from the $6$-point amplitude, we will also require the expression of the $5$-point MHV graviton amplitude. Using Hodge's formula this is given by
\begin{equation}
\label{hodges5pt}
\begin{split}
& \mathcal{A}_{5}  (1^{-},2^{-},3^{+},4^{+}, 5^{+})   =  \frac{\langle 1 2 \rangle^{8} \ \text{det}\left( \Phi^{123}_{345}\right)}{\langle 1 2 \rangle \langle 1 3 \rangle\langle 2 3 \rangle\langle 3 4 \rangle\langle 3 5 \rangle\langle 4 5 \rangle }
\end{split}
\end{equation}

Here we have chosen to remove the set of rows $\{i,j,k\}=\{1,2,3\}$ and columns $\{p,q,r\}=\{3,4,5\}$  from $\Phi$. In this case the minor $ \Phi^{123}_{345}$ is a $2\times 2$ matrix and its determinant is 
\begin{equation}
\label{hodges5pt1}
\begin{split}
  \text{det}\left( \Phi^{123}_{345}\right) & = \left( \frac{[14]}{\langle 14 \rangle} \frac{[25]}{\langle 25 \rangle} - \frac{[15]}{\langle 15 \rangle} \frac{[24]}{\langle 24 \rangle}\right)  
\end{split}
\end{equation}

Then using the parametrisation of null momentum in  \eqref{nullmom}, the $5$-point amplitude can be expressed as
\begin{equation}
\label{hodges5pt2}
\begin{split}
&  \mathcal{A}_{5}  (1^{-},2^{-},3^{+},4^{+}, 5^{+})   =  4 \ \epsilon_{1}\epsilon_{2}\epsilon_{4}\epsilon_{5} \ \frac{\omega_{1}^{3}\omega_{2}^{3}}{\omega_{3}^{2}\omega_{4}\omega_{5}} \ \frac{z_{12}^{8}}{z_{12}z_{13}z_{23}z_{34}z_{35}z_{45}}   \left( \frac{\bar{z}_{14}\bar{z}_{25}}{z_{14}z_{25}} -  \frac{\bar{z}_{15}\bar{z}_{24}}{z_{15}z_{24}}  \right)  
\end{split}
\end{equation}


\subsection{$6$-point Mellin Amplitude}
\label{6ptmellin}

The modified Mellin transform of the $6$-point MHV graviton amplitude in \eqref{hodges6pt1} is given by \cite{Banerjee:2018gce,Banerjee:2019prz}
\begin{equation}
\label{Hmellin6pt}
\begin{split}
 \mathcal{M}_{6} &= \left \langle  G^{-}_{\Delta_{1}}(1) G^{-}_{\Delta_{2}}(2) G^{+}_{\Delta_{3}}(3)  G^{+}_{\Delta_{4}}(4) G^{+}_{\Delta_{5}}(5)  G^{+}_{\Delta_{6}}  (6)\right\rangle  \\
 &  = \int_{0}^{\infty}  \prod_{i=1}^{6} d\omega_{i} \hspace{0.1cm} \omega_{i}^{\Delta_{i}-1}  \hspace{0.05cm} e^{- i\sum\limits_{i=1}^{6} \epsilon_{i}\omega_{i}u_{i}} \hspace{0.05cm} \mathcal{A}_{6}(\omega_{i}, z_{i},\bar{z}_{i}) \hspace{0.05cm} \delta^{(4)} \left( \sum_{i=1}^{6} \epsilon_{i} \omega_{i}q_{i}^{\mu}(z_{i},\bar{z}_{i})\right), \quad \Delta_i = 1 + i\lambda_i
 \end{split}
\end{equation}

where $G^{\pm}_{\Delta_{i}}(i)  \equiv G^{\pm}_{\Delta_{i}} (u_{i},z_{i},\bar{z}_{i}) $ is the primary operator dual to the $i$-\textit{th} external graviton in the $S$-matrix. The superscript denotes the spin $s_{i}=\pm 2$. 
Throughout the rest of this section we will take the gravitons $(5,6)$ in the amplitude $\mathcal{A}_{6}$ to be outgoing, i.e., $\epsilon_{5}=\epsilon_{6}=1$, but the $\varepsilon_{i}$'s for the remaining particles will be left unspecified.  

 \vskip 4pt
Now let us choose the pair of primary operators corresponding to gravitons $(5,6)$ in the amplitude for performing the celestial OPE.  The OPE can then be obtained by expanding the Mellin amplitude around $z_{56}= 0, \bar{z}_{56} =0,u_{56}=0$. In order to carry out this analysis it is first useful to perform the following change of integration variables 
\begin{equation}
\label{tomegapdef1}
\begin{split}
& \omega_{5}= \omega_{P}\hspace{0.04cm} t, \quad \omega_{6}= \omega_{P} (1-t)
\end{split}
\end{equation}

where $\omega_{P}\ge 0$ and $t\in[0,1]$. Then we get, 
\begin{equation}
\label{Hmellin6pt1}
\begin{split}
 \mathcal{M}_{6} &= \int_{0}^{1} dt  \hspace{0.06cm} t^{i\lambda_{5}} (1-t)^{i\lambda_{6}}  \int_{0}^{\infty}  d\omega_{P} \hspace{0.06cm} \omega_{P}^{i\lambda_{5}+i\lambda_{6}}   \int_{0}^{\infty}  \prod_{i=1}^{4} d\omega_{i} \hspace{0.1cm} \omega_{i}^{\Delta_{i}-1}   \hspace{0.05cm} e^{- i\sum\limits_{i=1}^{6} \epsilon_{i}\omega_{i}u_{i}} \\
 & \times \hspace{0.05cm} \mathcal{A}_{6}(\omega_{i}, z_{i},\bar{z}_{i})  \hspace{0.05cm} \delta^{(4)} \left( \sum_{i=1}^{6} \epsilon_{i} \omega_{i}q_{i}^{\mu}(z_{i},\bar{z}_{i})\right)
 \end{split}
\end{equation}

Now as shown in Section \ref{deltarep} of the Appendix, the delta function in the above integral can be written as
\begin{equation}
\label{delta6pt}
\begin{split}
& \delta^{(4)} \Big( \sum_{i=1}^{6} \epsilon_{i} \omega_{i} q_{i}^{\mu}(z_{i},\bar{z}_{i})\Big) = \frac{i}{4} \frac{1}{(r_{12,34}- \bar{r}_{12,34}) z_{13}\bar{z}_{13}z_{24}\bar{z}_{24}} \  \prod_{i=1}^{4} \delta(\omega_{i} - \omega_{i}^{*})
\end{split}
\end{equation}

where
\begin{equation}
\label{omegaist}
\begin{split}
& \omega_{i}^{*} =  \epsilon_{i}\hspace{0.04cm} \omega_{P} \left[ \sigma_{i,1} + t \left( z_{56}  \hspace{0.04cm}  \sigma_{i,2} + \bar{z}_{56} \hspace{0.04cm} \sigma_{i,3}+ z_{56} \bar{z}_{56} \hspace{0.04cm} \sigma_{i,4}\right) \right] , \quad i \in 1,2,3,4.
\end{split}
\end{equation}

and  $\sigma_{i,j}$ are functions of $z_{ij}, \bar{z}_{ij}$ with $(i,j)\in(1,2,3,4,6)$. Their explicit forms are given in equations \eqref{sigma1}-\eqref{sigmaderv} . Now  this representation of the delta function can be used to localise the integrals with respect to $\omega_{i}, i \in (1,2,3,4)$. We are then left with integrals with respect to the variables $\omega_{P}$ and $t$. Using the expression of the MHV amplitude  in \eqref{hodges6pt1}, the integral over $\omega_{P}$ is given by
\begin{equation}
\label{omegaPinteg}
\begin{split}
&   \int_{0}^{\infty}   d\omega_{P} \hspace{0.1cm} \omega_{P}^{3+i \Lambda} \exp\left[ -i \hspace{0.03cm} \omega_{P} \left( \mathcal{U}_{1} + z_{56} \hspace{0.03cm}t \hspace{0.04cm}  \mathcal{U}_{2}  + \bar{z}_{56} \hspace{0.03cm}t \hspace{0.04cm}  \mathcal{U}_{3} + z_{56} \bar{z}_{56}\hspace{0.03cm}t  \hspace{0.04cm}  \mathcal{U}_{4}  + t \hspace{0.04cm} u_{56}\right)\right] \\
& = \frac{\Gamma(4+i\Lambda)}{( i \hspace{0.05cm} \mathcal{U}_{1})^{4+i\Lambda}} \left[ 1+\frac{t}{\mathcal{U}_{1}} \left( z_{56} \hspace{0.04cm}\mathcal{U}_{2} + \bar{z}_{56}  \hspace{0.04cm}\mathcal{U}_{3} + z_{56} \bar{z}_{56}\hspace{0.04cm}\mathcal{U}_{4}  + u_{56}\right) \right]^{-4-i\Lambda}
\end{split}
\end{equation}

where $\Lambda = \sum_{i=1}^{6}\lambda_{i}$ and 
\begin{equation}
\label{Uidef}
\begin{split}
& \mathcal{U}_{1} = \sum_{i=1}^{4} \sigma_{i,1} u_{i6}, \quad   \mathcal{U}_{2}= \sum_{i=1}^{4} \sigma_{i,2} u_{i6}, \quad \mathcal{U}_{3}= \sum_{i=1}^{4} \sigma_{i,3} u_{i6}, \quad \mathcal{U}_{4}= \sum_{i=1}^{4} \sigma_{i,4} u_{i6}
\end{split}
\end{equation}

Then using \eqref{omegaPinteg} in \eqref{Hmellin6pt1}, the Mellin amplitude takes the  form
\begin{equation}
\label{Hmellin6pt2}
\begin{split}
& \mathcal{M}_{6}  =   \mathcal{N} \hspace{0.04cm} \mathcal{F} \int_{0}^{1} dt \hspace{0.1cm} t^{i\lambda_{5}-1}(1-t)^{i\lambda_{6}-1}  \prod_{i=1}^{4}\Theta\left(  \epsilon_{i} \left( \sigma_{i,1} + z_{56} \hspace{0.03cm} t \hspace{0.03cm}  \sigma_{i,2} + \bar{z}_{56} \hspace{0.03cm} t\hspace{0.03cm}  \sigma_{i,3}+ z_{56} \bar{z}_{56}\hspace{0.03cm} t \hspace{0.03cm}  \sigma_{i,4} \right) \right) \hspace{0.05cm} \mathcal{I}(t) 
\end{split}
\end{equation}

In the above expression, the integrand $\mathcal{I}(t)$ is given by
\begin{equation}
\label{Itdef1}
\begin{split}
\hspace{-0.5cm} \mathcal{I}(t)    & = \prod_{i=1}^{2}  \left( 1+ z_{56} \hspace{0.03cm} t \hspace{0.03cm} \frac{\sigma_{i,2}}{\sigma_{i,1} } + \bar{z}_{56} \hspace{0.03cm} t\hspace{0.03cm} \frac{\sigma_{i,3}}{\sigma_{i,1} }+ z_{56} \bar{z}_{56}\hspace{0.03cm} t \hspace{0.03cm} \frac{\sigma_{i,4}}{\sigma_{i,1} } \right)^{3+i\lambda_{i}}   \\
& \times \prod_{i=3}^{4}\left( 1+ z_{56} \hspace{0.03cm} t \hspace{0.03cm} \frac{\sigma_{i,2}}{\sigma_{i,1} } + \bar{z}_{56} \hspace{0.05cm} t\hspace{0.03cm} \frac{\sigma_{i,3}}{\sigma_{i,1} }+ z_{56} \bar{z}_{56}\hspace{0.05cm} t \hspace{0.03cm} \frac{\sigma_{i,4}}{\sigma_{i,1} } \right) ^{i\lambda_{i}-1}  \hspace{-0.1cm} \\ 
& \times \left[ 1+ \frac{t}{\mathcal{U}_{1}} \left( z_{56} \hspace{0.03cm} \mathcal{U}_{2} + \bar{z}_{56} \hspace{0.03cm} \mathcal{U}_{3}+z_{56}\bar{z}_{56} \hspace{0.03cm} \mathcal{U}_{4}+ u_{56}\right)\right]^{-4-i\Lambda} 
\end{split}
\end{equation}

The prefactors $\mathcal{N}$ and $\mathcal{F}$ are given by  
\begin{equation}
\label{ndef1}
\begin{split}
 \mathcal{N} &= - i \prod_{i=1}^{4} \epsilon_{i}  \prod_{j=1}^{2} ( \epsilon_{j} \sigma_{j,1})^{3+i\lambda_{j}}  \prod_{k=3}^{4}( \epsilon_{k}\sigma_{k,1})^{i\lambda_{k}-1}   \ \frac{z_{12}^{8}}{z_{12}z_{13}z_{14}z_{16}z_{23}z_{24}z_{26}z_{34}z_{36}z_{46}} \   \frac{\Gamma(4+i\Lambda)}{( i \hspace{0.05cm} \mathcal{U}_{1})^{4+i\Lambda}} 
\end{split}
\end{equation}

\begin{equation}
\label{Fdef1}
\begin{split}
 \mathcal{F} & =   \frac{1}{z_{46}z_{56}} \left( 1- \frac{z_{56}}{z_{46}}\right)^{-1} \sum_{i=1}^{3} z_{i4} \bar{z}_{i6} \hspace{0.05cm} \sigma_{i,1} \left( 1- \frac{\bar{z}_{56}}{\bar{z}_{i6}} \right) \left(1- \frac{z_{56}}{z_{i6}} \right)^{-1} 
\end{split}
\end{equation}

The theta functions 
\begin{equation}
\label{thetafunc}
\begin{split}
\Theta\left(  \epsilon_{i} \left( \sigma_{i,1} + z_{56} \hspace{0.03cm} t \hspace{0.03cm}  \sigma_{i,2} + \bar{z}_{56} \hspace{0.03cm} t\hspace{0.03cm}  \sigma_{i,3}+ z_{56} \bar{z}_{56}\hspace{0.03cm} t \hspace{0.03cm}  \sigma_{i,4} \right) \right), \quad i=1,2,3,4
\end{split}
\end{equation}

simply ensure that the delta function in \eqref{delta6pt} localises the integration variables $\omega_{i}, i \in (1,2,3,4)$ in \eqref{Hmellin6pt1} to only positive semi-definite values.  

 \vskip 4pt
Note that we have written the various terms in the R.H.S. of \eqref{Hmellin6pt2} in a form which is particularly convenient for expanding the Mellin amplitude around $z_{56}= 0, \bar{z}_{56}= 0, u_{56}=0 $ for the purposes of extracting the celestial OPE. 


\subsection{$5$-point Mellin Amplitude}
\label{5ptMellin}

In order to study the factorisation behaviour of the $6$-point Mellin amplitude in the OPE limit, we need to know the $5$-point Mellin amplitude. This is given by \cite{Banerjee:2018gce,Banerjee:2019prz}
\begin{equation}
\label{mellin5pt}
\begin{split}
\mathcal{M}_{5} &   = \left \langle  G^{-}_{\Delta_{1}}(1) G^{-}_{\Delta_{2}}(2) G^{+}_{\Delta_{3}}(3)  G^{+}_{\Delta_{4}}(4) G^{+}_{\Delta_{5}}(5) \right\rangle  \\
& =    \int  \prod_{i=1}^{5} d\omega_{i} \hspace{0.1cm} \omega_{i}^{i\lambda_{i}}  \hspace{0.05cm} e^{- i\sum\limits_{i=1}^{5} \epsilon_{i}\omega_{i}u_{i}} \hspace{0.05cm} \mathcal{A}_{5}(\omega_{i},z_{i},\bar{z}_{i}) \hspace{0.05cm} \delta^{(4)} \left( \sum_{i=1}^{5} \epsilon_{i} \omega_{i}q_{i}^{\mu}(z_{i},\bar{z}_{i})\right)
\end{split}
\end{equation}

where $\mathcal{A}_{5}$ is the $5$-point MHV graviton amplitude given by \eqref{hodges5pt2}. Now as shown in the Appendix, Section \ref{deltarep}, the delta function involved in the above integral can be represented as
\begin{equation}
\label{delta5pt}
\begin{split}
& \delta^{(4)} \Big( \sum_{i=1}^{5} \epsilon_{i} \omega_{i} q_{i}^{\mu}(z_{i},\bar{z}_{i})\Big) = \frac{i}{4} \frac{1}{(r_{12,34}- \bar{r}_{12,34}) z_{13}\bar{z}_{13}z_{24}\bar{z}_{24}} \  \prod_{i=1}^{4} \delta(\omega_{i} - \omega_{i}^{*})
\end{split}
\end{equation}

where\footnote{Note that $\sigma_{i,1}$ in \eqref{omegaist5pt} are identical to the $\sigma_{i,1}$ in \eqref{omegaist} upto the change of labels $ z_{6}\rightarrow z_{5}, \bar{z}_{6}\rightarrow z_{5}$.  }
\begin{equation}
\label{omegaist5pt}
\begin{split}
& \omega_{i}^{*} =  \epsilon_{i}\epsilon_{5}\hspace{0.04cm} \omega_{5} \hspace{0.04cm}  \sigma_{i,1} , \quad i \in 1,2,3,4.
\end{split}
\end{equation}

Due to \eqref{delta5pt}, the integrals with respect to $\omega_{i}, i \in (1,2,3,4)$ in \eqref{mellin5pt} can again be localised. Then plugging in the expression of $\mathcal{A}_{5}$ into \eqref{mellin5pt},  the remaining integral over $\omega_{5}$ gives
\begin{equation}
\label{omegaPinteg5pt}
\begin{split}
&   \int_{0}^{\infty}   d\omega_{5} \hspace{0.1cm} \omega_{5}^{2+i \Lambda'} \exp\left( -i \epsilon_{5} \hspace{0.03cm} \omega_{5} \sum_{i=1}^{4} \sigma_{i,1} u_{i5} \right)  = \frac{\Gamma(3+i\Lambda')}{( i \epsilon_{5} \hspace{0.05cm} \mathcal{U}_{1})^{3+i\Lambda'}} 
\end{split}
\end{equation}

where $\Lambda' = \sum_{i=1}^{5} \lambda_{i}$, $\mathcal{U}_{1}= \sum_{i=1}^{4}\sigma_{i,1} u_{i5}$.  Thus, the $5$-point Mellin amplitude can be written as
\begin{equation}
\label{Hmellin5pt1}
\begin{split}
 \mathcal{M}_{5}   & = i \prod_{i=1}^{5} \epsilon_{i}  \prod_{j=1}^{2} ( \epsilon_{j} \sigma_{j,1})^{3+i\lambda_{i}}  \prod_{k=3}^{4}( \epsilon_{k}\sigma_{k,1})^{i\lambda_{k}-1} \prod_{l=1}^{4} \Theta \left( \epsilon_{l} \sigma_{l,1}\right) \hspace{0.04cm} \frac{z_{12}^{8}}{z_{12}z_{13}z_{14}z_{15}z_{23}z_{24}z_{25}z_{34}z_{35}z_{45}}  \\
& \times \hspace{0.04cm}  \frac{\Gamma(3+i\Lambda')}{( i \epsilon_{5} \hspace{0.05cm} \mathcal{U}_{1})^{3+i\Lambda'}} 
\end{split}
\end{equation}

Now  let us note that the prefactor $\mathcal{N}$  in \eqref{Hmellin6pt2} is related to the $5$-point Mellin amplitude. In order to see this let us set $\epsilon_{5}=1$ and perform the following change of labels in \eqref{Hmellin5pt1} 
\begin{equation}
z_{5} \rightarrow z_{6}, \quad \bar{z}_{5} \rightarrow \bar{z}_{6}, \quad u_{5}\rightarrow u_{6}, \quad i\lambda_{5} \rightarrow i\lambda_{5}+i\lambda_{6}
\end{equation}

Then the $5$-point Mellin amplitude takes the form
\begin{equation}
\label{Hmellin5pt2}
\begin{split}
 \mathcal{M}_{5} & = \left \langle  G^{-}_{\Delta_{1}}(1) G^{-}_{\Delta_{2}}(2) G^{+}_{\Delta_{3}}(3)  G^{+}_{\Delta_{4}}(4) G^{+}_{\Delta_{5}+\Delta_{6}-1}(6) \right\rangle  \\
 & =   i \prod_{i=1}^{4} \epsilon_{i}  \prod_{j=1}^{2} ( \epsilon_{j} \sigma_{j,1})^{3+i\lambda_{i}}  \prod_{k=3}^{4}( \epsilon_{k}\sigma_{k,1})^{i\lambda_{k}-1} \prod_{l=1}^{4} \Theta \left( \epsilon_{l} \sigma_{l,1}\right) \hspace{0.04cm} \frac{z_{12}^{8}}{z_{12}z_{13}z_{14}z_{16}z_{23}z_{24}z_{26}z_{34}z_{36}z_{46}} \\
 & \times  \frac{\Gamma(3+i\Lambda)}{( i \hspace{0.05cm} \mathcal{U}_{1})^{3+i\Lambda}} 
\end{split}
\end{equation}

where $\Lambda= \sum_{i=1}^{6} \lambda_{i}$ and the $\sigma_{i,1}$'s are now given by \eqref{sigma1} to \eqref{sigma4}. Then from \eqref{ndef1} and \eqref{Hmellin5pt2} it is evident that
\begin{equation}
\label{u6derv5pt}
\begin{split}
&  -  \mathcal{N} \  \prod_{l=1}^{4} \Theta \left( \epsilon_{l} \sigma_{l,1}\right)= i \hspace{0.05cm} \partial_{u_{6}} \hspace{0.05cm} \mathcal{M}_{5} = \mathcal{P}_{-1,-1} \mathcal{M}_{5}
\end{split}
\end{equation}

In all subsequent references to the $5$-point Mellin amplitude in the rest of this section we will take it to be given by \eqref{Hmellin5pt2}. 


\subsection{OPE decomposition of $6$-point Mellin Amplitude}
\label{6ptOPEMHV}

We will now consider the celestial OPE decomposition of the $6$-point Mellin amplitude by expanding it around $z_{56}= 0, \bar{z}_{56}= 0, u_{56}=0$ while keeping the remaining $z_{ij},\bar{z}_{ij}, u_{ij}$ with $(i,j) \in (1,2,3,4,6)$ fixed and non-zero. Here we will ignore the delta function contributions that come from differentiating the theta functions in \eqref{Hmellin6pt2} w.r.t. $z_{56},\bar{z}_{56}$. The arguments of these delta functions involve $z_{ij},\bar{z}_{ij}$ with $(i,j)\in (1,2,3,4,6)$. Such contact terms can be neglected in the OPE regime where all operator insertions, except the pair whose OPE is being considered, are taken to be at separated points. 

 \vskip 4pt
Now the prefactor $\mathcal{N}$ in \eqref{Hmellin6pt2} is independent of $z_{56},\bar{z}_{56}$. Thus for extracting the OPE we only need to keep track of terms coming from expanding the other prefactor $\mathcal{F}$ and the integrand $\mathcal{I}(t)$. 


\subsubsection{Leading term}

It can be easily seen from \eqref{Itdef1} and \eqref{Fdef1} that the leading term in the Mellin amplitude in the OPE limit is of $\mathcal{O}\left(\frac{\bar{z}_{56}}{z_{56}}\right)$.  The $\bar{z}_{56}z_{56}^{-1}$ term comes entirely from\footnote{In \eqref{Fdef1} there is also apparently a term of $\mathcal{O}(z_{56}^{-1})$. Using the identities \eqref{sigmarel3} and \eqref{sigmarel4}, it can be straightforwardly checked that this term actually vanishes. }  $\mathcal{F}$ and is given by  
\begin{equation}
\label{ordzbarzinv}
\begin{split}
  \mathcal{F} & = -  \frac{\bar{z}_{56}}{z_{56}} \sum_{i=1}^{3} \frac{z_{i4}}{z_{46}} \hspace{0.05cm} \sigma_{i,1} + \cdots
\end{split}
\end{equation}

where the dots denote terms which are regular as $z_{56}\rightarrow 0$.  Then applying the identities \eqref{sigmarel1} and \eqref{sigmarel2} given in section \ref{deltarep} of the Appendix, \eqref{ordzbarzinv} simply becomes
\begin{equation}
\label{ordzbarzinv1}
\begin{split}
 \mathcal{F}& = -  \frac{\bar{z}_{56}}{z_{56}} + \cdots
\end{split}
\end{equation}

Note that $\mathcal{I}(t)$ is regular around $z_{56}= 0, \bar{z}_{56}= 0$ and at leading order we have $\mathcal{I}(t) \approx 1$. Thus the leading term in the Mellin amplitude in the OPE limit takes the form
\begin{equation}
\label{ordzbarzinv2}
\begin{split}
 \mathcal{M}_{6} &  =  - \frac{\bar{z}_{56}}{z_{56}}\  \mathcal{N} \int_{0}^{1} dt \hspace{0.1cm} t^{i\lambda_{5}-1}(1-t)^{i\lambda_{6}-1}  \prod_{i=1}^{4}\Theta\left(  \epsilon_{i} \sigma_{i,1} \right) \hspace{0.05cm} + \cdots \\
 & = - \frac{\bar{z}_{56}}{z_{56}} \ B(i\lambda_{5},i\lambda_{6}) \ \mathcal{P}_{-1,-1} \mathcal{M}_{5} + \cdots
\end{split}
\end{equation}

where we have used the relation \eqref{u6derv5pt} and $B(x,y)$ is the Euler Beta function. Now this result implies that the leading term in the celestial OPE  is given by
\begin{equation}
\label{leadcsope}
\begin{split}
& G^{+}_{\Delta_{5}} (z_{5},\bar{z}_{5}) G^{+}_{\Delta_{6}} (z_{6},\bar{z}_{6}) = - \frac{\bar{z}_{56}}{z_{56}} \  B(i\lambda_{5},i\lambda_{6}) \ P_{-1,-1} G^{+}_{\Delta_{5}+\Delta_{6}-1} (z_{6},\bar{z}_{6}) +\cdots
\end{split}
\end{equation}
This term exactly matches with \cite{Pate:2019lpp}.


\subsubsection{Subleading terms: $\mathcal{O}(z_{56}^0\bar z_{56}^0)$ }

Now let us study the first few subleading terms in the OPE decomposition.  We first consider the terms of $\mathcal{O}(1)$, which turn out to be non-trivial  here in contrast to the case of the $4$-point Mellin amplitude. The relevant contribution from $ \mathcal{F} $ at this order is
\begin{equation}
\label{ord1}
\begin{split}
 \mathcal{F}\Big|_{\mathcal{O}(1)} & =   \frac{1}{z_{46}}  \sum_{i=1}^{3} z_{i4}\frac{\bar{z}_{i6} }{z_{i6}} \hspace{0.05cm} \sigma_{i,1} +  \frac{1}{z^{2}_{46}} \sum_{i=1}^{3} z_{i4} \bar{z}_{i6} \hspace{0.05cm} \sigma_{i,1}\\
 &= -  \sum_{i=1}^{3} \frac{\bar{z}_{i6} }{z_{i6}} \hspace{0.05cm} \sigma_{i,1} +  \frac{1}{z^{2}_{46}} \sum_{i=1}^{3} z_{i6} \bar{z}_{i6} \hspace{0.05cm} \sigma_{i,1} 
\end{split}
\end{equation}

Using the identities \eqref{sigmarel3} and \eqref{sigmarel4} this can be simplified to give
\begin{equation}
\label{ord1a}
\begin{split}
 \mathcal{F}\Big|_{\mathcal{O}(1)}  &= -  \sum_{i=1}^{4} \frac{\bar{z}_{i6} }{z_{i6}} \hspace{0.05cm} \sigma_{i,1} 
\end{split}
\end{equation}

At this order we again have $\mathcal{I}(t) \approx 1$. Then after doing the $t$-integral we find the $\mathcal{O}(1)$ term from the Mellin amplitude to be given by
\begin{equation}
\label{ord1b}
\begin{split}
&  \mathcal{M}_{6} \Big|_{\mathcal{O}(1)}= - B(i\lambda_{5},i\lambda_{6})  \sum_{i=1}^{4} \frac{\bar{z}_{i6} }{z_{i6}} \hspace{0.05cm} \sigma_{i,1} \ \mathcal{P}_{-1,-1} \mathcal{A}_{5}  = B(i\lambda_{5},i\lambda_{6}) \ \mathcal{P}_{-2,0} \mathcal{M}_{5}
\end{split}
\end{equation}

where in obtaining the last equality above we have used \eqref{Pminn05pt} from Section \ref{desccorr} of the Appendix. 

 \vskip 4pt
Now let us take the subleading conformal soft limit $i\lambda_{5}\rightarrow -1$ in \eqref{ord1b}. This gives
\begin{equation}
\label{subsoftord1a}
\begin{split}
& \lim_{i\lambda_{5} \to -1} (1+i\lambda_{5})  \mathcal{M}_{6} \bigg|_{\mathcal{O}(1)}= -  (i\lambda_{6}-1) \mathcal{P}_{-2,0} \mathcal{M}'_{5}
\end{split}
\end{equation}

where $\mathcal{M}'_{5} = \mathcal{M}_{5}\Big|_{i\lambda_{5}=-1}$. But from the discussion in Section \ref{shope}, we know that the subleading conformal soft theorem \eqref{se} requires 
\begin{equation}
\label{subsoftord1b}
\begin{split}
& \lim_{i\lambda_{5} \to -1} (1+i\lambda_{5})  \mathcal{M}_{6} \bigg|_{\mathcal{O}(1)}= -   \mathcal{J}^{1}_{-1}\mathcal{P}_{-1,-1} \mathcal{M}'_{5}
\end{split}
\end{equation}

Then consistency of \eqref{subsoftord1a} with the \eqref{subsoftord1b} implies
\begin{equation}
\label{ord1nst}
\begin{split}
& \mathcal{J}^{1}_{-1}\mathcal{P}_{-1,-1} \mathcal{M}'_{5} =   (i\lambda_{6}-1) \mathcal{P}_{-2,0} \mathcal{M}'_{5}
\end{split}
\end{equation}

Shifting $i\lambda_{6}\rightarrow 1+ i\lambda_{5}+i\lambda_{6}$ in the above, we obtain the following differential equation
\begin{equation}
\label{ord1nst1}
\begin{split}
& \left[ \mathcal{J}^{1}_{-1}\mathcal{P}_{-1,-1} -  (i\lambda_{5}+ i\lambda_{6}) \mathcal{P}_{-2,0} \right]  \mathcal{M}_{5} =0
\end{split}
\end{equation} 

Equivalently this equation implies that the following descendant 
\begin{equation}
\label{ord1nst2}
\begin{split}
&\Phi^{+} = \left[ J^{1}_{-1}P_{-1,-1} - (\Delta-1) P_{-2,0}\right] G^{+}_{\Delta} 
\end{split}
\end{equation} 

is a null state. In section \ref{NULL} we will determine the null-state $\Phi^{+}$ using the extended symmetry algebra. We will also subsequently study in this paper the implications of the decoupling relation \eqref{ord1nst1} for the structure of the leading celestial OPE of gravitons. 


\subsubsection{Subleading terms: $\mathcal{O}(z_{56})$}

Next let us consider the terms of $\mathcal{O}(z_{56})$. In this case we have 
\begin{equation}
\label{ordz}
\begin{split}
 \mathcal{F}\Big|_{\mathcal{O}(z_{56})} & =   z_{56} \sum_{i=1}^{3} z_{i4} \bar{z}_{i6} \hspace{0.05cm} \sigma_{i,1} \left(\frac{1}{z^{2}_{i6}z_{46}}+ \frac{1}{z_{i6}z^{2}_{46}} +  \frac{1}{z^{3}_{46}} \right)\\
 & = z_{56} \left(  \frac{1}{z^{3}_{46}}\sum_{i=1}^{3}  \bar{z}_{i6} \hspace{0.05cm} \sigma_{i,1}  - \sum_{i=1}^{3} \frac{ \bar{z}_{i6}}{z^{2}_{i6}} \hspace{0.05cm} \sigma_{i,1} \right) = - z_{56} \sum_{i=1}^{4} \frac{ \bar{z}_{i6}}{z^{2}_{i6}} \hspace{0.05cm} \sigma_{i,1}
\end{split}
\end{equation}

where in arriving at the last equality above we have used  \eqref{sigmarel3}. Now at this order we have a non-trivial contribution from $\mathcal{I}(t)$ and this is given by 
\begin{equation}
\label{ordz1}
\begin{split}
 \mathcal{I}(t)\bigg|_{\mathcal{O}(z_{56})} & =  z_{56} \hspace{0.03cm} t \left(  (3+i\lambda_{1}) \frac{\sigma_{1,2}}{\sigma_{1,1} } +  (3+i\lambda_{2}) \frac{\sigma_{2,2}}{\sigma_{2,1} }  +  (i\lambda_{3}-1) \frac{\sigma_{3,2}}{\sigma_{3,1} }+ (i\lambda_{4}-1) \frac{\sigma_{4,2}}{\sigma_{4,1} } - \frac{(4+i\Lambda)}{\mathcal{U}_{1}} \hspace{0.03cm}  \mathcal{U}_{2} \right) \\
 & \equiv z_{56}  \hspace{0.03cm} t \hspace{0.05cm}  \mathcal{I}_{1,0}
\end{split}
\end{equation}

where the notation $ \mathcal{I}_{1,0}$ has been introduced for convenience. The subscripts $(1,0)$ can be regarded as keeping track of the associated orders of $z_{56},\bar{z}_{56}$ respectively.  Thus we have
\begin{equation}
\label{ordz2}
\begin{split}
  \mathcal{M}_{6} \bigg|_{\mathcal{O}(z_{56})} & = - z_{56} \int_{0}^{1} dt \hspace{0.05cm} t^{i\lambda_{5}-1}(1-t)^{i\lambda_{6}-1} \bigg[\mathcal{F}\Big|_{\mathcal{O}(z_{56})}  \mathcal{I}(t)\Big|_{\mathcal{O}(1)} + \mathcal{F}\Big|_{\mathcal{O}(1)}  \mathcal{I}(t)\Big|_{\mathcal{O}(z_{56})} \bigg] \mathcal{P}_{-1,-1} \mathcal{M}_{5} \\
  &=  - z_{56}  \hspace{0.03cm} B(i\lambda_{5},i\lambda_{6}) \left[  \sum_{i=1}^{4} \frac{\bar{z}_{i6} }{z^{2}_{i6}} \hspace{0.05cm} \sigma_{i,1}   + \frac{i\lambda_{5}}{i\lambda_{5}+i\lambda_{6}} \hspace{0.05cm}  \mathcal{I}_{1,0}\sum_{i=1}^{4} \frac{\bar{z}_{i6} }{z_{i6}} \hspace{0.05cm} \sigma_{i,1} \right] \mathcal{P}_{-1,-1} \mathcal{M}_{5} 
\end{split}
\end{equation}

Then using \eqref{Pminn05pt} and \eqref{Lmin1Pmin205pt2} given in Section \ref{desccorr} of the Appendix,  the above can be written as
\begin{equation}
\label{ordz3}
\begin{split}
& \mathcal{M}_{6} \bigg|_{\mathcal{O}(z_{56})} = z_{56} \ B(i\lambda_{5},i\lambda_{6}) \left[ \frac{i\lambda_{6}-i\lambda_{5}}{i\lambda_{5}+i\lambda_{6}} \ \mathcal{P}_{-3,0} + \frac{i\lambda_{5}}{i\lambda_{5}+i\lambda_{6}} \ \mathcal{L}_{-1} \mathcal{P}_{-2,0} \right] \mathcal{M}_{5}
\end{split}
\end{equation}

Now the subleading conformal soft theorem \eqref{se} requires that at this order we must have
\begin{equation}
\label{subsoftordz1}
\begin{split}
& \lim_{i\lambda_{5} \to -1} (1+i\lambda_{5}) \mathcal{M}_{6} \bigg|_{\mathcal{O}(z_{56})} = - z_{56} \ \mathcal{J}^{1}_{-2}\mathcal{P}_{-1,-1}\mathcal{M}'_{5}
\end{split}
\end{equation}

But from \eqref{ordz3} we get
\begin{equation}
\label{subsoftordz2}
\begin{split}
& \lim_{i\lambda_{5} \to -1} (1+i\lambda_{5}) \mathcal{M}_{6} \bigg|_{\mathcal{O}(z_{56})} = z_{56} \left[ -(1+i\lambda_{6}) \mathcal{P}_{-3,0} + \mathcal{L}_{-1} \mathcal{P}_{-2,0} \right]\mathcal{M}'_{5}
\end{split}
\end{equation}

Therefore in order for our result \eqref{ordz3} to be consistent with the subleading conformal soft theorem, we should have 
\begin{equation}
\label{ordznst}
\begin{split}
&  \mathcal{L}_{-1} \mathcal{P}_{-2,0} \mathcal{M}'_{5} = (1+i\lambda_{6})  \mathcal{P}_{-3,0}  \mathcal{M}'_{5}- \mathcal{J}^{1}_{-2}\mathcal{P}_{-1,-1} \mathcal{M}'_{5}
\end{split}
\end{equation}

Replacing $i\lambda_{6}\rightarrow 1+i\lambda_{5}+i\lambda_{6}$ in the above we can re-write it as
\begin{equation}
\label{ordznst1}
\begin{split}
& \left( \mathcal{L}_{-1} \mathcal{P}_{-2,0} + \mathcal{J}^{1}_{-2}\mathcal{P}_{-1,-1} -  (2+ i\lambda_{5}+i\lambda_{6})  \mathcal{P}_{-3,0} \right) \mathcal{M}_{5} = 0
\end{split}
\end{equation}

We have thus obtained another decoupling relation. Using this we can  also express \eqref{ordz3} as follows
\begin{equation}
\label{ordz4}
\begin{split}
& \mathcal{M}_{6} \Big|_{\mathcal{O}(z_{56})} =  z_{56} \ B(i\lambda_{5},i\lambda_{6}) \left[ (1+ i\lambda_{5})\mathcal{P}_{-3,0} -  \frac{i\lambda_{5}}{i\lambda_{5}+i\lambda_{6}} \  \mathcal{J}^{1}_{-2}\mathcal{P}_{-1,-1}  \right] \mathcal{M}_{5}
\end{split}
\end{equation}


\subsubsection{Subleading terms: $\mathcal{O}(\bar{z}_{56})$}


At $\mathcal{O}(\bar{z}_{56})$, we have the following contribution from $\mathcal{F}$
\begin{equation}
\label{ordzbar}
\begin{split}
 \mathcal{F}\Big|_{\mathcal{O}(\bar{z}_{56})} & =  - \bar{z}_{56} \sum_{i=1}^{3} z_{i4}  \hspace{0.05cm} \sigma_{i,1} \left(\frac{1}{z_{i6}z_{46}}+  \frac{1}{z^{2}_{46}} \right) =  \bar{z}_{56} \sum_{i=1}^{4} \frac{ \sigma_{i,1}}{z_{i6}}
\end{split}
\end{equation}

where again we have applied the identity \eqref{sigmarel2}. Then from $\mathcal{I}(t)$ we have
\begin{equation}
\label{ordzbar1}
\begin{split}
 \mathcal{I}(t)\Big|_{\mathcal{O}(\bar{z}_{56})} & =  \bar{z}_{56} \hspace{0.03cm} t \bigg(  (3+i\lambda_{1}) \frac{\sigma_{1,3}}{\sigma_{1,1} } +  (3+i\lambda_{2}) \frac{\sigma_{2,3}}{\sigma_{2,1} }  +  (i\lambda_{3}-1) \frac{\sigma_{3,3}}{\sigma_{3,1} }+ (i\lambda_{4}-1) \frac{\sigma_{4,3}}{\sigma_{4,1} } \\
 & - \frac{(4+i\Lambda)}{\mathcal{U}_{1}} \hspace{0.03cm}  \mathcal{U}_{3} \bigg) \\
 & \equiv \bar{z}_{56}  \hspace{0.03cm} t \hspace{0.05cm}  \mathcal{I}_{0,1}
\end{split}
\end{equation}

Besides the above we also need to include the relevant terms form the preceding orders, whose products can generate contributions at $\mathcal{O}(\bar{z}_{56})$.  Taking such terms into account we finally get
\begin{equation}
\label{ordzbar2}
\begin{split}
  \mathcal{M}_{6} \Big|_{\mathcal{O}(\bar{z}_{56})} &=     \bar{z}_{56}  \hspace{0.03cm} B(i\lambda_{5},i\lambda_{6}) \Bigg[ \frac{i\lambda_{5}}{i\lambda_{5}+i\lambda_{6}} \bigg(- \mathcal{I}_{0,1}\sum_{i=1}^{4} \frac{\bar{z}_{i6} }{z_{i6}} \hspace{0.05cm} \sigma_{i,1} -  \mathcal{I}_{1,0} \bigg)+   \sum_{i=1}^{4} \frac{\sigma_{i,1} }{z_{i6}}  \Bigg] \mathcal{P}_{-1,-1} \mathcal{M}_{5}  
\end{split}
\end{equation}

Now using the results in \eqref{Lmin1Pmin15pt1}, \eqref{Lbarmin1Pmin205pt1} and \eqref{Pminnmin15pt} from the Appendix, we can express the above as
\begin{equation}
\label{ordzbar3}
\begin{split}
&  \mathcal{M}_{6} \Big|_{\mathcal{O}(\bar{z}_{56})}  = \bar{z}_{56}  \hspace{0.03cm} B(i\lambda_{5},i\lambda_{6}) \left[ \frac{i\lambda_{5}}{i\lambda_{5}+i\lambda_{6}} \left( \bar{\mathcal{L}}_{-1}\mathcal{P}_{-2,0} -  \mathcal{L}_{-1}\mathcal{P}_{-1,-1} \right) + \frac{i\lambda_{5}-i\lambda_{6}}{i\lambda_{5}+i\lambda_{6}} \ \mathcal{P}_{-2,-1}  \right]\mathcal{M}_{5} 
\end{split}
\end{equation}

Once again let us consider the subleading conformal soft limit $i\lambda_{5}\rightarrow -1$ in the above result. This yields
\begin{equation}
\label{subsoftordzbar1}
\begin{split}
& \lim_{i\lambda_{5} \to -1} (1+i\lambda_{5}) \mathcal{M}_{6} \bigg|_{\mathcal{O}(\bar z_{56})} = \bar{z}_{56} \left[ \bar{\mathcal{L}}_{-1} \mathcal{P}_{-2,0} - \mathcal{L}_{-1} \mathcal{P}_{-1,-1} + (1+i\lambda_{6}) \mathcal{P}_{-2,-1}  \right]\mathcal{M}'_{5}
\end{split}
\end{equation}

But according to the subleading conformal soft theorem \eqref{se},  we should get
\begin{equation}
\label{subsoftordzbar2}
\begin{split}
& \lim_{i\lambda_{5} \to -1} (1+i\lambda_{5}) \mathcal{M}_{6} \bigg|_{\mathcal{O}(\bar z_{56})} = 2 \hspace{0.04cm} \bar{z}_{56} \ \mathcal{J}^{0}_{-1}\mathcal{P}_{-1,-1}\mathcal{M}'_{5}
\end{split}
\end{equation}

Comparing \eqref{subsoftordzbar1} and \eqref{subsoftordzbar2} and shifting $i\lambda_{6}\rightarrow 1+i\lambda_{5}+i\lambda_{6}$ we then get
\begin{equation}
\label{ordzbarnst}
\begin{split}
&  \( \mathcal L_{-1}\mathcal P_{-1,-1} + 2 \mathcal J^{0}_{-1} \mathcal P_{-1,-1} - (2 + i\lambda_5 + i\lambda_6) \mathcal P_{-2,-1} - \mathcal{\bar L}_{-1} \mathcal P_{-2,0}\)\mathcal{M}_{5} =0
\end{split}
\end{equation}

We will discuss the role of this decoupling relation in determining the celestial OPE coefficients of gravitons using the extended symmetry algebra in forthcoming sections of this paper. Let us also note that using this relation we can write \eqref{ordzbar3} as follows
\begin{equation}
\label{ordzbar4}
\begin{split}
  \mathcal{M}_{6} \bigg|_{\mathcal{O}(\bar{z}_{56})} & = \bar{z}_{56}  \hspace{0.03cm} B(i\lambda_{5},i\lambda_{6}) \left[ \frac{2 i\lambda_{5}}{i\lambda_{5}+i\lambda_{6}} \ \mathcal{J}^{0}_{-1} \mathcal{P}_{-1,-1}  - (1+i\lambda_{5}) \mathcal{P}_{-2,-1}  \right]\mathcal{M}_{5} 
\end{split}
\end{equation}


\subsubsection{Subleading terms: $\mathcal{O}(z_{56}\bar{z}_{56})$}

Finally let us consider the $\mathcal{O}(z_{56}\bar{z}_{56})$ terms in the OPE decomposition of the $6$-point amplitude.  Here we have from $\mathcal{F}$ 
\begin{equation}
\label{ordzzbar}
\begin{split}
 \mathcal{F}\Big|_{\mathcal{O}(z_{56}\bar{z}_{56})} & =  - z_{56} \bar{z}_{56} \sum_{i=1}^{3} z_{i4} \hspace{0.05cm} \sigma_{i,1} \left(\frac{1}{z^{2}_{i6}z_{46}}+ \frac{1}{z_{i6}z^{2}_{46}} +  \frac{1}{z^{3}_{46}} \right) = z_{56} \bar{z}_{56} \sum_{i=1}^{3} \frac{ \sigma_{i,1}}{z^{2}_{i6}} 
\end{split}
\end{equation}

The $\mathcal{O}(z_{56}\bar{z}_{56})$ contribution from $\mathcal{I}(t)$ is given by
\begin{equation}
\label{ordzzbar1}
\begin{split}
 \mathcal{I}(t)\Big|_{\mathcal{O}(z_{56}\bar{z}_{56})} &  = z_{56}\bar{z}_{56}  \left( t \hspace{0.05cm}  \mathcal{I}^{(1)}_{1,1} + t^{2} \hspace{0.05cm}  \mathcal{I}^{(2)}_{1,1}  \right)
\end{split}
\end{equation}

where
\begin{equation}
\label{ordzzbar2}
\begin{split}
& \mathcal{I}^{(1)}_{1,1}=   (3+i\lambda_{1}) \frac{\sigma_{1,4}}{\sigma_{1,1}}+ (3+i\lambda_{2}) \frac{\sigma_{2,4}}{\sigma_{2,1}}+ (i\lambda_{3}-1) \frac{\sigma_{3,4}}{\sigma_{3,1}} + (i\lambda_{4}-1) \frac{\sigma_{4,4}}{\sigma_{4,1}}- \frac{(4+i\Lambda)}{\mathcal{U}_{1}}   \hspace{0.03cm}  \mathcal{U}_{4} 
\end{split}
\end{equation}

and
\begin{equation}
\label{ordzzbar3}
\begin{split}
\mathcal{I}^{(2)}_{1,1}  = \mathcal{I}_{1,0}  \mathcal{I}_{0,1} - \Big[ & (3+i\lambda_{1}) \frac{\sigma_{1,2}\sigma_{1,3}}{\sigma^{2}_{1,1}} +  (3+i\lambda_{2}) \frac{\sigma_{2,2}\sigma_{2,3}}{\sigma^{2}_{2,1}} +  (i\lambda_{3}-1) \frac{\sigma_{3,2}\sigma_{3,3}}{\sigma^{2}_{3,1}}  \\
& + (i\lambda_{4}-1) \frac{\sigma_{4,2}\sigma_{4,3}}{\sigma^{2}_{4,1}}  - \frac{(4+i\Lambda)}{\mathcal{U}^{2}_{1}}  \ \mathcal{U}_{2} \hspace{0.03cm}  \mathcal{U}_{3} \Big]
\end{split}
\end{equation}

$\mathcal{I}_{1,0}, \mathcal{I}_{0,1}$ were defined in \eqref{ordz1} and \eqref{ordzbar1}. Now we also need to include the $\mathcal{O}(z^{2}_{56})$ term from $\mathcal{I}(t)$ since this can combine with the leading $\mathcal{O}(\bar{z}_{56}z^{-1}_{56})$ term from $\mathcal{F}$ and generate an $\mathcal{O}(\bar{z}_{56})$ contribution. So, we have
\begin{equation}
\label{ordzzbar4}
\begin{split}
 \mathcal{I}(t)\Big|_{\mathcal{O}(z^{2}_{56})}  &  = z^{2}_{56} \hspace{0.05cm} t^{2} \hspace{0.05cm}  \mathcal{I}_{2,0}
\end{split}
\end{equation}

where
\begin{equation}
\label{ordzzbar5}
\begin{split}
 \mathcal{I}_{2,0}  = & - \frac{1}{2} \bigg[(3+i\lambda_{1})  \frac{\sigma^{2}_{1,2}}{ \sigma^{2}_{1,1}} + (3+i\lambda_{2})\frac{\sigma^{2}_{2,2}}{ \sigma^{2}_{2,1}}  + (i\lambda_{3}-1) \frac{\sigma^{2}_{3,2}}{  \sigma^{2}_{3,1}} + (i\lambda_{4}-1) \frac{\sigma^{2}_{4,2}}{ \sigma^{2}_{4,1}}  - \frac{(4+i\Lambda)}{\mathcal{U}^{2}_{1}} \hspace{0.04cm}  \mathcal{U}^{2}_{2}\bigg] \\
 & + \frac{1}{2} \ \mathcal{I}^{2}_{1,0}
\end{split}
\end{equation}

Thus after accounting for all the relevant contributions from $\mathcal{F}$ and $\mathcal{I}(t)$, we get
\begin{equation}
\label{ordzzbar6}
\begin{split}
 \mathcal{M}_{6}\Big|_{\mathcal{O}(z_{56}\bar{z}_{56})} & = z_{56}\bar{z}_{56} \  B(i\lambda_{5},i\lambda_{6}) \Big[   \frac{i\lambda_{5}}{i\lambda_{5}+i\lambda_{6}} \left( \mathcal{I}_{0,1} \hspace{0.04cm}  \mathcal{P}_{-3,0} -    \mathcal{I}_{1,0} \hspace{0.04cm} \mathcal{P}_{-2,-1} + \mathcal{I}^{(1)}_{1,1} \hspace{0.04cm} \mathcal{P}_{-2,0} \right)\\
& +  \frac{i\lambda_{5}(i\lambda_{5}+1)}{(i\lambda_{5}+i\lambda_{6})(i\lambda_{5}+i\lambda_{6}+1)}  \left(\mathcal{I}^{(2)}_{1,1} \hspace{0.04cm} \mathcal{P}_{-2,0} - \mathcal{I}_{2,0}   \hspace{0.04cm}  \mathcal{P}_{-1,-1}  \right) -\mathcal{P}_{-3,-1}  \Big] \mathcal{M}_{5}
\end{split}
\end{equation}

Now we want to express the above completely in terms of symmetry generators acting on the $5$-point Mellin amplitude. To that end, using \eqref{Lmin1Pmin25pt1}, \eqref{Lbarmin1Pmin205pt1} and \eqref{Lbarmin1Pmin305pt1} from the Appendix, Section \eqref{desccorr}, it can be shown that
\begin{equation}
\label{ordzzbar7}
\begin{split}
& \left( \mathcal{I}_{0,1} \hspace{0.04cm}  \mathcal{P}_{-3,0} -    \mathcal{I}_{1,0} \hspace{0.04cm} \mathcal{P}_{-2,-1} + \mathcal{I}^{(1)}_{1,1} \hspace{0.04cm} \mathcal{P}_{-2,0} \right)\mathcal{M}_{5} \\
& =  \mathcal{L}_{-1}\mathcal{P}_{-2,-1}\mathcal{M}_{5}-  \bar{\mathcal{L}}_{-1}\mathcal{P}_{-3,0}\mathcal{A}_{5} + \mathcal{L}_{-1}\bar{\mathcal{L}}_{-1} \mathcal{P}_{-2,0}\mathcal{M}_{5} - \mathcal{I}^{(2)}_{1,1} \hspace{0.04cm}  \mathcal{P}_{-2,0} \mathcal{M}_{5}\\
&   - \mathcal{L}_{-1} \left(\sum_{i=1}^{4}\frac{1}{z_{i6}} \ \mathcal{P}_{-1,-1}\mathcal{M}_{5} \right)  - \sum_{i=1}^{4}\frac{\bar{z}_{i6}}{z_{i6}^{2}} \ \sigma_{i,3}\  \mathcal{P}_{-1,-1}\mathcal{M}_{5} - \sum_{i=1}^{4}\frac{ \sigma_{i,2}}{z_{i6}} \  \mathcal{P}_{-1,-1}\mathcal{M}_{5} \\
&+ \sum_{i=1}^{4}\frac{ 1}{z_{i6}} \  \mathcal{P}_{-2,-1}\mathcal{M}_{5} 
\end{split}
\end{equation}

The last three terms in \eqref{ordzzbar7} can be simplified using the following identity
\begin{equation}
\label{ordzzbar8}
\begin{split}
& - \sum_{i=1}^{4}\frac{\bar{z}_{i6}}{z_{i6}^{2}} \ \sigma_{i,3} - \sum_{i=1}^{4} \frac{\sigma_{i,2} }{z_{i6}} - \left( \sum_{i=1}^{4} \frac{1}{z_{i6}} \right)  \sum_{j=1}^{4} \frac{\sigma_{j,1}}{z_{j6}} =   \frac{1}{2} \left( \sum_{i=1}^{4}\frac{1}{z_{i6}} \right)^{2} + \sum_{i=1}^{4}\frac{1}{2 z^{2}_{i6}} 
\end{split}
\end{equation}

Let us also take note of the following relation
\begin{equation}
\label{ordzzbar9}
\begin{split}
 2  \hspace{0.04cm}  \mathcal{I}_{2,0} \hspace{0.04cm}  \mathcal{P}_{-1,-1}\mathcal{M}_{5} & =  \mathcal{L}^{2}_{-1} \mathcal{P}_{-1,-1}\mathcal{M}_{5}  - 2 \hspace{0.04cm} \mathcal{L}_{-1} \left( \sum_{i=1}^{4}\frac{1}{z_{i6}} \ \mathcal{P}_{-1,-1}\mathcal{M}_{5} \right) \\
 &  +\bigg[ \sum_{i=1}^{4}\frac{1}{z^{2}_{i6}}  + \bigg( \sum_{i=1}^{4}\frac{1}{z_{i6}} \bigg)^{2} \bigg]\mathcal{P}_{-1,-1}\mathcal{M}_{5} 
\end{split}
\end{equation}

Then using the last two equations \eqref{ordzzbar8} and \eqref{ordzzbar9} we get
\begin{equation}
\label{ordzzbar10}
\begin{split}
& \left( \mathcal{I}_{0,1} \hspace{0.04cm}  \mathcal{P}_{-3,0} -    \mathcal{I}_{1,0} \hspace{0.04cm} \mathcal{P}_{-2,-1} + \mathcal{I}^{(1)}_{1,1} \hspace{0.04cm} \mathcal{P}_{-2,0} \right)\mathcal{M}_{5} \\
& = \left(\mathcal{L}_{-1}\mathcal{P}_{-2,-1}-  \bar{\mathcal{L}}_{-1}\mathcal{P}_{-3,0} + \mathcal{L}_{-1}\bar{\mathcal{L}}_{-1} \mathcal{P}_{-2,0} - \frac{1}{2} \mathcal{L}^{2}_{-1} \mathcal{P}_{-1,-1}\right)\mathcal{M}_{5}  \\
& +  \left( \mathcal{I}_{2,0} \hspace{0.04cm}  \mathcal{P}_{-1,-1}  - \mathcal{I}^{(2)}_{1,1} \hspace{0.04cm}\mathcal{P}_{-2,0}\right) \mathcal{M}_{5} 
\end{split}
\end{equation}

Now in order to express the last set of terms in \eqref{ordzzbar10} in terms of the action of symmetry generators on the $5$-point amplitude, it is convenient to appeal to the subleading conformal soft theorem according to which we must have
\begin{equation}
\label{ordzzbar11}
\begin{split}
 \lim_{i\lambda_{5}\to -1} (1+i\lambda_{5}) \mathcal{M}_{6}\Big|_{\mathcal{O}(z_{56}\bar{z}_{56})} & = z_{56}\bar{z}_{56}  \ 2 \hspace{0.04cm} \mathcal{J}^{0}_{-2} \mathcal{P}_{-1,-1} \mathcal{M}'_{5}
 \end{split}
\end{equation}

This implies that upon using \eqref{ordzzbar10} in \eqref{ordzzbar6} we ought to get
\begin{equation}
\label{ordzzbar12}
\begin{split}
&\left(\mathcal{I}^{(2)}_{1,1} \hspace{0.04cm}\mathcal{P}_{-2,0} - \mathcal{I}_{2,0} \hspace{0.04cm}  \mathcal{P}_{-1,-1}  \right)\mathcal{M}'_{5}\\
& = \bigg( \mathcal{L}_{-1}\mathcal{P}_{-2,-1} -  \bar{\mathcal{L}}_{-1}\mathcal{P}_{-3,0}+ \mathcal{L}_{-1}\bar{\mathcal{L}}_{-1} \mathcal{P}_{-2,0} - \frac{1}{2} \mathcal{L}^{2}_{-1} \mathcal{P}_{-1,-1} \\
& \hspace{1.0cm}+ ( i\lambda_{6}-1) \mathcal{P}_{-3,-1} \mathcal{M}'_{5} - 2 \hspace{0.04cm} \mathcal{J}^{0}_{-2}  \mathcal{P}_{-1,-1} \bigg)\mathcal{M}'_{5}
\end{split}
\end{equation}

Then assembling the above results and shifting $i\lambda_{6}\rightarrow i\lambda_{5}+i\lambda_{6}+1$ in \eqref{ordzzbar12} we finally obtain
\begin{equation}
\label{ordzzbar13}
\begin{split}
&  \mathcal{M}_{6}\Big|_{\mathcal{O}(z_{56}\bar{z}_{56})} \\
& =  z_{56}\bar{z}_{56} \  B(i\lambda_{5},i\lambda_{6}) \Bigg[    \frac{2  \hspace{0.03cm}  i\lambda_{5} \hspace{0.03cm} i\lambda_{6}}{(i\lambda_{5}+i\lambda_{6})(i\lambda_{5}+i\lambda_{6}+1)}  \hspace{0.04cm} \mathcal{J}^{0}_{-2} \mathcal{P}_{-1,-1} - \left( 1+ \frac{i\lambda_{5}\hspace{0.03cm} i \lambda_{6} }{i\lambda_{5}+i\lambda_{6}+1}  \right)  \mathcal{P}_{-3,-1} \\ 
& +  \frac{i\lambda_{5}(1+i\lambda_{5})}{(i\lambda_{5}+i\lambda_{6})(i\lambda_{5}+i\lambda_{6}+1)} \left( \mathcal{L}_{-1}\mathcal{P}_{-2,-1} -  \bar{\mathcal{L}}_{-1}\mathcal{P}_{-3,0}+ \mathcal{L}_{-1}\bar{\mathcal{L}}_{-1} \mathcal{P}_{-2,0} - \frac{1}{2} \mathcal{L}^{2}_{-1} \mathcal{P}_{-1,-1} \right)\Bigg] \mathcal{M}_{5} 
\end{split}
\end{equation}


\section{Summary : Celestial OPE from MHV Mellin amplitude}

Using the results obtained in \eqref{leadcsope}, \eqref{ord1b}, \eqref{ordz4}, \eqref{ordzbar4} and \eqref{ordzzbar13}, we can now extract the celestial OPE for outgoing positive helicity gravitons. This takes the following form
\begin{equation}
\label{csope6ptfinal}
\begin{split}
  & G^{+}_{\Delta_{5}} (z_{5},\bar{z}_{5}) G^{+}_{\Delta_{6}} (z_{6},\bar{z}_{6}) \Big|_{MHV}  \\
  &=  B(i\lambda_{5},i\lambda_{6}) \bigg[ - \frac{\bar{z}_{56}}{z_{56}}  \hspace{0.04cm} P_{-1,-1}  +  P_{-2,0} + z_{56} \bigg\{ (1+i\lambda_{5})P_{-3,0} - \frac{i\lambda_{5}}{i\lambda_{5}+i\lambda_{6}} \hspace{0.04cm} J^{1}_{-2}P_{-1,-1} \bigg\}\\
 &  + \bar{z}_{56} \bigg\{\frac{2i\lambda_{5}}{i\lambda_{5}+i\lambda_{6}} \hspace{0.04cm} J^{0}_{-1}P_{-1,-1} - (1+i\lambda_{5})P_{-2,-1}  \bigg\} \\
 &+ z_{56}\bar{z}_{56} \bigg\{  \frac{2  \hspace{0.03cm}  i\lambda_{5} \hspace{0.03cm} i\lambda_{6}}{(i\lambda_{5}+i\lambda_{6})(i\lambda_{5}+i\lambda_{6}+1)}  \hspace{0.04cm} J^{0}_{-2} P_{-1,-1} -  \bigg( 1+ \frac{i\lambda_{5}\hspace{0.03cm} i \lambda_{6} }{i\lambda_{5}+i\lambda_{6}+1}  \bigg)  P_{-3,-1} \\
 & + \frac{i\lambda_{5}(1+i\lambda_{5})}{2(i\lambda_{5}+i\lambda_{6})(i\lambda_{5}+i\lambda_{6}+1)} \bigg( 2 L_{-1}P_{-2,-1} - 2 \bar{L}_{-1}P_{-3,0}+ 2L_{-1}\bar{L}_{-1} P_{-2,0} -  L^{2}_{-1} P_{-1,-1} \bigg) \bigg\} \\
 & +\cdots \bigg] G^{+}_{\D} (z_{6},\bar{z}_{6}) 
\end{split}
\end{equation}

where $\D = \D_5 + \D_6 -1$ and the dots denote higher order terms in the OPE which we have not considered here. The subscript $MHV$ in \eqref{csope6ptfinal} denotes that this particular representation of the celestial OPE in terms of the extended symmetry algebra discussed in this paper, holds as an operator statement only within MHV graviton amplitudes. 

Now the other main result from this section involves the following differential equations 
\begin{equation}
\begin{split}
& \left[ \mathcal{J}^{1}_{-1}(6)\mathcal{P}_{-1,-1}(6) -  (\D -1) \mathcal{P}_{-2,0}(6) \right] \left \langle  G^{-}_{\Delta_{1}}(1) G^{-}_{\Delta_{2}}(2) G^{+}_{\Delta_{3}}(3)  G^{+}_{\Delta_{4}}(4) G^{+}_{\D}  (6)\right\rangle =0
\end{split}
\end{equation} 

\begin{equation}
\begin{gathered}
 \left[ \mathcal{L}_{-1}(6) \mathcal{P}_{-2,0}(6) + \mathcal{J}^{1}_{-2}(6)\mathcal{P}_{-1,-1}(6) -   (\Delta+1) \mathcal{P}_{-3,0}(6) \right] \\
\times  \left \langle  G^{-}_{\Delta_{1}}(1) G^{-}_{\Delta_{2}}(2) G^{+}_{\Delta_{3}}(3)  G^{+}_{\Delta_{4}}(4) G^{+}_{\D}  (6)\right\rangle = 0
\end{gathered}
\end{equation}

and
\begin{equation}
\begin{gathered}
 \[ \mathcal L_{-1}(6)\mathcal P_{-1,-1}(6) + 2 \mathcal J^{0}_{-1}(6) \mathcal P_{-1,-1}(6) - (\D+1) \mathcal P_{-2,-1}(6) - \mathcal{\bar L}_{-1}(6) \mathcal P_{-2,0}(6)\] \\ \times \left \langle  G^{-}_{\Delta_{1}}(1) G^{-}_{\Delta_{2}}(2) G^{+}_{\Delta_{3}}(3) G^{+}_{\Delta_{4}}(4) G^{+}_{\D}  (6)\right\rangle =0
\end{gathered}
\end{equation}

where $\Delta= 1+i\lambda_{5}+i\lambda_{6}$. These differential equations, which are equivalent to null state relations, played a crucial role in arriving at the above structure of the celestial OPE in \eqref{csope6ptfinal}. 
In subsequent sections of this paper we will illustrate how the leading as well as as subleading OPE coefficients can be obtained using these differential equations for MHV amplitudes and the extended symmetry algebra. The results of this section will then serve as an important check of the symmetry based analysis.


\section{Null states and differential equations for MHV amplitudes}\label{NULL}
Let us consider the graviton primary $G^{\sigma}_{\D}(z,\bar z)$ where $\sigma=\pm 2$ is the helicity. This is a conformal primary as well as a primary under the $\overline{SL(2,\mathbb{C})}$ current algebra and supertranslations, obtained from positive helicity soft graviton. Using the commutation relations and definition of a primary state one can check that this is also true for the operator $P_{-1,-1}G^{\sigma}_{\D}(z,\bar z)$. 

\subsection{$\Phi^{\sigma}$}
We now consider the following descendants of $G^{\sigma}_{\D}(z,\bar z)$ given by
\be
\phi_1 = J^1_{-1}P_{-1,-1}G^{\sigma}_{\D}(z,\bar z) , \quad  \phi_2 = P_{-2,0}G^{\sigma}_{\D}(z,\bar z)
\ee

They have the scaling dimensions $\(h + 3/2, \bar h - 1/2\)$. One can easily check that these are the only possible descendants with the scaling dimensions $\(h + 3/2, \bar h - 1/2\)$. 

 \vskip 4pt
Now $\phi_1$ and $\phi_2$ are primaries of the Poincare group, i.e, 
\be
L_1\phi_i = \bar L_1\phi_i = P_{0,-1}\phi_i = P_{-1,0}\phi_{i} = 0, \quad i =1, 2
\ee

but not of the extended symmetry algebra. So we take a linear combination of them and construct the state,
\be
\Phi^{\sigma} = \phi_1 + c \phi_2
\ee

Now we impose the conditions
\be
J^a_n\Phi^{\sigma} =0, \quad n\ge 1
\ee

Using the commutators and the fact that $G^{\sigma}_{\D}$ is a primary we get,
\be
J^1_n\Phi^{\sigma} = J^0_n\Phi^{\sigma} = 0, \quad n\ge 1
\ee

Similarly, applying $J^{-1}_1$ we get,
\be
J^{-1}_1\Phi^{\sigma} = 0 \Longrightarrow \[ c + (2\bar h+1)\] P_{-1,-1}G^{\sigma}_{\D}(z,\bar z) = 0 \Longrightarrow c = -\(2\bar h+1\)
\ee

So the primary descendant $\Phi^{\sigma}$ is given by, 
\be\label{null1}
\boxed{
\Phi^{\sigma}(z,\bar z) = \[J^1_{-1}P_{-1,-1} - (2\bar h +1) P_{-2,0} \] G^{\sigma}_{\D}(z,\bar z)}
\ee

The other generators $J^a_n$ for $n\ge 2$ automatically annihilate the state. Also we have not imposed the vanishing of the supertranslation generators of the form $P_{n,-1}, n\ge 1$ and $P_{n,0}, n\ge 0$, because the commutators 
\be
[J^0_m, P_{0,-1}] = \frac{1}{2} P_{m,-1}, \quad \[J^0_m, P_{-1,0}\] = - \frac{1}{2}P_{m-1,0}
\ee 
together with $P_{-1,0}\Phi^{\sigma} = P_{0,-1}\Phi^{\sigma} = J^0_{m\ge 1}\Phi^{\sigma} = 0$, imply this. $\Phi^{\sigma}(z,\bar z)$ is the most interesting null-state because its existence solely requires the \textit{local current algebra symmetry}. 

 \vskip 4pt
Finally, we set \eqref{null1} to zero inside a MHV amplitude,
\be\label{de1mhv}
\bigg\langle{ \[J^1_{-1}P_{-1,-1} - (2\bar h +1) P_{-2,0} \] G^{\sigma}_{\D}(z,\bar z) \prod_i G^{\sigma_i}_{\D_i}(z_i,\bar z_i)}\bigg\rangle_{MHV} = 0, \quad \sigma = \pm 2
\ee

Now using the fact that the correlation functions of the descendants can be written in terms of correlation functions of primaries only, we get the following equation
\be\label{de1}
\boxed{
\[\mathcal J^1_{-1}\mathcal P_{-1,-1} - (2\bar h +1) \mathcal P_{-2,0} \] \bigg\langle{G^{\sigma}_{\D}(z,\bar z) \prod_i G^{\sigma_i}_{\D_i}(z_i,\bar z_i)}\bigg\rangle_{MHV} = 0, \quad \sigma = \pm 2}
\ee

where
\be
\mathcal J^1_{-1} =  - \sum_{i} \frac{2\bar h_i(\bar z_i - \bar z) + (\bar z_i - \bar z)^2 \bar\partial_i}{z_i - z}, \quad \mathcal P_{-2,0} = - \sum_{i} \frac{\bar z_i - \bar z}{z_i - z} \epsilon_i \mathcal P_i
\ee

where $\epsilon_i = \pm 1$ for an outgoing (incoming) particle and $\mathcal P_i G^{\sigma_j}_{\D_j}(z,\bar z)=\delta_{ij} G^{\sigma_j}_{\D_j +1}(z_j,\bar z_j)$.

 \vskip 4pt
For an explicit check of the equation \eqref{de1} for $n$ point MHV scattering amplitude we refer the reader to the Appendix \eqref{hodge}. 

\subsection{$\Psi$}
Let us consider the following descendants of $G^{\sigma}_{\D}(z,\bar z)$ \footnote{To be precise, we should write $J^{-1}_0$ instead of $\bar L_{-1}$. But for the sake of clarity we use the more familiar notation $\bar L_{-1}$.} 
\be
L_{-1}P_{-1,-1}G^{\sigma}_{\D}(z,\bar z), \quad J^0_{-1}P_{-1,-1}G^{\sigma}_{\D}(z,\bar z), \quad P_{-2,-1} G^{\sigma}_{\D}(z,\bar z), \quad \bar L_{-1}P_{-2,0}G^{\sigma}_{\D}(z,\bar z)
\ee

They all have the same scaling dimensions given by $(h+3/2, \bar h + 1/2)$. In the above list we could have added one more term given by $\bar L_{-1}J^1_{-1}P_{-1,-1}G^{\sigma}_{\D}(z,\bar z)$, but note that this term is proportional to $\bar L_{-1}P_{-2,0}G^{\sigma}_{\D}(z,\bar z)$ because of the vanishing condition \eqref{de1mhv}. Now we construct a state $\Psi$ which is a linear combination of the above four states and also a primary under the extended symmetry algebra. So we write
\be
\Psi = \( L_{-1}P_{-1,-1} + c_1 J^{0}_{-1} P_{-1,-1} + c_2 P_{-2,-1} + c_3 \bar L_{-1} P_{-2,0}\) G^{\sigma}_{\D}(z,\bar z)
\ee

where $c_1,c_2,c_3$ are constants to be determined. We fist impose the conditions,
\be
L_1\Psi = \bar L_1\Psi = P_{0,-1}\Psi= P_{-1,0}\Psi=0
\ee

These conditions, together with the vanishing condition \eqref{de1mhv}\footnote{We need to use the vanishing of the primary descendant $\Phi^{\sigma}$ because $\bar L_1$ acting on $\Psi$ produces a state proportional to $J^1_{-1}P_{-1,-1}G^{\sigma}_{\D}(z,\bar z)$. We then replace the this state with $(2\bar h+1) P_{-2,0}G^{\sigma}_{\D}(z,\bar z)$ using \eqref{de1mhv}.}, give the following result
\be
c_1 = 2, \quad c_2 = -(\D+1), \quad c_3 = -1, \qquad \sigma = +2
\ee

Here $\sigma = +2$ means that the \textit{primary descendant $\Psi$ can exist only for \underline{positive} helicity gravitons $G^{+}_{\D}(z,\bar z)$}. Now, it turns out that for the above values of $c_1,c_2,c_3$ and $\sigma=+2$, the operator $\Psi$ is also annihilated by $J^a_{n>0}$, $P_{n\ge 0, -1}$ and $P_{n\ge -1, 0}$. Therefore
\be\label{null2}
\boxed{
\Psi = \( L_{-1}P_{-1,-1} + 2 J^{0}_{-1} P_{-1,-1} - (\D+1) P_{-2,-1} - \bar L_{-1} P_{-2,0}\) G^{+}_{\D}(z,\bar z)}
\ee 

is a primary descendant of the extended symmetry algebra. 

 \vskip 4pt
Finally we can set the state $\Psi$ to zero inside a MHV amplitude
\be\label{de2mhv}
\bigg\langle{\( L_{-1}P_{-1,-1} + 2 J^{0}_{-1} P_{-1,-1} - (\D+1) P_{-2,-1} - \bar L_{-1} P_{-2,0}\) G^{+}_{\D}(z,\bar z)}\prod_iG^{\sigma_i}_{\D_i}(z_i,\bar z_i)\bigg\rangle_{MHV} = 0
\ee

In terms of differential operators the vanishing condition \eqref{de2mhv} can be written as,
\be\label{de2}
\hspace{-0.1cm}
\boxed{
\( \mathcal L_{-1}\mathcal P_{-1,-1} + 2 \mathcal J^{0}_{-1} \mathcal P_{-1,-1} - (\D+1) \mathcal P_{-2,-1} - \mathcal{\bar L}_{-1} \mathcal P_{-2,0}\)\bigg\langle{G^{+}_{\D}(z,\bar z)}\prod_iG^{\sigma_i}_{\D_i}(z_i,\bar z_i)\bigg\rangle_{MHV} = 0}
\ee

where
\be
\begin{gathered}
\mathcal L_{-1} = \frac{\partial}{\partial z} , \quad \mathcal{ \bar L}_{-1} = \frac{\partial}{\partial \bar z} \\
\mathcal J^0_{-1} = - \sum_{i} \frac{\bar h_i +( \bar z_i - \bar z) \bar\partial_i}{z_i - z}, \quad 
\mathcal P_{-2,-1} = - \sum_{i} \frac{1}{z_i - z} \epsilon_i \mathcal P_i \\
\mathcal P_{-2,0} = - \sum_{i} \frac{\bar z_i - \bar z}{z_i - z} \epsilon_i \mathcal P_i
\end{gathered}
\ee

Here $\epsilon_i = \pm 1$ for an outgoing (incoming) particle and $\mathcal P_i G^{\sigma_j}_{\D_j}(z,\bar z)=\delta_{ij} G^{\sigma_j}_{\D_j +1}(z_j,\bar z_j)$. For a direct check of the equation \eqref{de2} for $5$ graviton MHV amplitude we refer the refer to Appendix \eqref{5direct}. We have also checked \eqref{de2} numerically for $6$ graviton MHV amplitude. 

 \vskip 4pt
Equation \eqref{de2} is a linear first order partial differential equation. For an $(n+2)$ point MHV amplitude with $n$ positive and two negative helicity gravitons, there are $n$ such equations. Now due to conformal invariance the $(n+2)$ point amplitude depends on $(n-1)$ variables because we can take three points to be fixed at $(1,0,\infty)$. So it appears that one may be able to solve \eqref{de1} and \eqref{de2} together and get the MHV amplitudes, at least in principle.  

 \vskip 4pt
We can also see that if we take the three point function $\langle{- - +}\rangle$ then the equation \eqref{de2} reduces to \eqref{fde} which we obtained earlier by requiring consistency with the subleading soft theorem. This happens because $P_{-2,0}$ annihilates the three point function as shown in \eqref{ls}.  

 \vskip 4pt
Let us now make few comments about the nature of these equations. First of all, note that both the holomorphic and the antiholomorphic derivatives of $G^+_{\D}(z,\bar z)$ appear in \eqref{de2}. This is reflection of the fact that the underlying infinite-dimensional symmetry algebra is not homomorphically factorizable.

\vskip 4pt
Secondly, the two sets of equations 
\be\label{e1}
\hspace{-0.1cm}
\( \mathcal L_{-1}\mathcal P_{-1,-1} + 2 \mathcal J^{0}_{-1} \mathcal P_{-1,-1} - (\D+1) \mathcal P_{-2,-1} - \mathcal{\bar L}_{-1} \mathcal P_{-2,0}\)\bigg\langle{G^{+}_{\D}(z,\bar z)}\prod_iG^{\sigma_i}_{\D_i}(z_i,\bar z_i)\bigg\rangle_{MHV} = 0
\ee  

and 
\be\label{e2}
\[\mathcal J^1_{-1}\mathcal P_{-1,-1} - (\D -1) \mathcal P_{-2,0} \] \bigg\langle{G^{+}_{\D}(z,\bar z) \prod_i G^{\sigma_i}_{\D_i}(z_i,\bar z_i)}\bigg\rangle_{MHV} = 0
\ee

are not independent. Here we have substituted $2\bar h + 1 = \D -1$ for a positive helicity graviton. The first set \eqref{e1} implies the second set \eqref{e2} due to special conformal invariance. In order to see this let us note that
\be
\[\bar L_1, L_{-1}P_{-1,-1} + 2 J^{0}_{-1} P_{-1,-1} - (\D+1) P_{-2,-1} - \bar L_{-1} P_{-2,0}\] = 2 \(J^1_{-1}P_{-1,-1} - (\D-1)P_{-2,0}\) 
\ee

This, together with \eqref{e1} and special conformal invariance, implies \eqref{e2}. In any case we will now see that these two equations can be used to obtain all the informations about the leading term in the OPE in the holomorphic collinear limit. 

\section{Differential equations for Fock space MHV amplitudes} \label{fspde}
As expected, the differential equations we have obtained for the Mellin space amplitudes can be transformed back to Fock space by making the following replacements,
\be
\D_i \rightarrow - \omega_i \frac{\partial}{\partial\omega_i}, \quad \mathcal P_i \rightarrow \omega_i, \quad \mathcal P_{-1,-1} \rightarrow \epsilon \hspace{0.03cm} \omega
\ee

So let us transform the equation \eqref{de2} to Fock space. We write the Fock space MHV amplitude as,
\be
\langle{a(\epsilon\omega, z,\bar z, \sigma=+2)\prod_{i} a(\epsilon_i\omega_i, z_i,\bar z_i, \sigma_i)}\rangle_{MHV}
\ee

where $\epsilon_k = \pm 1$ for an outgoing (incoming) particle. For an outgoing particle $a_k(\omega_k, z_k,\bar z_k,\sigma_k)$ is an annihilation operator whereas for an incoming particle $a_k(- \omega_k, z_k,\bar z_k,\sigma_k)$ is a creation operator. With this notation we can write the differential equation \eqref{de2} as,
\be
\begin{gathered}
\( \mathcal L_{-1}\mathcal P_{-1,-1} + 2 \mathcal J^{0}_{-1} \mathcal P_{-1,-1} - (\D+1) \mathcal P_{-2,-1} - \mathcal{\bar L}_{-1} \mathcal P_{-2,0}\) \\
\times \bigg\langle{a(\epsilon\omega, z,\bar z, \sigma=+2)\prod_{i} a(\epsilon_i\omega_i, z_i,\bar z_i, \sigma_i)}\bigg\rangle_{MHV} = 0
\end{gathered}
\ee

where 
\be
\begin{gathered}
\mathcal L_{-1} \mathcal P_{-1,-1} = \epsilon \omega \frac{\partial}{\partial z} \\
\mathcal J^0_{-1}\mathcal P_{-1,-1} = - \(\sum_{i} \frac{\frac{1}{2}\(-\omega_i \frac{\partial}{\partial\omega_i} - \sigma_i\) +( \bar z_i - \bar z) \bar\partial_i}{z_i - z}\) \omega \\
(\D + 1)\mathcal P_{-2,-1} = - \(- \omega \frac{\partial}{\partial\omega} +1\) \(\sum_{i} \frac{\epsilon_i\omega_i}{z_i - z}\) \\
\mathcal{\bar L}_{-1} \mathcal P_{-2,0} = - \frac{\partial}{\partial\bar z} \(\sum_{i} \frac{\bar z_i - \bar z}{z_i - z} \epsilon_i \omega_i\)
\end{gathered}
\ee

For an $(n+2)$ point MHV amplitude with two negative helicity gravitons there are $n$ such equations corresponding to $n$ positive helicity gravitons. Using the same prescription we can also transfer the differential equation \eqref{de1} to Fock space. 

\section{General comments on the structure of the OPE} \label{genope}
Consider two primaries $\phi_1$ and $\phi_2$. We write the contribution of a single primary $\phi_3$ to the $\phi_1\phi_2$ OPE as, 
\be\label{OPE}
\phi_1(z, \bar z) \phi_2(0) = \sum_{p,\bar p} C_{p\bar p} z^p \bar z^{\bar p} O_{p\bar p}(0) = z^{h_3- h_1-h_2} \bar z^{\bar h_3 - \bar h_1- \bar h_2} C_{123} \phi_3(0) + {\sum_{p,\bar p}}' z^p \bar z^{\bar p} C'_{p\bar p} O'_{p\bar p}(0)
\ee

where the prime over the summation means that the contribution of the primary operator $\phi_3$ has been subtracted. So $O'_{p\bar p}$ is a sum of descendants. For our purpose it is convenient to normalise the operator $O'_{p\bar p}$ in such a way that the coefficient $C'_{p\bar p}$ is either $0$ or $1$.  

The descendants are created by the operators $J^a_{n<0}$, $P_{n\le -2,0}$, $P_{n\le -1,-1}$ and $L_{-1}$ with scaling dimensions given by,
\be
J^a_{n}:(-n,-a), \quad P_{n,0}:\(-n - 1/2, -1/2\), \quad P_{n,-1}: \(-n - 1/2, 1/2\), \quad L_{-1}:(1,0)
\ee

We can see that they all have positive holomorphic scaling dimension and so \textit{the sum over $p$ is bounded from below}. The lowest value of $p$ corresponds to the primary operator and is given by,
\be
p_{min} = h_3 - h_1 - h_2
\ee

where $h_3$ is the scaling dimension of the primary. 

 \vskip 4pt
Now, a priori, there is no reason for the sum over $\bar p$ to be bounded from below because the generators $J^1_{n\le 0}$ and $P_{n\le -2,0}$ have \textit{negative} antiholomorphic scaling dimension. So starting from a primary one can create states with arbitrarily negative antiholomorphic scaling dimension. This in particular will require poles of arbitrarily high order in $\bar z$. This is of course not what we find if we start from the MHV amplitude. So we need to prove a stronger statement that the sum over $\bar p$ is not only bounded from below but $\bar p_{min}\ge 0$. We will now provide strong evidence that this in fact is true if we assume that \textit{both sides of the OPE transform in the same way under the local symmetry algebra}. 

 \vskip 4pt
We start with the generator $J^{-1}_1$ with scaling dimensions $(-1,1)$, which acts on a primary as
\be\label{action}
\[J^{-1}_1, \phi(z,\bar z)\] = z \bar\partial \phi(z,\bar z) \Rightarrow \[J^{-1}_1, \phi(0)\] = 0 
\ee
Note that the second equation in \eqref{action} follows from the definition of a primary operator. Now applying $J^{-1}_{1}$ to both sides of \eqref{OPE} we get, 
\be
z\bar \partial \phi_1(z,\bar z) \phi_2(0) = \sum_{p\bar p} \bar p C_{p\bar p} z^{p+1}\bar z^{\bar p -1} O_{p\bar p}(0)= \sum_{p\bar p} C_{p\bar p} z^p\bar z^{\bar p} \[J^{-1}_1, O_{p\bar p}(0)\]
\ee

Comparing powers of $z^p\bar z^{\bar p}$ on both sides we get the following recursion relation,
\be\label{rec}
\boxed{
C_{p\bar p}\[J^{-1}_1, O_{p\bar p}(0)\] = (\bar p +1) C_{p-1,\bar p +1} O_{p-1, \bar p +1}(0)} 
\ee

\vskip 4pt
In general equation \eqref{rec} is a system of equations because $O_{p\bar p}$ is a sum of descendants. In this paper we will not try to give a general proof that no poles in $\bar z$ arise, but merely satisfy ourselves by studying a concrete example. 

 \vskip 4pt
So consider $\phi_1$ and $\phi_2$ to be outgoing gravitons with $\phi_1$ being of positive helicity. In this case we can write, 
\be
\begin{gathered}
\phi_1(z,\bar z) \phi_2(0) = \frac{\bar z}{z} C_{123} P_{-1,-1}\phi_3(0) + \cdots \\
+ C_{0,0} P_{-2,0}\phi_3(0) + \cdots \\
+ \frac{z}{\bar z} C_{1,-1} O_{1,-1}(0) + \cdots
\end{gathered}
\ee

We want to show that $C_{1,-1}=0$. Now one can easily check that the only candidate for $O_{1,-1}(0)$ is
\be\label{propto}
O_{1,-1}(0) = J^{1}_{-1} P_{-2,0}\phi_3(0)
\ee

We can now use \eqref{rec} with $p=1$ and $\bar p = -1$ and get, 
\be
C_{1,-1}\[J^{-1}_1, O_{1,-1}(0)\] = 0
\ee

So $C_{1,-1}=0$ if $\[J^{-1}_1, O_{1,-1}(0)\]\ne 0$. Now using the commutation relations we get,
\be
\[J^{-1}_1, O_{1,-1}(0)\] = -2(\D_3 -2) P_{-2,0}\phi_3(0) \ne 0 \Longrightarrow C_{1,-1} = 0
\ee

Therefore the unwanted term in the OPE with a pole in $\bar z$ is ruled out by the local symmetry algebra. It is very likely that this method of proof can be generalised to arbitrary order but we leave that to future work. An all order proof will also imply that the sum over $\bar p$ is bounded from below. 

 \vskip 4pt
With a more refined understanding of the representation theory of this infinite dimensional symmetry algebra one should be able to derive this result from \textit{unitarity}, which is not at all clear at present. 

\subsection{More examples} 
In the above example the operator multiplying $z/\bar z$ was a descendant. But now let us consider cases where the operator multiplying $z/\bar z$ is a primary. So consider the following OPEs
\be\label{+-}
G^+_{\D_1}(z) G^-_{\D_2}(0) \sim \frac{z}{\bar z} C_1 G^+_{\D_1 + \D_2}(0)
\ee

and 
\be\label{--}
G^-_{\D_1}(z) G^-_{\D_2}(0) \sim \frac{z}{\bar z} C_2 G^-_{\D_1 + \D_2}(0)
\ee

According to equation \eqref{mhvope}, in the MHV sector of the Celestial CFT we should have $C_1=C_2=0$. It is simple to see by again applying $J^{-1}_{1}$ to both sides of the OPE that $C_1=C_2=0$. For example, applying this to \eqref{+-} we get,
\be
- C_1 \frac{z^2}{\bar z^2} G^+_{\D_1+\D_2} = 0 + \cdots
\ee

where we have used \eqref{action} and the fact that in the OPE \eqref{+-} double or higher order poles in $\bar z$ do not appear. So we have $C_1=0$ and similarly $C_2=0$. 

 \vskip 4pt
So we can see that any term with a pole in $\bar z$ cannot appear in the OPE if it is to be invariant under the $\overline{SL(2,\mathbb C)}$ current algebra. This is what we expect if we extract the OPE from the MHV amplitudes. This is consistent with the fact that the MHV sector of the Celestial CFT can be abstractly defined by declaring $\overline{SL(2,\mathbb C)}$ current algebra and supertranslations as the underlying symmetry. 


\section{Leading OPE coefficients from differential equations} \label{leadopede}

\subsection{Two outgoing (incoming) gravitons}

In this section we want to determine the leading OPE coefficients starting from the differential equations for the MHV amplitudes. We start with two outgoing gravitons one of which is positive helicity, denoted by $G^{+}_{\D}(z,\bar z)$, and the other one is $G^{\sigma_1}_{\D_1}(z_1,\bar z_1)$. The general OPE can be written as, 
\be\label{ope}
\begin{split}
G^+_{\D}(z,\bar z) G^{\sigma_1}_{\D_1}(z_1,\bar z_1) &= C_{pq}(\D,\D_1, \sigma_1) (z- z_1)^p(\bar z - \bar z_1)^q G^{\sigma_2}_{\D_2}(z_1,\bar z_1) \\ 
&+ \sum_{m,n}\sum_{i=1}^{a(m,n)} (z-z_1)^m (\bar z - \bar z_1)^n C^{i}_{mn} O^i_{mn}(z_1,\bar z_1)
\end{split}
\ee 

where $C$s are the OPE coefficients. The index $i$ distinguishes between different descendants of $G^{\sigma_2}_{\D_2}$ and $a(m,n)$ is the number of descendants  which is a function of $m$ and $n$. Here we have kept only the contribution of the primary $G^{\sigma_2}_{\D_2}$ to the OPE. 

 \vskip 4pt
According to the discussion in the last section $p$ should be the \textit{smallest} power of $(z-z_1)$ occurring in the OPE \eqref{ope} because the generators, which create descendants, all have positive holomorphic scaling dimension. Now there are an infinite number of terms in \eqref{ope} with coefficients of the form $(z-z_1)^p (\bar z - \bar z_1)^r$ but, since $G^{\sigma_1}_{\D_1}(z_1,\bar z_1)$ is a primary operator, $r$ starts from $q$. In other words, $q$ is the smallest power of $(\bar z- \bar z_1)$ occurring in the sum $(z-z_1)^p \sum_r \sum_{i=1}^{a(p,r)} (\bar z - \bar z_1)^r C^i_{pr}O_{pr}(z_1,\bar z_1)$. Note that $q$ is \textit{not} necessarily the smallest power of $(\bar z - \bar z_1)$ occurring in the OPE \eqref{ope}. Let us now write down the differential equations explicitly. 

\vskip 4pt
We have two differential equations 
\be\label{e11}
\begin{gathered}
 \left( \mathcal L_{-1}\mathcal P_{-1,-1} + 2 \mathcal J^{0}_{-1} \mathcal P_{-1,-1} - (\D+1) \mathcal P_{-2,-1} - \mathcal{\bar L}_{-1} \mathcal P_{-2,0}\right) \\
 \times \bigg\langle{G^{+}_{\D}(z,\bar z)} G^{\sigma_1}_{\D_1}(z_1,\bar z_1)\prod_{i\ne 1}G^{\sigma_i}_{\D_i}(z_i,\bar z_i)\bigg\rangle_{MHV} = 0
 \end{gathered}
\ee  

and
\be\label{e22}
\left(\mathcal J^1_{-1}\mathcal P_{-1,-1} - (\D -1) \mathcal P_{-2,0} \right) \bigg\langle{G^{+}_{\D}(z,\bar z) G^{\sigma_1}_{\D_1}(z_1,\bar z_1) \prod_{i\ne 1} G^{\sigma_i}_{\D_i}(z_i,\bar z_i)}\bigg\rangle_{MHV} = 0
\ee

where the differential operators are given by,
\be\label{1}
\mathcal J^0_{-1} = - \frac{\bar h_1 + (\bar z_1- \bar z) \bar\partial_1}{z_1 - z} - \sum_{i\ne 1} \frac{\bar h_i +( \bar z_i - \bar z) \bar\partial_i}{z_i - z}
\ee

\be\label{2}
\mathcal P_{-2,-1} = - \frac{1}{z_1 - z} \epsilon_1 \mathcal P_1 - \sum_{i\ne 1} \frac{1}{z_i - z} \epsilon_i \mathcal P_i
\ee

\be\label{3}
\mathcal P_{-2,0} = - \frac{\bar z_1 - \bar z}{z_1 - z} \epsilon_1 \mathcal P_1 - \sum_{i\ne 1} \frac{\bar z_i - \bar z}{z_i - z} \epsilon_i \mathcal P_i
\ee

\be\label{4}
\mathcal J^1_{-1} =  - \frac{2\bar h_1(\bar z_1 - \bar z) + (\bar z_1-\bar z)^2 \bar\partial_1}{z_1 - z} - \sum_{i\ne 1} \frac{2\bar h_i(\bar z_i - \bar z) + (\bar z_i - \bar z)^2 \bar\partial_i}{z_i - z}
\ee

\be\label{5}
\mathcal L_{-1} = \frac{\partial}{\partial z} , \quad \mathcal{ \bar L}_{-1} = \frac{\partial}{\partial \bar z}
\ee

Here $\epsilon = \pm 1$ for an outgoing (incoming) particle and $\mathcal P_j G^{\sigma_k}_{\D_k}(z,\bar z)=\delta_{jk} G^{\sigma_k}_{\D_k +1}(z,\bar z)$. 

 \vskip 4pt
Let us start with the second differential equation \eqref{e22}. We can write this equation as,
\be\label{aa}
\begin{gathered}
\mathcal J^1_{-1}\bigg\langle{G^{+}_{\D+1}(z,\bar z) G^{\sigma_1}_{\D_1}(z_1,\bar z_1) \prod_{i\ne 1} G^{\sigma_i}_{\D_i}(z_i,\bar z_i)}\bigg\rangle_{MHV} \\ = (\D -1) \mathcal P_{-2,0} \bigg\langle{G^{+}_{\D}(z,\bar z) G^{\sigma_1}_{\D_1}(z_1,\bar z_1) \prod_{i\ne 1} G^{\sigma_i}_{\D_i}(z_i,\bar z_i)}\bigg\rangle_{MHV}
\end{gathered}
\ee 

where we have used the fact that $G^+_{\D}(z,\bar z)$ is outgoing and so $P_{-1,-1}G^+_{\D}(z,\bar z)=G^+_{\D+1}(z,\bar z)$. 

 \vskip 4pt
Now we take the holomorphic OPE limit $z \rightarrow z_1$ and in this limit we can keep only the singular terms in the differential operators \eqref{3} and \eqref{4}. We also have to substitute the OPE \eqref{ope} inside the correlator. We can keep only the leading term of the OPE. Now by matching the coefficients of $(z-z_1)^{p-1}(\bar z - \bar z_1)^{q+1}$ on both sides of \eqref{aa}, we get the following recursion relation, 
\be\label{r1}
(\D-1)C_{pq}(\D, \D_1+1,\sigma_1) = (\D_1 - \sigma_1 + q) C_{pq}(\D+1,\D_1,\sigma_1)
\ee

We can follow identical procedure for the other equation \eqref{e11} and obtain the equation,
\be\label{r2}
(\D - q) C_{pq}(\D, \D_1+1,\sigma_1) = (\D_1 - \sigma_1 + 2q + p) C_{pq}(\D+1, \D_1, \sigma_1)
\ee

The equations \eqref{r1} and \eqref{r2} have non-trivial solutions iff, 
\be\label{det}
\frac{\D-1}{\D-q} = \frac{\D_1 - \sigma_1 + q}{\D_1 - \sigma_1 + 2q + p}
\ee

Now note that the null-states exist for arbitrary values of dimension $\D$ and we can vary $\D$ and $\D_1$  independently. So the only nontrivial solution of \eqref{det} can be, 
\be
q =1, \quad p = -q = -1
\ee

Therefore, the leading term in the OPE must have the structure,
\be
G^+_{\D}(z,\bar z) G^{\sigma_1}_{\D_1}(z_1, \bar z_1) \sim C_{-1,1}(\D,\D_1,\sigma_1) \frac{\bar z - \bar z_1}{z - z_1} G^{\sigma_2}_{\D_2}(z_1,\bar z_1)
\ee

This also immediately tells us that,
\be
\D_2 = \D + \D_1, \quad \sigma_2 = \sigma_1
\ee

So finally we have,
\be\label{lope}
\boxed{
G^+_{\D}(z,\bar z) G^{\sigma_1}_{\D_1}(z_1, \bar z_1) \sim C_{-1,1}(\D,\D_1,\sigma_1) \frac{\bar z - \bar z_1}{z - z_1} G^{\sigma_1}_{\D + \D_1}(z_1,\bar z_1)}
\ee

which is the expected answer \cite{Pate:2019lpp}. This in particular means that the bulk theory which is dual to the MHV sector of the Celestial CFT must be a two derivative theory of gravity \cite{Pate:2019lpp}. This is a consistency check. Let us now find out the OPE coefficient. 

 \vskip 4pt
From \eqref{r1} or \eqref{r2} we get, 
\be\label{r3}
(\D-1)C_{-1,1}(\D, \D_1+1,\sigma_1) = (\D_1 - \sigma_1 + 1) C_{-1,1}(\D+1,\D_1,\sigma_1)
\ee

We also have the following recursion from global time translation invariance,
\be\label{r4}
C_{-1,1}(\D,\D_1,\sigma_1) = C_{-1,1}(\D+1, \D_1, \sigma_1) + C_{-1,1}(\D, \D_1 +1, \sigma_1)
\ee

Combining \eqref{r3} and \eqref{r4} we get
\be\label{s1}
C_{-1,1}(\D+1, \D_1,\sigma_1) = \frac{\D-1}{\D + \D_1 - \sigma_1} C_{-1,1}(\D,\D_1, \sigma_1)
\ee

and
\be\label{s2}
C_{-1,1}(\D, \D_1+1,\sigma_1) = \frac{\D_1 - \sigma_1 +1}{\D + \D_1 - \sigma_1} C_{-1,1}(\D,\D_1, \sigma_1)
\ee

The solution to the recursion relations \eqref{s1} and \eqref{s2} are given by \cite{Pate:2019lpp},
\be
C_{-1,1}(\D,\D_1,\sigma_1) = \alpha B(\D-1, \D_1 - \sigma_1 +1)
\ee

where $B(x,y)$ is the Euler Beta function and $\alpha$ is a constant. Now matching with the leading conformal soft limit $\D\rightarrow 1$ we get $\alpha=-1$.  So the leading term of the OPE comes out to be,
\be\label{lo}
\boxed{
G^+_{\D}(z,\bar z) G^{\sigma_1}_{\D_1}(z_1, \bar z_1) \sim - B(\D-1, \D_1 - \sigma_1 +1) \frac{\bar z - \bar z_1}{z - z_1} G^{\sigma_1}_{\D + \D_1}(z_1,\bar z_1)}
\ee

This is precisely the answer obtained in \cite{Pate:2019lpp}. Now we can start with two incoming gravitons. The differential equations \eqref{e11} and \eqref{e22} do not change because they are determined by symmetry algebra. The only thing that changes is that now we have 
\be
P_{-1,-1} G^+_{\D}(z,\bar z) = - G^+_{\D+1}(z,\bar z)
\ee

instead of $P_{-1,-1} G^+_{\D}(z,\bar z) = G^+_{\D+1}(z,\bar z)$. This, together with the fact that $G^{\sigma_1}_{\D_1}(z_1,\bar z_1)$ is also incoming, lead to identical recursion relations. So for two incoming gravitons also we get the same leading OPE \eqref{lo}. 

\subsection{Outgoing-Incoming OPE}

In this section, to distinguish between incoming and outgoing graviton primaries, we denote them by $G^{\sigma,\epsilon}_{\D}(z,\bar z)$ where $\epsilon=\pm 1$ for an outgoing (incoming) particle. 

 \vskip 4pt
Let us consider $G^+_{\D}(z,\bar z)$ to be outgoing and $G^{\sigma_1}_{\D_1}(z_1,\bar z_1)$ to be incoming. In this case two different primaries can appear on the R.H.S of the OPE \cite{Pate:2019lpp}, 
\be\label{out}
G^{+,\epsilon}_{\D}(z,\bar z) G^{\sigma_1, - \epsilon}_{\D_1}(z_1,\bar z_1) \sim C^{\epsilon}_{pq}(\D,\D_1, \sigma_1) (z- z_1)^p(\bar z - \bar z_1)^q G^{\sigma_2, \epsilon}_{\D_2}(z_1,\bar z_1)
\ee

and
\be\label{in}
G^{+,\epsilon}_{\D}(z,\bar z) G^{\sigma_1, - \epsilon}_{\D_1}(z_1,\bar z_1) \sim C^{-\epsilon}_{pq}(\D,\D_1, \sigma_1) (z- z_1)^p(\bar z - \bar z_1)^q G^{\sigma_2, - \epsilon}_{\D_2}(z_1,\bar z_1)
\ee

where $\epsilon = 1$ because we are taking $G^+_{\D}(z,\bar z)$ to be outgoing. Proceeding in the same way as in the case of two outgoing gravitons we get the following recursion relations from the differential equations \eqref{e11} and \eqref{e22},
\be\label{rec1}
(\D_1 - \sigma_1 + p + 2q) C^{\pm\epsilon}_{pq}(\D+1,\D_1, \sigma_1) + (\D-q) C^{\pm\epsilon}_{pq}(\D,\D_1 +1,\sigma_1) = 0
\ee
\be\label{rec2}
(\D_1 - \sigma_1 +  q) C^{\pm\epsilon}_{pq}(\D+1,\D_1, \sigma_1) + (\D-1) C^{\pm\epsilon}_{pq}(\D,\D_1 +1,\sigma_1) = 0
\ee

Now \eqref{rec1} and \eqref{rec2} together imply that,
\be
q =1, \quad p =-1 \quad  \Longrightarrow \quad  \sigma_2 = \sigma_1, \quad \D_2 = \D+\D_1
\ee

So we can write a single recursion relation as follows,
\be\label{srec}
\boxed{
(\D_1 - \sigma_1 +  1) C^{\pm\epsilon}_{pq}(\D+1,\D_1, \sigma_1) + (\D-1) C^{\pm\epsilon}_{pq}(\D,\D_1 +1,\sigma_1) = 0}
\ee

Now the recursion relation following from global time translation invariance is given by,
\be\label{rect}
\boxed{
C^{\pm\epsilon}_{pq}(\D+1,\D_1, \sigma_1) - C^{\pm\epsilon}_{pq}(\D,\D_1 +1,\sigma_1) = \pm C^{\pm\epsilon}_{pq}(\D,\D_1,\sigma_1)}
\ee

Then \eqref{srec} and $\eqref{rect}$ can be combined to give, 
\be\label{fr1}
C^{\pm\epsilon}_{-1,1}(\D+1, \D_1, \sigma_1) = \pm \frac{\D-1}{\D+\D_1-\sigma_1} C^{\pm\epsilon}_{-1,1}(\D,\D_1,\sigma_1)
\ee

and
\be\label{fr2}
C^{\pm\epsilon}_{-1,1}(\D, \D_1+1, \sigma_1) = \mp \frac{\D_1 - \sigma_1 +1}{\D+\D_1-\sigma_1} C^{\pm\epsilon}_{-1,1}(\D,\D_1,\sigma_1)
\ee

For concreteness let us start with $\sigma_1 = 2$. In this case the relations \eqref{fr1} and \eqref{fr2} give, 
\be
(\D + \D_1 -2)C^{\epsilon}_{-1,1}(\D+1,\D_1, \sigma_1 = 2) = (\D-1) C^{\epsilon}_{-1,1}(\D, \D_1,\sigma_1 =2) 
\ee

\be
(\D + \D_1 -2)C^{\epsilon}_{-1,1}(\D,\D_1 +1, \sigma_1=2) = - (\D_1-1) C^{\epsilon}_{-1,1}(\D, \D_1,\sigma_1=2) 
\ee

and 
\be
(\D + \D_1 -2)C^{-\epsilon}_{-1,1}(\D+1,\D_1, \sigma_1 = 2) = - (\D-1) C^{-\epsilon}_{-1,1}(\D, \D_1,\sigma_1 =2) 
\ee

\be
(\D + \D_1 -2)C^{-\epsilon}_{-1,1}(\D,\D_1 +1, \sigma_1=2) =  (\D_1-1) C^{-\epsilon}_{-1,1}(\D, \D_1,\sigma_1=2) 
\ee

The solutions of these recursion relations are given by, 
\be
C^{\epsilon}_{-1,1}(\D, \D_1,\sigma_1 =2) = \alpha B(\D_1 -1, 3 - \D - \D_1)
\ee

\be
C^{-\epsilon}_{-1,1}(\D, \D_1,\sigma_1=2) = \beta B(\D-1, 3 - \D - \D_1)
\ee

where $\alpha$ and $\beta$ are constants to be determined. So we can write,
\be
G^{+,\epsilon}_{\D}(z,\bar z) G^{+, - \epsilon}_{\D_1}(z_1,\bar z_1) \sim \alpha B(\D_1 -1, 3 - \D - \D_1) \frac{\bar z- \bar z_1}{z - z_1} G^{+, \epsilon}_{\D+\D_1}(z_1,\bar z_1)
\ee

and 
\be
G^{+,\epsilon}_{\D}(z,\bar z) G^{+, - \epsilon}_{\D_1}(z_1,\bar z_1) \sim \beta B(\D-1, 3 - \D - \D_1) \frac{\bar z- \bar z_1}{z - z_1} G^{+, - \epsilon}_{\D+\D_1}(z_1,\bar z_1)
\ee

where $\epsilon = 1$. The constants $\a$ and $\b$ can be obtained by taking the leading conformal soft limit. To determine $\a$ we can make $G^{+, - \epsilon}_{\D_1}(z_1,\bar z_1)$ conformally soft by taking $\D_1\rightarrow 1$. This gives $\a =1$. Similarly to determine $\b$ we can make $G^{+,\epsilon}_{\D}(z,\bar z)$ conformally soft by taking $\D\rightarrow 1$ and this gives $\b =1$. So the final answer becomes,
\be
\boxed{
G^{+,\epsilon}_{\D}(z,\bar z) G^{+, - \epsilon}_{\D_1}(z_1,\bar z_1) \sim B(\D_1 -1, 3 - \D - \D_1) \frac{\bar z- \bar z_1}{z - z_1} G^{+, \epsilon}_{\D+\D_1}(z_1,\bar z_1)}
\ee

and 
\be
\boxed{
G^{+,\epsilon}_{\D}(z,\bar z) G^{+, - \epsilon}_{\D_1}(z_1,\bar z_1) \sim B(\D-1, 3 - \D - \D_1) \frac{\bar z- \bar z_1}{z - z_1} G^{+, - \epsilon}_{\D+\D_1}(z_1,\bar z_1)}
\ee
These answers also match with those obtained in \cite{Pate:2019lpp}.

 \vskip 4pt
In a similar way we can find out the OPE coefficients $C^{\pm\epsilon}_{pq}(\D,\D_1, \sigma_1 = -2)$ starting from the recursion relations \eqref{fr1} and \eqref{fr2}. 

\section{Subleading OPE coefficients from symmetry} \label{subleadopede}

Having determined the leading celestial OPE coefficients, we now turn to the subleading OPE coefficients. These can be obtained by solving recursion relations that follow from demanding the invariance of the OPE under the extended symmetry algebra. We will illustrate this here by considering the $\mathcal O(z^0\bar z^0)$ and $\mathcal{O}(\bar{z})$ terms in the OPE for outgoing gravitons. The OPE coefficients at other subleading orders as well as the case of incoming-outgoing gravitons can be worked out in the same fashion.

\subsection{Recursion relation for $\mathcal O(z^0\bar z^0)$ term}

We have seen that in both the $(++)$ and $(+-)$ OPEs the leading term is 
\be
G^+_{\D_1}(z,\bar{z}) G^{\pm}_{\D_2}(0,0) \sim \frac{\bar z}{z} C_{\pm} P_{-1,-1} G^{\pm}_{\D_3}(0,0), \quad \D_3 = \D_1 + \D_2 -1
\ee

where $C_{\pm}$ are given by \eqref{lo}
\be\label{cpm}
C_{+} = - B(\D_1 -1, \D_2 -1), \quad C_{-} = - B(\D_1-1, \D_2 + 3)
\ee

Now at $\mathcal O(z^0\bar z^0)$ there are only two possible descendants
\be
P_{-2,0}G^{\pm}_{\D_3}(0), \quad J^1_{-1}P_{-1,-1}G^{\pm}_{\D_3}(0)
\ee

Due to the null state decoupling relation \eqref{de1mhv} these two states are not independent and we can keep only $P_{-2,0}G^{\pm}_{\D_3}(0)$. This allows us to write the OPE as, 
\be\label{order1}
G^+_{\D_1}(z,\bar{z}) G^{\pm}_{\D_2}(0,0) = \frac{\bar z}{z} C_{\pm} P_{-1,-1} G^{\pm}_{\D_3}(0,0) + C'_{\pm} P_{-2,0}G^{\pm}_{\D_3}(0,0) + \cdots
\ee

To determine $C'_{\pm}$ we apply the generator $J^{-1}_{1}$ to both sides of \eqref{order1}. We have to use the relations,
\be\label{p1}
\[ J^{-1}_1 , \phi_{h,\bar h}(z,\bar z)\] = z \bar\partial \phi_{h,\bar h}(z,\bar z) \Longrightarrow \[ J^{-1}_1 , \phi_{h,\bar h}(0)\] = 0
\ee

and 
\be
\[ J^{-1}_1, P_{-2,0}\] = - P_{-1,-1}
\ee

where $\phi_{h,\bar h}(z,\bar z)$ is a primary. Note that the second condition in \eqref{p1} is essentially the definition of a primary operator as stated earlier. Now using these relations and equating the coefficient of the $\mathcal O(z^0\bar z^0)$ term on both sides we get,
\be
C'_{\pm} = - C_{\pm}
\ee

This precisely matches with the coefficient of the $P_{-2,0}G^{\pm}_{\D_3}(0)$ term obtained from the MHV amplitudes. See for example \eqref{csope6ptfinal}. Therefore \eqref{order1} becomes
\be
G^+_{\D_1}(z,\bar{z}) G^{\pm}_{\D_2} (0,0)= \frac{\bar z}{z} C_{\pm}  P_{-1,-1} G^{\pm}_{\D_3}(0,0) - C_{\pm} P_{-2,0}G^{\pm}_{\D_3}(0,0) + \cdots, \quad \D_3 = \D_1 + \D_2 -1
\ee

where $C_{\pm}$ are given by \eqref{cpm}. 

\subsection{Recursion relations for $\mathcal{O}(\bar{z})$ terms in $(+,+)$ OPE }

Let us consider two outgoing graviton primary operators $G^{+}_{\Delta_{1}}(z,\bar{z})$ and $G^{+}_{\Delta_{2}}(0,0) $.  At $\mathcal{O}(\bar{z})$ the operators that can appear in their celestial OPE are
\begin{equation}
\label{zbarops}
\begin{split}
J^{0}_{-1}P_{-1,-1} G^{+}_{\Delta}, \quad P_{-2,-1}G^{+}_{\Delta}, \quad L_{-1}P_{-1,-1}G^{+}_{\Delta}, \quad \bar{L}_{-1} P_{-2,0}G^{+}_{\Delta}, \quad  \bar{L}_{-1} J^{1}_{-1} P_{-1,-1}G^{+}_{\Delta}
\end{split}
\end{equation}

where $\Delta= \Delta_{1}+\Delta_{2}-1= 1+i\lambda_{1}+i\lambda_{2}$. But using the decoupling relations \eqref{de1mhv} and \eqref{de2} we can eliminate the descendants $L_{-1}P_{-1,-1}G^{+}_{\Delta}$ and $\bar{L}_{-1} J^{1}_{-1} P_{-1,-1}G^{+}_{\Delta}$ from the above list.  Then the general form of the OPE at $\mathcal{O}(\bar{z})$ can be written as
\begin{equation}
\label{zbarope}
\begin{split}
& G^{+}_{\Delta_{1}}(z_{1},\bar{z}_{1}) G^{+}_{\Delta_{2}}(0,0) \\
& \supset  \bar{z} \ B(i\lambda_{1},i\lambda_{2}) \left[  \alpha_{1}  \hspace{0.04cm} J^{0}_{-1}P_{-1,-1} + \alpha_{2} \ P_{-2,-1} +  \alpha_{3} \ \bar{L}_{-1} P_{-2,0} \right] G^{+}_{\Delta}(0,0)
\end{split}
\end{equation}

Using the result from the previous section that the leading OPE coefficient is given by the Euler Beta function, we have chosen to include in \eqref{zbarope} an overall factor of $B(i\lambda_{1},i\lambda_{2})$.  

 \vskip 4pt
Now let us impose that both sides of \eqref{zbarope} transform in the same way under the action of the extended symmetry algebra generators. In particular let us first take the commutator of both sides w.r.t $J^{0}_{1}$. Using the following commutation relation for primary operators
\begin{equation}
\label{J0min1comm}
\begin{split}
& [J^{0}_{1}, G^{\sigma}_{\Delta}(z,\bar{z}) ] =  z(\bar{z} \partial_{\bar{z}} + \bar{h}) G^{\sigma}_{\Delta}(z,\bar{z}), \quad \sigma=\pm 2
\end{split}
\end{equation}

we get
\begin{equation}
\label{l1commlhs}
\begin{split}
& [J^{0}_{1}, G^{+}_{\Delta_{1}}(z_{1},\bar{z}_{1}) G^{+}_{\Delta_{2}}(0,0) ] = z(\bar{z} \partial_{\bar{z}} + \bar{h}_{1}) G^{+}_{\Delta_{1}}(z,\bar{z}) G^{+}_{\Delta_{2}}(0,0)
\end{split}
\end{equation}

Now to match with the order on the R.H.S. of \eqref{zbarope} we insert the leading OPE in the above. This gives, 
\begin{equation}
\label{l1commlhs1}
\begin{split}
& [J^{0}_{1}, G^{+}_{\Delta_{1}}(z_{1},\bar{z}_{1}) G^{+}_{\Delta_{2}}(0,0) ] \supset  - \bar{z} \ B(i\lambda_{1},i\lambda_{2}) \left( \bar{h}_{1}+1\right)P_{-1,-1} G^{+}_{\Delta}(0,0)
\end{split}
\end{equation}

Then using the commutation relations for the generators of the extended symmetry algebra given in Section \ref{extalgebsum}, we get from the R.H.S. of \eqref{zbarope}
\begin{equation}
\label{l1commrhs}
\begin{split}
& [J^{0}_{1}, G^{+}_{\Delta_{1}}(z_{1},\bar{z}_{1}) G^{\sigma}_{\Delta_{2}}(z_{2},\bar{z}_{2}) ]  \\
& \supset  \bar{z}\ B(i\lambda_{1},i\lambda_{2}) \left(  \frac{\alpha_{2}}{2}  - \alpha_{3}\right)P_{-1,-1} G^{\sigma}_{\Delta}(0,0)
\end{split}
\end{equation}

In obtaining the above we have also used the fact that $P_{-1,0}G^{\sigma}_{\Delta}(0,0) = P_{0,-1}G^{\sigma}_{\Delta}(0,0)=0$. Then comparing \eqref{l1commlhs1} and \eqref{l1commrhs} we get the following equation
\begin{equation}
\label{recrel1}
\begin{split}
& \alpha_{2}  - 2  \hspace{0.04cm}  \alpha_{3} = -(2 \bar{h}_{1}+2)
\end{split}
\end{equation}

Next let us consider the commutator of both sides of the OPE \eqref{zbarope} with $\bar{L}_{1}$. This will generate the operator  $J^{1}_{-1}P_{-1,-1}G^{+}_{\Delta}(0,0)$ on the R.H.S. We can  use the decoupling relation to write it then in terms of $P_{-2,0}G^{+}_{\Delta}(0,0)$. Then following the same procedure as before we find 
\begin{equation}
\label{recrel2}
\begin{split}
& (2\bar{h}+1) \alpha_{1} +  \alpha_{2}  +(2\bar{h}-1) \alpha_{3} =  2 \bar{h}_{1}
\end{split}
\end{equation}

Similarly demanding that both sides of the OPE transform in the same fashion under the action of $L_{1}$ we get
\begin{equation}
\label{recrel3}
\begin{split}
& (2\bar{h}+1) \alpha_{1} +  4 \alpha_{2}  =  - 2 (2 h_{1}-1)
\end{split}
\end{equation}

Simultaneously solving the above set of equations \eqref{recrel1}, \eqref{recrel2}, \eqref{recrel3}, we get 
\begin{equation}
\label{solrecrel}
\begin{split}
& \alpha_{1} = \frac{2 i\lambda_{1}}{ i\lambda_{1}+i\lambda_{2}}, \quad \alpha_{2} = -(1+i\lambda_{1}), \quad \alpha_{3} = 0
\end{split}
\end{equation}

These coefficients match precisely with the ones obtained from the OPE decomposition of the $6$-point Mellin amplitude in Section \ref{6ptmellin}.


\subsection{Recursion relations for $\mathcal{O}(\bar{z})$ terms in $(+,-)$ OPE }

Now let us deal with the mixed helicity OPE. In this case the general form of the OPE between graviton primaries $G^{+}_{\Delta_{1}}(z,\bar{z})$ and $G^{-}_{\Delta_{2}}(0,0) $ can be written as
\begin{equation}
\label{zbarpmope}
\begin{split}
& G^{+}_{\Delta_{1}}(z_{1},\bar{z}_{1}) G^{-}_{\Delta_{2}}(0,0) \\
& \supset  \bar{z} \ B(i\lambda_{1},i\lambda_{2}+4) \left[  \beta_{1}  \hspace{0.04cm} J^{0}_{-1}P_{-1,-1} + \beta_{2} \ P_{-2,-1} +  \beta_{3} \ L_{-1}P_{-1,-1}+  \beta_{4} \ \bar{L}_{-1} P_{-2,0} \right] G^{-}_{\Delta}(0,0)
\end{split}
\end{equation}

Here also we have used the decoupling relation \eqref{de1mhv} to eliminate the operator \\
$\bar{L}_{-1} J^{1}_{-1} P_{-1,-1}G^{-}_{\Delta}$ which would otherwise be allowed simply on grounds of dimensional analysis. But unlike the previous case the null state relation \eqref{de2} does not exist for a graviton primary with spin $\sigma=-2$. Thus we have to keep all  $4$ operators in the above OPE. 

 \vskip 4pt
Now we can obtain recursion relations for the  OPE coefficients in \eqref{zbarpmope} by following exactly the same procedure as in the case of the $(+,+)$ OPE.  The commutator with $J^{0}_{1}$ in this case leads to
\begin{equation}
\label{pmrecrel1}
\begin{split}
& \beta_{2} + (4+i\lambda_{1}+i\lambda_{2})\beta_{3}  - 2  \hspace{0.04cm}  \beta_{4}= -(1+i\lambda_{1})
\end{split}
\end{equation}

Then invariance of the OPE under the action of $\bar{L}_{1}$ requires
\begin{equation}
\label{pmrecrel2}
\begin{split}
& (4+ i\lambda_{1}+i\lambda_{2}) \beta_{1} + \beta_{2} +  (2+i\lambda_{1}+i\lambda_{2})\beta_{4} = i\lambda_{1}-1
\end{split}
\end{equation}

Under the action of $L_{1}$  both sides of the OPE \eqref{zbarpmope} transform identically iff
\begin{equation}
\label{pmrecrel3}
\begin{split}
& (4+ i\lambda_{1}+i\lambda_{2}) \beta_{1} + 4  \hspace{0.04cm}  \beta_{2} +  (i\lambda_{1}+i\lambda_{2})\beta_{3} = -2  (2+i\lambda_{1})
\end{split}
\end{equation}

Now to be able to determine all the $4$ coefficients in \eqref{zbarpmope} we need another equation. For this let us take the commutator of \eqref{zbarpmope} with $P_{0,-1}$. In this case both sides of the OPE transform in the same manner provided the following equation holds
\begin{equation}
\label{recrel4}
\begin{split}
& - \beta_{1} +  2 \beta_{3}  =  - \frac{2i\lambda_{1}}{i\lambda_{1}+i\lambda_{2}+4}
\end{split}
\end{equation}

Solving the above set of equations \eqref{pmrecrel1}, \eqref{pmrecrel2} and \eqref{pmrecrel3} we get
\begin{equation}
\label{pmsolrecrel}
\begin{split}
& \beta_{1} = \frac{2 \hspace{0.04cm} i\lambda_{1}}{4+ i\lambda_{1}+i\lambda_{2}}, \quad \beta_{2} = -(1+ i \lambda_{1}), \quad \beta_{3}=0, \quad \beta_{4}=0
\end{split}
\end{equation}

We have checked that these coefficients also precisely match with the corresponding results from the Mellin amplitude. This calculation can be done in exactly the same fashion as performed for the $(+,+)$ OPE in Section \ref{6ptOPEMHV}. 

\section{Acknowledgements}
We would like to thank Arjun Bagchi for very helpful discussions. The work of SB is partially supported by the Science and Engineering Research Board (SERB) grant MTR/2019/000937 (Soft-Theorems, S-matrix and Flat-Space Holography). SG would like to thank Yasha Neiman and Vyacheslav Lysov for useful discussions. The work of SG is supported by the Quantum Gravity Unit of the Okinawa Institute of Science and Technology Graduate University (OIST), Japan. The work of PP is partially supported by a grant to CMI from the Infosys Foundation.

\section{Appendix}\label{appendix}
\subsection{Brief review of Celestial or Mellin amplitudes for massless particles}\label{review}

The Celestial or Mellin amplitude for massless particles in four dimensions is defined as the Mellin transformation of the $S$-matrix element, given by \cite{Pasterski:2016qvg,Pasterski:2017kqt}
\be\label{mellin}
\mathcal M_n\big(\{z_i, \bar z_i, h_i, \bar h_i\}\big) = \prod_{i=1}^{n} \int_{0}^{\infty} d\omega_i \ \omega_i^{\D_i -1} S_n\big(\{\omega_i,z_i,\bar z_i, \sigma_i\}\big)
\ee 

where $\sigma_i$ denotes the helicity of the $i$-th particle and the on-shell momenta are parametrized as,
\be\label{para}
p_i = \omega_i (1+z_i\bar z_i, z_i + \bar z_i , -i(z_i - \bar z_i), 1- z_i \bar z_i), \quad p_i^2 = 0
\ee

The scaling dimensions $(h_i,\bar h_i)$ are defined as,
\be
h_i = \frac{\D_i + \sigma_i}{2}, \quad \bar h_i = \frac{\D_i - \sigma_i}{2}
\ee

The Lorentz group $SL(2,\mathbb C)$ acts on the celestial sphere as the group of global conformal transformations and the Mellin amplitude $\mathcal M_n$ transforms as,
\be
\mathcal M_n\big(\{z_i, \bar z_i, h_i, \bar h_i\}\big) = \prod_{i=1}^{n} \frac{1}{(cz_i + d)^{2h_i}} \frac{1}{(\bar c \bar z_i + \bar d)^{2\bar h_i}} \mathcal M_n\bigg(\frac{az_i+b}{cz_i+d} \ ,\frac{\bar a \bar z_i + \bar b}{\bar c \bar z_i + \bar d} \ , h_i,\bar h_i\bigg)
\ee

This is the familiar transformation law for the correlation function of primary operators of weight $(h_i,\bar h_i)$ in a $2$-D CFT under the global conformal group $SL(2,\mathbb C)$.

 \vskip 4pt
In Einstein gravity, the Mellin amplitude as defined in \eqref{mellin} usually diverges. This divergence can be regulated by defining a modified Mellin amplitude as \cite{Banerjee:2018gce,Banerjee:2019prz}, 
\be\label{mellinmod}
\mathcal M_n\big(\{u_i,z_i, \bar z_i, h_i, \bar h_i\}\big) = \prod_{i=1}^{n} \int_{0}^{\infty} d\omega_i \ \omega_i^{\D_i -1} e^{-i\sum_{i=1}^n \epsilon_i \omega_i u_i} S_n\big(\{\omega_i,z_i,\bar z_i, \sigma_i\}\big)
\ee 

where $u$ can be thought of as a time coordinate and $\epsilon_i = \pm 1$ for an outgoing (incoming) particle. Under (Lorentz) conformal tranansformation the modified Mellin amplitude $\mathcal M_n$ transforms as,
\be
\mathcal M_n\big(\{u_i,z_i, \bar z_i, h_i, \bar h_i\}\big) = \prod_{i=1}^{n} \frac{1}{(cz_i + d)^{2h_i}} \frac{1}{(\bar c \bar z_i + \bar d)^{2\bar h_i}} \mathcal M_n\bigg(\frac{u_i}{|cz_i + d|^2} \ , \frac{az_i+b}{cz_i+d} \ ,\frac{\bar a \bar z_i + \bar b}{\bar c \bar z_i + \bar d} \ , h_i,\bar h_i\bigg)
\ee

Under global space-time translation, $u \rightarrow u + A + Bz + \bar B\bar z + C z\bar z$, the modified amplitude is invariant, i.e, 
\be
\mathcal M_n\big(\{u_i + A + Bz_i + \bar B\bar z_i + C z_i\bar z_i ,z_i, \bar z_i, h_i, \bar h_i\}\big) = \mathcal M_n\big(\{u_i,z_i, \bar z_i, h_i, \bar h_i\}\big)
\ee

Now in order to make manifest the conformal nature of the dual theory living on the celestial sphere it is useful to write the (modified) Mellin amplitude as a correlation function of conformal primary operators. So let us define a generic conformal primary operator as, 
\be
\label{confprim}
\phi^{\epsilon}_{h,\bar h}(z,\bar z) = \int_{0}^{\infty} d\omega \  \omega^{\D-1} a(\epsilon\omega, z, \bar z, \sigma)
\ee

where $\epsilon=\pm 1$ for an annihilation (creation) operator of a massless particle of helicity $\sigma$. Under (Lorentz) conformal transformation the conformal primary transforms like a primary operator of scaling dimension $(h,\bar h)$
\be
\phi'^{\epsilon}_{h,\bar h}(z,\bar z) = \frac{1}{(cz + d)^{2h}} \frac{1}{(\bar c \bar z + \bar d)^{2\bar h}} \mathcal \phi^{\epsilon}_{h,\bar h}\bigg(\frac{az+b}{cz+d} \ ,\frac{\bar a \bar z + \bar b}{\bar c \bar z + \bar d}\bigg)
\ee

Similarly in the presence of the time coordinate $u$ we have,
\be
\label{confprimu}
\phi^{\epsilon}_{h,\bar h}(u,z,\bar z) = \int_{0}^{\infty} d\omega \ \omega^{\D-1} e^{-i \epsilon \omega u} a(\epsilon\omega, z, \bar z, \sigma)
\ee

Under (Lorentz) conformal transformations 
\be
\phi'^{\epsilon}_{h,\bar h}(u,z,\bar z) = \frac{1}{(cz + d)^{2h}} \frac{1}{(\bar c \bar z + \bar d)^{2\bar h}} \mathcal \phi^{\epsilon}_{h,\bar h}\bigg(\frac{u}{|cz+d|^2},\frac{az+b}{cz+d} \ ,\frac{\bar a \bar z + \bar b}{\bar c \bar z + \bar d}\bigg)
\ee

In terms of \eqref{confprim},  the Mellin amplitude can be written as the correlation function of conformal primary operators
\be
\mathcal M_n = \bigg\langle{\prod_{i=1}^n \phi^{\epsilon_i}_{h_i,\bar h_i}(z_i,\bar z_i)}\bigg\rangle
\ee

Similarly using \eqref{confprimu}, the modified Mellin amplitude can be written as,
\be
\mathcal M_n = \bigg\langle{\prod_{i=1}^n \phi^{\epsilon_i}_{h_i,\bar h_i}(u_i,z_i,\bar z_i)}\bigg\rangle
\ee 

\subsubsection{Comments on notation in the paper}
Note that conformal primaries carry an extra index $\epsilon$ which distinguishes between an incoming and an outgoing particle. In the paper, for notational simplicity, we omit this additional index unless this plays an important role. So in most places we simply write the (modified) Mellin amplitude as,
\be
\mathcal M_n = \bigg\langle{\prod_{i=1}^n \phi_{h_i,\bar h_i}(z_i,\bar z_i)}\bigg\rangle
\ee

or
\be
\mathcal M_n = \bigg\langle{\prod_{i=1}^n \phi_{h_i,\bar h_i}(u_i,z_i,\bar z_i)}\bigg\rangle
\ee 

Similarly in many places in the paper we denote a graviton primary of weight $\D = h+\bar h$ by $G^{\sigma}_\D$ where $\sigma = \pm 2$ is the helicity (= $h-\bar h$). Since we are considering pure gravity, we can further simplify the notation to $G^{\pm}_{\D}$ by omitting the $2$. 


\subsection{Subleading conformal soft limit}\label{sstt}

Let us consider a correlation function of the form 
\be
\big\langle{G^+_{\D}(u,z,\bar z)} \prod_{i=1}^n \phi_{h_i,\bar h_i}(u_i,z_i,\bar z_i)\big\rangle
\ee 

where $G^{+}_{\D}$ is a positive helicity graviton primary with weight $\D$ and $\phi_{h_i,\bar h_i}$ is a generic conformal primary with weight $(h_i,\bar h_i)$. Now let us consider the subleading conformal soft limit given by,
\be\label{css}
\lim_{\D\rightarrow 0} \D \big\langle{G^+_{\D}(u,z,\bar z)} \prod_{i=1}^n \phi_{h_i,\bar h_i}(u_i,z_i,\bar z_i)\big\rangle = \big\langle{\tilde S^+_1(u,z,\bar z)} \prod_{i=1}^n \phi_{h_i,\bar h_i}(u_i,z_i,\bar z_i)\big\rangle
\ee

where $\tilde S^+_1(u,z,\bar z)$ is the subleading \textit{conformally soft} graviton operator which, in the presence of the time coordinate $u$, is \textit{different} from the subleading \textit{energetically soft} graviton operator $S^+_1(z,\bar z)$, defined as \cite{Kapec:2016jld,Kapec:2017gsg},
\be
S^+_1(z,\bar z) = \lim_{\om\rightarrow 0} \(1+ \om \frac{d}{d\om}\) G^{+}(\omega, z,\bar z)
\ee 

Here $G^+(\omega,z,\bar z)$ is the (creation) annihilation operator for a positive helicity graviton. Now we want to compute the R.H.S of \eqref{css} and along the way we will also obtain an expression for $\tilde S^+_1(u,z,\bar z)$ in terms of leading and subleading energetically soft gravitons. 

 \vskip 4pt
In order to do this, we start from the definition of the modified Mellin amplitude in terms of the $S$-matrix element \cite{Banerjee:2018gce,Banerjee:2019prz}, 
\be
\begin{gathered}
\big\langle{G^+_{\D}(u,z,\bar z)} \prod_{i=1}^n \phi_{h_i,\bar h_i}(u_i,z_i,\bar z_i)\big\rangle \\ = \int_{0}^\infty d\om \om^{\Delta -1} e^{-i\om u} \int_0^{\infty} \prod_{i=1}^n d\om_i \om_i^{\D_i-1} e^{- \epsilon_i \omega_i u_i} S_{n+1}\(\om,z,\bar z,\sigma=+2; \{\om_i,z_i,\bar z_i,\sigma_i\}\)
\end{gathered}
\ee

Here $\epsilon=\pm 1$ for an outgoing (incoming) particle and we have assumed that the graviton $G^+_{\D}$ is outgoing. Now to take the subleading conformal soft limit given in \eqref{css} we use the following identity \cite{Guevara:2019ypd}, 
\be\label{id}
\alpha^{\D-2} \Theta(\alpha) \sim \frac{\delta(\alpha)}{\D-1} - \frac{\delta'(\alpha)}{\D} + \frac{1}{2} \frac{\delta''(\alpha)}{\D+1} +....
\ee

where $\sim$ means that only pole terms in $\D$ are shown on the R.H.S of \eqref{id}. Now,
\be\label{alf}
\begin{gathered}
\lim_{\D\rightarrow 0} \D \int_{0}^\infty d\om \om^{\Delta -1} e^{-i\om u} S_{n+1}\(\om,z,\bar z,\sigma=+2; \{\om_i,z_i,\bar z_i,\sigma_i\}\) \\ = \lim_{\D\rightarrow 0} \D \int_{-\infty}^\infty d\om \ \Theta(\om) \om^{\Delta -2} e^{-i\om u} \om S_{n+1}\(\om,z,\bar z,\sigma=+2; \{\om_i,z_i,\bar z_i,\sigma_i\}\) \\ = - \int_{-\infty}^\infty d\om \ \ \delta'(\om) e^{-i\om u} \om S_{n+1}\(\om,z,\bar z,\sigma=+2; \{\om_i,z_i,\bar z_i,\sigma_i\}\)
\end{gathered}
\ee

Because of the delta function the integrand in \eqref{alf} is supported near $\om=0$ and so we can do a soft expansion of the $S$-matrix element in $\om$,
\be
S_{n+1}\(\om,z,\bar z,\sigma=+2; \{\om_i,z_i,\bar z_i,\sigma_i\}\) = \(\frac{s_0}{\om} + s_1 + \om s_2 +...\)S_{n}\(\{\om_i,z_i,\bar z_i,\sigma_i\}\)
\ee

where $s_0$ and $s_1$ are the leading and subleading soft factors, respectively. Now substituting this \eqref{alf} and doing integration by parts we get, 
\be
\begin{gathered}
\lim_{\D\rightarrow 0} \D \int_{0}^\infty d\om \om^{\Delta -1} e^{-i\om u} S_{n+1}\(\om,z,\bar z,\sigma=+2; \{\om_i,z_i,\bar z_i,\sigma_i\}\) \\ = -iu s_0 S_{n}\(\{\om_i,z_i,\bar z_i,\sigma_i\}\) + s_1 S_{n}\(\{\om_i,z_i,\bar z_i,\sigma_i\}\)
\end{gathered} 
\ee

Therefore we can write \eqref{css} as, 
\be\label{fa}
\begin{gathered}
\big\langle{\tilde S^+_1(u,z,\bar z)} \prod_{i=1}^n \phi_{h_i,\bar h_i}(u_i,z_i,\bar z_i)\big\rangle = \lim_{\D\rightarrow 0} \D \big\langle{G^+_{\D}(u,z,\bar z)} \prod_{i=1}^n \phi_{h_i,\bar h_i}(u_i,z_i,\bar z_i)\big\rangle \\ = -iu \int_{0}^\infty \prod_{i=1}^n d\om_i \om_i^{\D_i-1} e^{- \epsilon_i \omega_i u_i} s_0 S_{n}\(\{\om_i,z_i,\bar z_i,\sigma_i\}\) \\ + \int_{0}^\infty \prod_{i=1}^n d\om_i \om_i^{\D_i-1} e^{- \epsilon_i \omega_i u_i} s_1 S_{n}\(\{\om_i,z_i,\bar z_i,\sigma_i\}\) \\ = -iu \big\langle{S^+_0(z,\bar z)} \prod_{i=1}^n \phi_{h_i,\bar h_i}(u_i,z_i,\bar z_i)\big\rangle + \big\langle{S^+_1(z,\bar z)} \prod_{i=1}^n \phi_{h_i,\bar h_i}(u_i,z_i,\bar z_i)\big\rangle
\end{gathered}
\ee

where $S^+_0(z,\bar z)$ and $S^+_1(z,\bar z)$ are the leading and subleading (energentically) soft graviton operators. From \eqref{fa} we can write,
\be
\lim_{\D\rightarrow 0} \D G^+_{\D}(u,z,\bar z) = \tilde S^+_1(u,z,\bar z) = -iu S^+_0(z,\bar z) + S^+_1(z,\bar z)
\ee

We can see that in the presence of the time coordinate $u$, the subleading conformally soft graviton operator $\tilde S^+_1(u,z,\bar z)$ does not coincide with the energetically soft graviton operator $S^+_1(z,\bar z)$. The additional piece proportional to the leading soft operator $S^+_0(z,\bar z)$ can be projected out, say by setting $u=0$, but this turns out to be inconvenient because this breaks manifest time translation invariance. So in the presence of the time coordinate it is natural to work with the conformally soft subleading operator $\tilde S^+_1(u,z,\bar z)$.  

 \vskip 4pt
Now one can check that \cite{Banerjee:2018fgd,Partha2,Banerjee:2020kaa},
\be
\big\langle{S^+_0(z,\bar z)} \prod_{i=1}^n \phi_{h_i,\bar h_i}(u_i,z_i,\bar z_i)\big\rangle = - \(\sum_{k=1}^n \frac{\bar z - \bar z_k}{z - z_k} i\frac{\partial}{\partial u_k}\) \big\langle{\prod_{i=1}^n \phi_{h_i,\bar h_i}(u_i,z_i,\bar z_i)}\big\rangle
\ee

and 
\be\label{ocss}
\begin{gathered}
\big\langle{S^+_1(z,\bar z)} \prod_{i=1}^n \phi_{h_i,\bar h_i}(u_i,z_i,\bar z_i)\big\rangle = \sum_{k=1}^n \frac{(\bar z- \bar z_k)^2}{z-z_k}  \[\frac{2\bar h'_k}{\bar z- \bar z_k} - \bar\partial_k\] \langle{\prod_{i=1}^n \phi_{h_i, \bar h_i}(u_i,z_i,\bar z_i)}\rangle
\end{gathered}
\ee

where 
\be
2\bar h'_k = \D_k -\sigma_k + u_k \frac{\partial}{\partial u_k}
\ee

Combining these results we can write \eqref{fa} as, 
\be\label{csso}
\big\langle{\tilde S^+_1(u,z,\bar z)} \prod_{i=1}^n \phi_{h_i,\bar h_i}(u_i,z_i,\bar z_i)\big\rangle =  \sum_{k=1}^n \frac{(\bar z- \bar z_k)^2}{z-z_k}  \[\frac{2\bar h_k}{\bar z- \bar z_k} - \bar\partial_k\] \langle{\prod_{i=1}^n \phi_{h_i, \bar h_i}(u_i,z_i,\bar z_i)}\rangle
\ee

where
\be
\bar h_k = \frac{\D_k -\sigma_k}{2} +\frac{1}{2}\(u_k- u\) \frac{\partial}{\partial u_k}
\ee

This is precisely the result quoted in \eqref{replace}.



\subsection{Delta function representation for $n=6$ particles} 
\label{deltarep}
In this appendix we derive the representation of the delta function, used in Section \ref{6ptOPEMHV}, which imposes overall energy-momentum conservation for $n=6$ massless particles. Suppose we want to take the celestial OPE between the primary operators corresponding to gravitons labelled by $(5,6)$ in the S-matrix.  Then it is convenient to  parametrise their energies $\omega_{5}, \omega_{6}$ as 
\begin{equation}
\label{tomegapdef}
\begin{split}
& \omega_{5}= \omega_{P}\hspace{0.04cm} t, \quad \omega_{6}= \omega_{P} (1-\epsilon_{5}\epsilon_{6} \hspace{0.04cm}t)
\end{split}
\end{equation}

where $\epsilon_{5},\epsilon_{6}=\pm 1$ for outgoing (incoming) particles. Now for $n=6$ particles,  the constraints of energy-momentum conservation in $4$-dimensions yield $4$ equations for the $6$ energy variables. Thus we can solve for $4$ of the energy variables $\omega_{i}, i \in (1,2,3,4)$ in terms of $\omega_{5},\omega_{6}$ or equivalently $\omega_{p}$ and  $t$ after using \eqref{tomegapdef}. A convenient way of doing this is to use the spinor helicity variables in terms of which overall energy-momentum conservation implies
\begin{equation}
\label{momconsv}
\begin{split}
& \sum_{i=1}^{6} \langle q i \rangle [i k] =0
\end{split}
\end{equation}

where $q, k$ denote reference spinors. The spinor helicity brackets can be written as 
\begin{equation}
\label{angsqbrack}
\begin{split}
\langle i j \rangle = -2 \hspace{0.04cm} \epsilon_{i}\epsilon_{j} \sqrt{\omega_{i}\omega_{j}} \hspace{0.04cm}z_{ij}, \quad [ i j ] = 2 \sqrt{\omega_{i}\omega_{j}} \hspace{0.04cm} \bar{z}_{ij}
\end{split}
\end{equation}  

where we have used the following parametrisation of null momenta 
\begin{equation}
\begin{split}
 p^{\mu} = \epsilon \hspace{0.03cm}  \omega \hspace{0.03cm} q^{\mu}(z,\bar{z}), \quad  q^{\mu}(z,\bar{z})= (1+z \bar{z}, z+\bar{z}, -i(z-\bar{z}), 1- z\bar{z})
\end{split}
\end{equation} 

Now let us choose $q=3, k=4 $ in \eqref{momconsv} . Then we get
\begin{equation}
\label{q3k4}
\begin{split}
\epsilon_{1} \omega_{1} z_{13} \bar{z}_{14} + \epsilon_{2} \omega_{2} z_{23} \bar{z}_{24}  = - \epsilon_{6}  \omega_{P} \left[z_{36} \bar{z}_{46} - \epsilon_{5}  \epsilon_{6}  t \left( z_{36} \bar{z}_{56}+z_{56} \bar{z}_{46}-z_{56} \bar{z}_{56} \right) \right]
\end{split}
\end{equation}

Similarly for $q=4, k=3$ we get from \eqref{momconsv} 
\begin{equation}
\label{q4k3}
\begin{split}
\epsilon_{1} \omega_{1} z_{14} \bar{z}_{13} + \epsilon_{2} \omega_{2} z_{24} \bar{z}_{23}  = - \epsilon_{6}  \omega_{P} \left[z_{46} \bar{z}_{36} - \epsilon_{5}  \epsilon_{6}  t \left( z_{56} \bar{z}_{36}+z_{46} \bar{z}_{56}-z_{56} \bar{z}_{56} \right) \right]
\end{split}
\end{equation}

Now we can simultaneously solve the above equations \eqref{q3k4} and \eqref{q4k3}  and obtain $\omega_{1}, \omega_{2}$ in terms of $\omega_{P}, t$ and the $z_{ij},\bar{z}_{ij}$'s.  Implementing this procedure for other choices of the reference spinors, we can easily solve for $\omega_{3}$ and $\omega_{4}$ as well. Finally we get
\begin{equation}
\label{omegast}
\begin{split}
& \omega_{i}^{*} =  \epsilon_{i}\epsilon_{6}\hspace{0.04cm} \omega_{P} \left[ \sigma_{i,1} + \epsilon_{5}\epsilon_{6}\hspace{0.04cm} t \left( z_{56}  \hspace{0.04cm}  \sigma_{i,2} + \bar{z}_{56} \hspace{0.04cm} \sigma_{i,3}+ z_{56} \bar{z}_{56} \hspace{0.04cm} \sigma_{i,4}\right) \right] , \quad i \in 1,2,3,4.
\end{split}
\end{equation}

where 
\begin{equation}
\label{sigma1}
\begin{split}
& \sigma_{1,1}  =  \frac{  z_{36} \bar{z}_{36}}{z_{13} \bar{z}_{13}} \ \frac{\left(r_{23,46}-\bar{r}_{23,46}\right)}{\left(r_{12,34}-\bar{r}_{12,34}\right)} 
\end{split}
\end{equation} 

\begin{equation}
\label{sigma2}
\begin{split}
& \sigma_{2,1}  = - \frac{ z_{14} \bar{z}_{14} z_{36} \bar{z}_{36}}{z_{13} \bar{z}_{13} z_{24} \bar{z}_{24}} \ \frac{\left(r_{13,46}-\bar{r}_{13,46}\right)}{\left(r_{12,34}-\bar{r}_{12,34}\right)}
\end{split}
\end{equation} 

\begin{equation}
\label{sigma3}
\begin{split}
& \sigma_{3,1}  =  \frac{z_{14}\bar{z}_{14}z_{26} \bar{z}_{26}}{z_{13} \bar{z}_{13}z_{24}\bar{z}_{24}} \ \frac{\left(r_{12,46}-\bar{r}_{12,46}\right)}{\left(r_{12,34}-\bar{r}_{12,34}\right)}
\end{split}
\end{equation} 

\begin{equation}
\label{sigma4}
\begin{split}
& \sigma_{4,1}  = - \frac{  z_{26} \bar{z}_{26}}{z_{24} \bar{z}_{24}} \ \frac{\left(r_{12,36}-\bar{r}_{12,36}\right)}{\left(r_{12,34}-\bar{r}_{12,34}\right)}
\end{split}
\end{equation} 

and 
\begin{equation}
\label{sigmaderv}
\begin{split}
& \sigma_{i,2}  = \frac{\partial \hspace{0.04cm} \sigma_{i,1}}{\partial z_{6}}, \quad \sigma_{i,3} =  \frac{\partial \hspace{0.04cm} \sigma_{i,1}}{\partial \bar{z}_{6}}, \quad \sigma_{i,4} =  \frac{\partial^{2} \hspace{0.04cm} \sigma_{i,1}}{\partial z_{6}\partial \bar{z}_{6}}, \quad  \forall i = 1,2,3,4
\end{split}
\end{equation} 

In the above expressions, $r_{ij,kl}, \bar{r}_{ij,kl}$  denote holomorphic and antiholomorphic cross ratios and are given by
\begin{equation}
\label{crossratios}
\begin{split}
& r_{ij,kl} = \frac{z_{ij}z_{kl}}{z_{ik}z_{jl}}, \quad \bar{r}_{ij,kl} = \frac{\bar{z}_{ij}\bar{z}_{kl}}{\bar{z}_{ik}\bar{z}_{jl}}
\end{split}
\end{equation} 

Using the above results, the delta function for $n=6$ particles can now be expressed as
\begin{equation}
\label{delta}
\begin{split}
& \delta^{(4)} \Big( \sum_{i=1}^{6} \epsilon_{i} \omega_{i} q_{i}^{\mu}\Big) = \frac{i}{4} \frac{1}{(r_{12,34}- \bar{r}_{12,34}) z_{13}\bar{z}_{13}z_{24}\bar{z}_{24}} \  \prod_{i=1}^{4} \delta(\omega_{i} - \omega_{i}^{*})
\end{split}
\end{equation}

where $\omega^{*}_{i}$ are defined in \eqref{omegast}. The prefactor in \eqref{delta} is simply the Jacobian for the change of variables that we have performed here. Now it is easy to verify that  \eqref{delta} is equivalent to the representation of the delta function given in \cite{Fan:2019emx} on the locus of energy-momentum conservation. However, the form that we have presented here is better suited  for our purposes of performing the OPE decomposition of the Mellin amplitude in the $(5,6)$ channel. Representations which are convenient for doing the OPE in other channels can be easily worked out from \eqref{delta} by appropriate change of labels. 

 \vskip 4pt
Let us also note the following identities  which are useful for simplifying intermediate stages in computations that lead to some of the results obtained in Section \ref{6ptOPEMHV}. 
\begin{equation}
\label{sigmarel1}
\begin{split}
& \sum_{i=1}^{4} \sigma_{i,1} + 1=0
\end{split}
\end{equation}

\vspace{-0.5cm}
\begin{equation}
\label{sigmarel2}
\begin{split}
& \sum_{i=1}^{4} z_{i6} \hspace{0.05cm} \sigma_{i,1} =0
\end{split}
\end{equation}

\vspace{-0.5cm}
\begin{equation}
\label{sigmarel3}
\begin{split}
\sum_{i=1}^{4} \bar{z}_{i6} \hspace{0.05cm}  \sigma_{i,1} =0
\end{split}
\end{equation}

\vspace{-0.5cm}
\begin{equation}
\label{sigmarel4}
\begin{split}
& \sum_{i=1}^{4} z_{i6}\bar{z}_{i6} \hspace{0.05cm}   \sigma_{i,1} =0
\end{split}
\end{equation}

Before ending this section let us also note that  the limit $t\rightarrow 0$ in \eqref{omegast} and \eqref{delta}, yields a representation of the delta function that imposes energy-momentum conservation for $n=5$ particles.  Changing labels as $\epsilon_{6} \rightarrow \epsilon_{5}, \omega_{P}\rightarrow \omega_{5}, z_{6}\rightarrow z_{5}, \bar{z}_{6}\rightarrow \bar{z}_{5}$ we then have 
\begin{equation}
\label{delta5points}
\begin{split}
& \delta^{(4)} \Big( \sum_{i=1}^{5} \epsilon_{i} \omega_{i} q_{i}^{\mu}\Big) = \frac{i}{4} \frac{1}{(r_{12,34}- \bar{r}_{12,34}) z_{13}\bar{z}_{13}z_{24}\bar{z}_{24}} \  \prod_{i=1}^{4} \delta(\omega_{i} - \omega_{i}^{*})
\end{split}
\end{equation}

where
\begin{equation}
\label{omegast5pt}
\begin{split}
& \omega_{i}^{*} =  \epsilon_{i}  \epsilon_{5}  \hspace{0.04cm} \omega_{5} \hspace{0.04cm}  \sigma_{i,1} , \quad i \in 1,2,3,4.
\end{split}
\end{equation}

Here the $ \sigma_{i,1}$ are identical to \eqref{sigma1}-\eqref{sigma4}, upto the change of labels $z_{6}\rightarrow z_{5}, \bar{z}_{6}\rightarrow \bar{z}_{5}$.


\subsection{Delta function representation for $n(\ge 5)$ particles }\label{Delta}

In this section we note down the representation of the delta function that imposes overall energy-momentum conservation  for $n\ge 5$ massless particles which was obtained in \cite{Fan:2019emx}. This will turn out to be useful for the analysis in subsection \ref{hodge}. 

Let us define the cross-ratio $t_k$ as
\be
t_k = \frac{z_{12}z_{3k}}{z_{13}z_{2k}}
\ee

Then, following \cite{Fan:2019emx}, the momentum-conserving delta function, for $n\ge 5$, can be written as
\be
\d^{(4)}\(\sum_{i=1}^{n}\e_i \om_i q^{\mu}(z_i,\bar{z}_i)\)=\f{i}{4}\f{(1-t_4)(1-\bar{t}_4)}{t_4-\bar{t}_4}\f{1}{z_{14}\bar{z}_{14}z_{23}\bar{z}_{23}}\prod_{i=1}^4\d(\om_i-\om_i^*) = \mathcal J \prod_{i=1}^4\d(\om_i-\om_i^*)
\ee

where
\be
\mathcal{J} = \f{i}{4}\f{(1-t_4)(1-\bar{t}_4)}{t_4-\bar{t}_4}\f{1}{z_{14}\bar{z}_{14}z_{23}\bar{z}_{23}}
\ee

\be
\om_i^* = \sum_{j=5}^n f_{ij}\omega_j, \quad i =1,2,3,4,   \quad j= 5,6,....,n
\ee

and 
\be
f_{1j} = t_4 \bigg | \frac{z_{24}}{z_{12}}\bigg |^2 \frac{(1- t_4)(1-\bar t_4)}{t_4 - \bar t_4}\epsilon_1\epsilon_j\frac{t_j - \bar t_j}{(1-t_j)(1-\bar t_j)} - \epsilon_1\epsilon_j t_j \bigg | \frac{z_{2j}}{z_{12}}\bigg |^2
\ee

\be
f_{2j} = - \frac{1-t_4}{t_4} \bigg | \frac{z_{34}}{z_{23}}\bigg |^2 \frac{(1- t_4)(1-\bar t_4)}{t_4 - \bar t_4}\frac{\epsilon_1\epsilon_j}{\epsilon_1\epsilon_2}\frac{t_j - \bar t_j}{(1-t_j)(1-\bar t_j)} + \frac{\epsilon_1\epsilon_j}{\epsilon_1\epsilon_2}\frac{1-t_j}{t_j} \bigg | \frac{z_{3j}}{z_{23}}\bigg |^2
\ee

\be
f_{3j} = (1-t_4) \bigg | \frac{z_{24}}{z_{23}}\bigg |^2 \frac{(1- t_4)(1-\bar t_4)}{t_4 - \bar t_4}\frac{\epsilon_1\epsilon_j}{\epsilon_1\epsilon_3}\frac{t_j - \bar t_j}{(1-t_j)(1-\bar t_j)} - \frac{\epsilon_1\epsilon_j}{\epsilon_1\epsilon_3}(1-t_j) \bigg | \frac{z_{2j}}{z_{23}}\bigg |^2
\ee

\be
f_{4j} = - \frac{(1- t_4)(1-\bar t_4)}{t_4 - \bar t_4}\frac{\epsilon_1\epsilon_j}{\epsilon_1\epsilon_4}\frac{t_j - \bar t_j}{(1-t_j)(1-\bar t_j)}\bigg | \frac{z_{1j}}{z_{14}}\bigg |^2
\ee

\subsection{Descendant correlation functions}
\label{desccorr}

Here we note down the action of some of the extended symmetry algebra generators on the $5$-point Mellin amplitude that we have used in Section \ref{6ptOPEMHV} of this paper. The $5$-point Mellin amplitude was evaluated in Section \ref{5ptMellin} and is given by
\begin{equation}
\label{5ptcorr}
\begin{split}
 \mathcal{M}_{5}  &  =  \left \langle  G^{-}_{\Delta_{1}}(1) G^{-}_{\Delta_{2}}(2) G^{+}_{\Delta_{3}}(3)  G^{+}_{\Delta_{4}}(4) G^{+}_{\Delta}  (6)\right\rangle \\
 & =  i \prod_{i=1}^{4} \epsilon_{i} \prod_{j=1}^{2} ( \epsilon_{j} \sigma_{j,1})^{3+i\lambda_{i}}  \prod_{k=3}^{4}( \epsilon_{k}\sigma_{k,1})^{i\lambda_{k}-1} \prod_{l=1}^{4} \Theta \left( \epsilon_{l} \sigma_{l,1}\right) \hspace{0.04cm} \frac{z_{12}^{8}}{z_{12}z_{13}z_{14}z_{16}z_{23}z_{24}z_{26}z_{34}z_{36}z_{46}}  \\
 &  \times  \frac{\Gamma(3+i\Lambda)}{( i \hspace{0.05cm} \mathcal{U}_{1})^{3+i\Lambda}}   
\end{split}
\end{equation}

where $G^{s_{i}}_{\Delta_{i}}(i)  \equiv G^{s_{i}}_{\Delta_{i}} (u_{i},z_{i},\bar{z}_{i}) $, $s_{i}= \pm 2$; $i \in (1,2,3,4)$, $\Delta = 1+i\lambda_{5}+i\lambda_{6}$ and $\Lambda=\sum_{i=1}^{6}\lambda_{i}$. The $\sigma_{i,1}$'s are defined in equations \eqref{sigma1} to \eqref{sigma4}. We have also taken here $G^{+}_{\Delta}  (6)$ to correspond to an outgoing graviton. 

\subsubsection{$ \mathcal{P}_{-n,-1}\mathcal{M}_{5}$}

First let us consider the following correlation function involving the insertion of the descendant $P_{-n,-1}G^{+}_{\Delta-1}(6)$ with $n\ge 2$
\begin{equation}
\begin{split}
  \mathcal{P}_{-n,-1}\mathcal{M}_{5}  &  =  \left \langle  G^{-}_{\Delta_{1}}(1) G^{-}_{\Delta_{2}}(2) G^{+}_{\Delta_{3}}(3)  G^{+}_{\Delta_{4}}(4)   (P_{-n,-1} G^{+}_{\Delta-1} ) (6)\right\rangle  \\
  & = - \sum_{k=1}^{4} \frac{1}{z^{n-1}_{k6}} \ i \frac{ \partial}{ \partial u _{k}} \mathcal{M}_{5}
\end{split}
\end{equation}

Using the expression of the $5$-point function in \eqref{5ptcorr}, the above correlator becomes
\begin{equation}
\label{Pminnmin15pt}
\begin{split}
  \mathcal{P}_{-n,-1}\mathcal{M}_{5}  =  - \sum_{k=1}^{4} \frac{\sigma_{k,1}}{z^{n-1}_{k6}}  \  \mathcal{P}_{-1,-1}\mathcal{M}_{5}
\end{split}
\end{equation} 


\subsubsection{$ \mathcal{P}_{-n, 0}\mathcal{M}_{5}$}

Now let us consider the $5$-point correlation function with the insertion of $P_{-n, 0}G^{+}_{\Delta-1}(6)$ for $n\ge 2$. This is given by
\begin{equation}
\label{Pminn05pt}
\begin{split}
  \mathcal{P}_{-n,0}\mathcal{M}_{5}  &  =  \left \langle  G^{-}_{\Delta_{1}}(1) G^{-}_{\Delta_{2}}(2) G^{+}_{\Delta_{3}}(3)  G^{+}_{\Delta_{4}}(4)   (P_{-2,0} G^{+}_{\Delta-1} ) (6)\right\rangle  \\
  & = - \sum_{k=1}^{4} \frac{\bar{z}_{k6}}{z^{n-1}_{k6}} \ i \frac{ \partial}{ \partial u _{k}} \mathcal{M}_{5} = - \sum_{k=1}^{4} \frac{\bar{z}_{k6}}{z^{n-1}_{k6}} \  \sigma_{k,1} \  \mathcal{P}_{-1,-1}\mathcal{M}_{5}
  \end{split}
\end{equation}


\subsubsection{$\mathcal{L}_{-1} \mathcal{P}_{-1,-1}\mathcal{M}_{5}$}

The action of $\mathcal{L}_{-1} \mathcal{P}_{-1,-1}$ on the $5$-point Mellin amplitude is given by the following correlation function 
\begin{equation}
\label{Lmin1Pmin15pt}
\begin{split}
 \mathcal{L}_{-1} \mathcal{P}_{-1,-1}\mathcal{M}_{5}   &  =  \left \langle  G^{-}_{\Delta_{1}}(1) G^{-}_{\Delta_{2}}(2) G^{+}_{\Delta_{3}}(3)  G^{+}_{\Delta_{4}}(4)   (L_{-1}P_{-1,-1} G^{+}_{\Delta-1} ) (6)\right\rangle  \\
 & = \partial_{z_{6}}  \left( i  \partial_{u_{6}}  \mathcal{M}_{5}    \right)
\end{split}
\end{equation}

Then using the explicit form of $\mathcal{M}_{5}$ and also \eqref{sigmaderv} we get
\begin{equation}
\label{Lmin1Pmin15pt1}
\begin{split}
& \mathcal{L}_{-1} \mathcal{P}_{-1,-1}\mathcal{M}_{5}   \\
& = \left[(3+i\lambda_{1}) \frac{\sigma_{1,2}}{\sigma_{1,1}} + (3+i\lambda_{2}) \frac{\sigma_{2,2}}{\sigma_{2,1}} +(i\lambda_{3}-1) \frac{\sigma_{3,2}}{\sigma_{3,1}} +(i\lambda_{4}-1) \frac{\sigma_{4,2}}{\sigma_{4,1}} - \frac{(4+i\Lambda)}{\mathcal{U}_{1}}\  \mathcal{U}_{2}\right]  \mathcal{P}_{-1,-1}\mathcal{M}_{5} \\
&+ \sum_{i=1}^{4} \frac{1}{z_{i6}} \ \mathcal{P}_{-1,-1}\mathcal{M}_{5} 
\end{split}
\end{equation}

We refer the reader to \eqref{Uidef} for the definition of  $\mathcal{U}_{1}$ and $\mathcal{U}_{2}$. Now it is important to note that in deriving the above result we have not applied the $z_{6}$-derivative to the theta function in \eqref{5ptcorr}. This is because doing so yields contact terms and here will be only considering correlation functions where all operator insertions shall be kept at separated points.  Also note that the term inside the square brackets in the above has been denoted by $\mathcal{I}_{1,0}$ in Section \ref{6ptOPEMHV}. We will also use this convenient notation in relevant expressions throughout the rest of this section of the Appendix. 


\subsubsection{$\mathcal{L}_{-1} \mathcal{P}_{-2,-1}\mathcal{M}_{5}$}

Here the correlation function of interest is given by
\begin{equation}
\label{Lmin1Pmin25pt}
\begin{split}
 \mathcal{L}_{-1}\mathcal{P}_{-2,-1}\mathcal{M}_{5}  & =   \left \langle  G^{-}_{\Delta_{1}}(1) G^{-}_{\Delta_{2}}(2) G^{+}_{\Delta_{3}}(3)  G^{+}_{\Delta_{4}}(4)   (L_{-1}P_{-2,-1} G^{+}_{\Delta-1} ) (6)\right\rangle  
\end{split}
\end{equation}

Using \eqref{Pminnmin15pt} for $n=2$ and the Ward identity for global $z$-translations in the above we obtain the following result
\begin{equation}
\label{Lmin1Pmin25pt1}
\begin{split}
& \mathcal{L}_{-1}\mathcal{P}_{-2,-1}\mathcal{M}_{5} = \mathcal{P}_{-3,-1}\mathcal{M}_{5} - \sum_{i=1}^{4}\frac{ \sigma_{i,2}}{z_{i6}} \  \mathcal{P}_{-1,-1}\mathcal{M}_{5} + \sum_{i=1}^{4}\frac{ 1}{z_{i6}} \  \mathcal{P}_{-2,-1}\mathcal{M}_{5} + \mathcal{I}_{1,0} \hspace{0.04cm} \mathcal{P}_{-2,-1}  \mathcal{M}_{5} 
\end{split}
\end{equation}


\subsubsection{$\mathcal{L}_{-1} \mathcal{P}_{-2,0}\mathcal{M}_{5}$}

Now let us evaluate the action of $\mathcal{L}_{-1} \mathcal{P}_{-2,0}$ on the $5$-point Mellin amplitude.  This is given by the following correlation function 
\begin{equation}
\label{Lmin1Pmin205pt}
\begin{split}
 \mathcal{L}_{-1} \mathcal{P}_{-2,0}\mathcal{M}_{5} & = \left \langle  G^{-}_{\Delta_{1}}(1) G^{-}_{\Delta_{2}}(2) G^{+}_{\Delta_{3}}(3)  G^{+}_{\Delta_{4}}(4)   (L_{-1}P_{-2,0} G^{+}_{\Delta-1} ) (6)\right\rangle  
\end{split}
\end{equation}

Applying  \eqref{Pminn05pt} for $n=2$ in the above we get
\begin{equation}
\label{Lmin1Pmin205pt1}
\begin{split}
& \mathcal{L}_{-1} \mathcal{P}_{-2,0}\mathcal{M}_{5}  =  \mathcal{I}_{1,0} \mathcal{P}_{-2,0} \mathcal{M}_{5} +  \mathcal{P}_{-3,0} \mathcal{M}_{5} -  \sum_{i=1}^{4} \frac{\bar{z}_{i6}}{z_{i6}} \ \sigma_{i,2} \ \mathcal{P}_{-1,-1} \mathcal{M}_{5} + \sum_{i=1}^{4} \frac{1}{z_{i6}}  \ \mathcal{P}_{-2,0} \mathcal{M}_{5}
\end{split}
\end{equation}

Then let us take note of the following identity 
\begin{equation}
\label{sigmaid2}
\begin{split}
 \sum_{i=1}^{4} \frac{\bar{z}_{i6}}{z_{i6}} \ \sigma_{i,2} + \sum_{i=1}^{4} \frac{1}{z_{i6}} \bigg( \sum_{j=1}^{4} \frac{\bar{z}_{j6}}{z_{j6}} \ \sigma_{j,1}\bigg) =  \sum_{i=1}^{4} \frac{\bar{z}_{i6}}{z^{2}_{i6}} \ \sigma_{i,1}
\end{split}
\end{equation}

This identity can be easily proved using the set of relations given in \eqref{sigmarel1} to \eqref{sigmarel4}. 
Using this we can write \eqref{Lmin1Pmin205pt1} as
\begin{equation}
\label{Lmin1Pmin205pt2}
\begin{split}
& \mathcal{L}_{-1} \mathcal{P}_{-2,0}\mathcal{M}_{5}  =  \mathcal{I}_{1,0} \mathcal{P}_{-2,0} \mathcal{M}_{5} + 2\hspace{0.03cm} \mathcal{P}_{-3,0} \mathcal{M}_{5} 
\end{split}
\end{equation}


\subsubsection{$\bar{\mathcal{L}}_{-1} \mathcal{P}_{-1,-1}\mathcal{M}_{5}$}

Next we want to consider the following descendant correlation function 
\begin{equation}
\label{Lbarmin1Pmin15pt}
\begin{split}
 \bar{\mathcal{L}}_{-1} \mathcal{P}_{-1,-1}\mathcal{M}_{5} &   =  \left \langle  G^{-}_{\Delta_{1}}(1) G^{-}_{\Delta_{2}}(2) G^{+}_{\Delta_{3}}(3)  G^{+}_{\Delta_{4}}(4)   (\bar{L}_{-1}P_{-1,-1} G^{+}_{\Delta-1} ) (6)\right\rangle \\
 & = \partial_{\bar{z}_{6}}  \left( i  \partial_{u_{6}}  \mathcal{M}_{5}    \right)
\end{split}
\end{equation}

Again using the expression of $\mathcal{M}_{5}$ given in \eqref{5ptcorr} it can be easily shown that
\begin{equation}
\label{Lbarmin1Pmin15pt1}
\begin{split}
& \bar{\mathcal{L}}_{-1} \mathcal{P}_{-1,-1}\mathcal{M}_{5} \\
&  = \left[(3+i\lambda_{1}) \frac{\sigma_{1,3}}{\sigma_{1,1}} + (3+i\lambda_{2}) \frac{\sigma_{2,3}}{\sigma_{2,1}} +(i\lambda_{3}-1) \frac{\sigma_{3,3}}{\sigma_{3,1}} +(i\lambda_{4}-1) \frac{\sigma_{4,3}}{\sigma_{4,1}} - \frac{(4+i\Lambda)}{\mathcal{U}_{1}} \ \mathcal{U}_{3}\right]  \mathcal{P}_{-1,-1}\mathcal{M}_{5} 
\end{split}
\end{equation}

Note that the term inside the square brackets above has been assigned the notation $\mathcal{I}_{0,1}$ in Section \ref{6ptOPEMHV}.  We will employ this notation here as well in some of the expressions to be considered next. 


\subsubsection{$\bar{\mathcal{L}}_{-1} \mathcal{P}_{-2,0}\mathcal{M}_{5}$}

The action $\bar{\mathcal{L}}_{-1} \mathcal{P}_{-2,0}$ on the $5$-point Mellin amplitude is given by the correlator
\begin{equation}
\label{Lbarmin1Pmin205pt}
\begin{split}
 \bar{\mathcal{L}}_{-1} \mathcal{P}_{-2,0}\mathcal{A}_{5} & = \left \langle  G^{-}_{\Delta_{1}}(1) G^{-}_{\Delta_{2}}(2) G^{+}_{\Delta_{3}}(3)  G^{+}_{\Delta_{4}}(4)   (\bar{L}_{-1}P_{-2,0} G^{+}_{\Delta-1} ) (6)\right\rangle  
\end{split}
\end{equation}

Using  \eqref{Pminn05pt} in the above we get
\begin{equation}
\label{Lbarmin1Pmin205pt1}
\begin{split}
 \bar{\mathcal{L}}_{-1} \mathcal{P}_{-2,0}\mathcal{M}_{5} &  = \partial_{\bar{z}_{6}} \bigg( \sum_{i=1}^{4}\frac{\bar{z}_{i6}}{z_{i6}} \ \sigma_{i,1}\ \mathcal{P}_{-1,-1}\mathcal{M}_{5} \bigg)   \\
& = - \mathcal{P}_{-2,-1}\mathcal{M}_{5}  - \sum_{i=1}^{4} \frac{\bar{z}_{i6}}{z_{i6}}\ \sigma_{i,3} \ \mathcal{P}_{-1,-1}\mathcal{M}_{5} -  \sum_{i=1}^{4}\frac{\bar{z}_{i6}}{z_{i6}} \ \sigma_{i,1} \ \bar{\mathcal{L}}_{-1} \mathcal{P}_{-1,-1}\mathcal{M}_{5} \\
& = - 2 \hspace{0.03cm} \mathcal{P}_{-2,-1}\mathcal{M}_{5} + \sum_{i=1}^{4} \frac{1}{z_{i6}}\ \mathcal{P}_{-1,-1}\mathcal{M}_{5} -  \sum_{i=1}^{4}\frac{\bar{z}_{i6}}{z_{i6}} \ \sigma_{i,1} \ \bar{\mathcal{L}}_{-1} \mathcal{P}_{-1,-1}\mathcal{M}_{5} 
\end{split}
\end{equation}

where in obtaining the last line above we used the identity
\begin{equation}
\label{sigmaid1}
\begin{split}
 &  \sum_{i=1}^{4} \frac{\bar{z}_{i6} }{z_{i6}} \hspace{0.05cm} \sigma_{i,3} +  \sum_{i=1}^{4} \frac{1 }{z_{i6}} = -     \sum_{i=1}^{4} \frac{\sigma_{i,1} }{z_{i6}}
\end{split}
\end{equation}

This identity can also be easily checked using \eqref{sigmarel1} to \eqref{sigmarel4}.


\subsubsection{$\bar{\mathcal{L}}_{-1} \mathcal{P}_{-3,0}\mathcal{M}_{5}$}

In order to evaluate the action of $\bar{\mathcal{L}}_{-1} \mathcal{P}_{-3,0}$ on $\mathcal{M}_{5}$ we consider the correlation function
\begin{equation}
\label{Lbarmin1Pmin305pt}
\begin{split}
 \bar{\mathcal{L}}_{-1}\mathcal{P}_{-3,0}\mathcal{M}_{5}  & =   \left \langle  G^{-}_{\Delta_{1}}(1) G^{-}_{\Delta_{2}}(2) G^{+}_{\Delta_{3}}(3)  G^{+}_{\Delta_{4}}(4)   (\bar{L}_{-1}P_{-3,0} G^{+}_{\Delta-1} ) (6)\right\rangle  
\end{split}
\end{equation}

Applying \eqref{Pminn05pt} we find the above to be given by
\begin{equation}
\label{Lbarmin1Pmin305pt1}
\begin{split}
 \bar{\mathcal{L}}_{-1}\mathcal{P}_{-3,0}\mathcal{M}_{5} = - \mathcal{P}_{-3,-1}\mathcal{M}_{5}-  \sum_{i=1}^{4}\frac{\bar{z}_{i6}}{z_{i6}^{2}} \ \sigma_{i,3}\  \mathcal{P}_{-1,-1}\mathcal{M}_{5} + \mathcal{I}_{0,1} \hspace{0.04cm}\mathcal{P}_{-3,0}  \mathcal{M}_{5}
\end{split}
\end{equation}


\subsubsection{$\mathcal{J}^{0}_{-1} \mathcal{P}_{-1,-1}\mathcal{M}_{5}$}

Finally let us compute the action of  $\mathcal{J}^{0}_{-1} \mathcal{P}_{-1,-1}$ on $\mathcal{M}_{5}$. This is given by
\begin{equation}
\label{J0min1Pmin15pt}
\begin{split}
 \mathcal{J}^{0}_{-1}\mathcal{P}_{-1,-1}\mathcal{M}_{5}  & =   \left \langle  G^{-}_{\Delta_{1}}(1) G^{-}_{\Delta_{2}}(2) G^{+}_{\Delta_{3}}(3)  G^{+}_{\Delta_{4}}(4)   (J^{0}_{-1}P_{-2,0} G^{+}_{\Delta-1} ) (6)\right\rangle  \\
 & = - \sum_{k=1}^{4}\frac{2 \hspace{0.03cm} \bar{h}_{k} + u_{k6} \partial_{u_{k}}+ 2\hspace{0.03cm}  \bar{z}_{k6} \partial_{\bar{z}_{k}}}{2\hspace{0.03cm}  z_{k6}} \ \mathcal{P}_{-1,-1}\mathcal{M}_{5}
\end{split}
\end{equation}

Evaluating the action of the differential operators in the above we get
\begin{equation}
\label{J0min1Pmin15pt1}
\begin{split}
& \mathcal{J}^{0}_{-1}\mathcal{P}_{-1,-1}\mathcal{M}_{5} \\ 
&= - \frac{1}{2} \left[ \left( \frac{3+i\lambda_{1}}{z_{16}} + \frac{3+i\lambda_{2}}{z_{26}} +\frac{i\lambda_{3}-1}{z_{36}} + \frac{i\lambda_{4}-1}{z_{46}}  \right) - \frac{(4+i\Lambda)}{\mathcal{U}_{1}} \sum_{i=1}^{4} \frac{\sigma_{i,1}u_{i6}}{z_{i6}} \right] \mathcal{P}_{-1,-1}\mathcal{A}_{5}\\
& -  \sum_{i=1}^{4}\frac{\bar{z}_{i6}}{z_{i6}} \ \sigma_{i,1} \ \bar{\mathcal{L}}_{-1} \mathcal{P}_{-1,-1}\mathcal{M}_{5} 
\end{split}
\end{equation}

We can express the above result in an equivalent form using \eqref{Lbarmin1Pmin205pt1} as follows
\begin{equation}
\label{J0min1Pmin15pt2}
\begin{split}
& \mathcal{J}^{0}_{-1}\mathcal{P}_{-1,-1}\mathcal{M}_{5} \\ 
&= - \frac{1}{2} \left[ \left( \frac{3+i\lambda_{1}}{z_{16}} + \frac{3+i\lambda_{2}}{z_{26}} +\frac{i\lambda_{3}-1}{z_{36}} + \frac{i\lambda_{4}-1}{z_{46}}  \right) - \frac{(4+i\Lambda)}{\mathcal{U}_{1}} \sum_{i=1}^{4} \frac{\sigma_{i,1}u_{i6}}{z_{i6}} \right] \mathcal{P}_{-1,-1}\mathcal{A}_{5}\\
& -\sum_{i=1}^{4} \frac{1}{z_{i6}} \ \mathcal{P}_{-1,-1}\mathcal{A}_{5} + \bar{\mathcal{L}}_{-1}\mathcal{P}_{-2,0} \mathcal{A}_{5} + 2 \hspace{0.03cm} \mathcal{P}_{-2,-1}\mathcal{A}_{5} 
\end{split}
\end{equation}

This representation of $ \mathcal{J}^{0}_{-1}\mathcal{P}_{-1,-1}\mathcal{M}_{5}$ will be useful in the next section where we explicitly verify the decoupling relation \eqref{de2} for the $5$-point Mellin amplitude.


\subsection{Direct check of decoupling relation \eqref{de2} for $5$-point MHV amplitude}\label{5direct}

In this section of the Appendix, we prove that the $5$-point Mellin amplitude in \eqref{Hmellin5pt1}  satisfies the following differential equation
\begin{equation}
\label{nst2}
\begin{split}
 \left( \bar{\mathcal{L}}_{-1}\mathcal{P}_{-2,0} + (2+i\lambda_{5}+i\lambda_{6}) \mathcal{P}_{-2,-1}  -  \mathcal{L}_{-1}\mathcal{P}_{-1,-1}  - 2\hspace{0.03cm} \mathcal{J}^{0}_{-1} \mathcal{P}_{-1,-1}  \right)\mathcal{M}_{5} =0
\end{split}
\end{equation}

where $\mathcal{M}_{5}$ is the Mellin transform of the $5$-point MHV amplitude. In order to prove \eqref{nst2} let us first consider the following combination of differential operators acting on $\mathcal{M}_{5}$
\begin{equation}
\label{nst2proof1}
\begin{split}
& \left(\bar{\mathcal{L}}_{-1} \mathcal{P}_{-2,0}  + 2 \hspace{0.04cm} \mathcal{P}_{-2,-1} -\mathcal{L}_{-1} \mathcal{P}_{-1,-1}  -  2 \hspace{0.04cm} \mathcal{J}^{0}_{-1}\mathcal{P}_{-1,-1} \right) \mathcal{M}_{5} 
\end{split}
\end{equation}

Using \eqref{Pminnmin15pt},  \eqref{Lmin1Pmin15pt1}, \eqref{Lbarmin1Pmin15pt1}, \eqref{Lbarmin1Pmin205pt1},  and \eqref{J0min1Pmin15pt2} we find that \eqref{nst2proof1} evaluates to 
\begin{equation}
\label{nst2proof2}
\begin{split}
 &  \Bigg( \frac{(3+i\lambda_{1})}{\sigma_{1,1}} \left[ \frac{\sigma_{1,1}}{z_{16}} + \sigma_{1,3} \left( \sum_{i=1}^{4}\frac{\bar{z}_{i6}}{z_{i6}} \ \sigma_{i,1}  \right) - \sigma_{1,2}\right] + \frac{(3+i\lambda_{2})}{\sigma_{2,1}} \left[ \frac{\sigma_{2,1}}{z_{26}} + \sigma_{2,3} \left( \sum_{i=1}^{4}\frac{\bar{z}_{i6}}{z_{i6}} \ \sigma_{i,1}  \right) - \sigma_{2,2}\right] \\
 & + \frac{(i\lambda_{3}-1)}{\sigma_{3,1}} \left[ \frac{\sigma_{3,1}}{z_{36}} + \sigma_{3,3} \left( \sum_{i=1}^{4}\frac{\bar{z}_{i6}}{z_{i6}} \ \sigma_{i,1}  \right) - \sigma_{3,2}\right] + \frac{(i\lambda_{4}-1)}{\sigma_{4,1}} \left[ \frac{\sigma_{4,1}}{z_{46}} + \sigma_{4,3} \left( \sum_{i=1}^{4}\frac{\bar{z}_{i6}}{z_{i6}} \ \sigma_{i,1}  \right) - \sigma_{4,2}\right] \\
 & - \frac{(4+i\Lambda)}{\mathcal{U}_{1}} \sum_{i=1}^{4} \left[ \frac{\sigma_{i,1}}{z_{16}} + \sigma_{i,3} \left( \sum_{j=1}^{4}\frac{\bar{z}_{j6}}{z_{j6}} \ \sigma_{j,1}  \right) - \sigma_{i,2} \right] u_{i6} \Bigg) \mathcal{P}_{-1,-1} \mathcal{M}_{5} 
\end{split}
\end{equation}

Now let us note the following identity 
\begin{equation}
\label{nst2proof3}
\begin{split}
 & \frac{\sigma_{i,1}}{z_{i6}} + \sigma_{i,3} \left( \sum_{k=1}^{4}\frac{\bar{z}_{k6}}{z_{k6}} \ \sigma_{k,1}  \right) - \sigma_{i,2} =  - \sigma_{i,1} \left( \sum_{j=1}^{4}\frac{\sigma_{j,1}}{z_{j6}}  \right) , \quad \forall i=1,2,3,4
\end{split}
\end{equation}

Applying the above identity in \eqref{nst2proof2} we end up with
\begin{equation}
\begin{split}
& \left(\bar{\mathcal{L}}_{-1} \mathcal{P}_{-2,0}  + 2 \hspace{0.04cm} \mathcal{P}_{-2,-1} -\mathcal{L}_{-1} \mathcal{P}_{-1,-1}  -  2 \hspace{0.04cm} \mathcal{J}^{0}_{-1}\mathcal{P}_{-1,-1} \right) \mathcal{M}_{5} \\
& = - \left[ (4+i\lambda_{1}+i\lambda_{2}+i\lambda_{3}+i\lambda_{4} ) \sum_{j=1}^{4}\frac{\sigma_{j,1}}{z_{j6}}   + \frac{(4+i\Lambda)}{\mathcal{U}_{1}} \left( - \sum_{i=1}^{4} \sigma_{i,1} u_{i6} \right) \left( \sum_{j=1}^{4}\frac{\sigma_{j,1}}{z_{j6}}  \right) \right]  \mathcal{P}_{-1,-1} \mathcal{M}_{5}
\end{split}
\end{equation}

Using $\Lambda= \sum_{i=1}^{6}\lambda_{i}$ and the definition of $\mathcal{U}_{1}$ given in \eqref{Uidef}, we then obtain
\begin{equation}
\begin{split}
& \left(\bar{\mathcal{L}}_{-1} \mathcal{P}_{-2,0}  + 2 \hspace{0.04cm} \mathcal{P}_{-2,-1} -\mathcal{L}_{-1} \mathcal{P}_{-1,-1}  -  2 \hspace{0.04cm} \mathcal{J}^{0}_{-1}\mathcal{P}_{-1,-1} \right) \mathcal{M}_{5} \\
& = (i\lambda_{5}+i\lambda_{6}) \sum_{j=1}^{4}\frac{\sigma_{j,1}}{z_{j6}} \ \mathcal{P}_{-1,-1} \mathcal{M}_{5} = -  (i\lambda_{5}+i\lambda_{6})  \mathcal{P}_{-2,-1} \mathcal{M}_{5}
\end{split}
\end{equation}

From the above result it then follows that
\begin{equation}
\begin{split}
& \left(\bar{\mathcal{L}}_{-1} \mathcal{P}_{-2,0}  + (2 +i\lambda_{5}+i\lambda_{6}) \mathcal{P}_{-2,-1} -\mathcal{L}_{-1} \mathcal{P}_{-1,-1}  -  2 \hspace{0.04cm} \mathcal{J}^{0}_{-1}\mathcal{P}_{-1,-1} \right) \mathcal{M}_{5} =0
\end{split}
\end{equation}

This is the result that we wanted to prove. 


\subsection{Proof of null state decoupling relation \eqref{de1} from Hodges' formula}\label{hodge}

We start with the Hodges' formula \cite{Hodges:2011wm, Hodges:2012ym} for the $n$ - graviton MHV scattering amplitude given by,
\begin{equation}
\label{hodge2}
\begin{split}
& \mathcal{A}_{n} (1^{-},2^{-},3^{+},\ldots,n^{+})=  \langle 1 2 \rangle^{8} \ \frac{ \text{det}( \Phi^{i j k}_{p q r })}{\langle i j \rangle \langle i k \rangle\langle j k \rangle\langle p q \rangle\langle p r \rangle\langle q r \rangle }
\end{split}
\end{equation}

Here $\Phi^{ijk}_{pqr}$ is a $(n-3)\times (n-3)$ matrix obtained by deleting the set of rows $\{i,j,k\}$ and columns $\{p,q,r\}$ from a $n\times n$ matrix  $\Phi$ whose elements are defined as follows
  \begin{equation}
  \label{offd}
    \Phi_{ij}= 
\begin{cases}
    \frac{[i j]}{\langle i j \rangle},&  i\ne j\\
    - \sum\limits_{k\ne i} \frac{[i k]\langle x k \rangle \langle y k \rangle }{\langle i k \rangle \langle x i \rangle \langle y i \rangle},              & i=j
\end{cases}
\end{equation}  

where $x, y$ denote reference spinors. 

\vskip 4pt
At first we want to study the leading and subleading conformal soft limits of the MHV amplitude \eqref{hodge2} in Mellin space when the outgoing graviton $(n-1)^+$ becomes soft. After that we take the OPE limit where the soft graviton $(n-1)^+$ is brought close to the hard graviton $n^+$. For simplicity we also assume that the graviton $n^+$ is outgoing.
 
\vskip 4pt
In order to do this let us make the following choices
\be
\{i,j,k\} = \{1,2,3\}, \quad \{p,q,r\} = \{ n-2, n-1, n\}, \quad x= n-1, \quad y = n
\ee

We also have to express the angle and square brackets in terms of $(\omega, z , \bar z)$ by using the formulas, 
\be
\langle{ij\rangle} = -2 \epsilon_i \epsilon_j \sqrt{\omega_i\omega_j} z_{ij}, \quad \[ij\] = 2 \sqrt{\omega_i\omega_j} \bar z_{ij}
\ee

where $\epsilon_i = \pm 1$ for an outgoing (incoming) particle. Since the gravitons $(n-1)^+$ and $n^+$ are assumed to be outgoing, $\epsilon_{n-1}= \epsilon_n =1$. 

\vskip 4pt
Now using these various definitions, the Hodges' formula \eqref{hodge2} for the MHV amplitude can be written as,
\be\label{hsh}
\begin{split}
&  \mathcal{A}_{n} (1^{-},2^{-},3^{+},\ldots,n^{+}) \\
 &=  \langle 1 2 \rangle^{8} \ \frac{ \text{det}\left( \Phi^{123}_{n-2 \ n-1 \ n }\right)}{\langle 12 \rangle \langle 13 \rangle\langle 23 \rangle\langle n-2 \ n-1 \rangle\langle n-2 \  n \rangle\langle n-1 \  n \rangle } \\
 & = \frac{4 \omega_1^4 \omega_2^4}{\omega_1\omega_2\omega_3 \omega_{n-2}\omega_{n-1}\omega_n}\frac{z^8_{12}}{z_{12}z_{13}z_{23} z_{n-2 \ n-1}z_{n-2 \ n}z_{n-1 \ n}} \text{det}( \Phi^{123}_{n-2 \ n-1 \ n })
 \end{split}
\ee

\vskip 4pt
Let us now go to the Mellin space. The Mellin transformation of the scattering amplitude $\mathcal A_n$ is given by,
\be
\begin{split}
\mathcal M_n &= \langle{G^-_{\D_1}(1)G^-_{\D_2}(2)\prod_{i=3}^{n}G^+_{\D_i}(i)}\rangle \\
& = \prod_{i=1}^n \int_{0}^{\infty} d\omega_i \hspace{0.1cm} \omega_i^{\D_i-1} \hspace{0.05cm}\mathcal A_n \delta^{(4)}\(\sum_{i=1}^n \epsilon_i \omega_i q^{\mu}(z_i,\bar z_i)\) \\
&= \frac{4 z^8_{12}}{z_{12}z_{13}z_{23} z_{n-2 \ n-1}z_{n-2 \ n}}   \prod_{i=1}^{n-2} \int_{0}^{\infty} d\omega_i \omega_i^{\D_i-1} \int_{0}^{\infty}d\omega_{n-1}\int_{0}^{\infty}d\omega_n \ \omega_{n-1}^{\D_{n-1}-2} \ \omega_{n}^{\D_{n}-2} \\
& \times \frac{\omega_1^4 \omega_2^4}{\omega_1\omega_2\omega_3 \omega_{n-2}} \frac{\text{det}( \Phi^{123}_{n-2 \ n-1 \ n })}{z_{n-1 \ n}}  \delta^{(4)}\(\sum_{i=1}^n \epsilon_i \omega_i q^{\mu}(z_i,\bar z_i)\)
\end{split}
\ee

Now we make a change of variable 
\be
\omega_{n-1} = t \omega, \quad \omega_{n} = (1-t)\omega, \quad d\omega_{n-1} d\omega_n = \omega d\omega dt 
\ee

and also note that the delta function can be written as \cite{Fan:2019emx}, 
\be
\delta^{(4)}\(\sum_{i=1}^n \epsilon_i \omega_i q^{\mu}(z_i,\bar z_i)\) = \mathcal J \prod_{i=1}^4 \delta(\omega_i - \omega_i^*)
\ee

where
\be\label{theta}
\omega^*_i = \sum_{k=5}^n f_{ik}\omega_k, \quad i =1,2,3,4
\ee
and $\mathcal J$ is a Jacobian factor which depends only on $\(z_{i=1,2,3,4}, \bar z_{i=1,2,3,4}\)$. Neither $\mathcal J$ nor $f_{ik}$ depend on $\{\omega_{j=1,..,n}\}$. Please see Appendix \eqref{Delta} for explicit expressions for $f_{ik}$ and $\mathcal J$.

\vskip 4pt
Now in terms of $\omega$ and $t$ we can write,
\be\label{omega*}
\omega^*_i = \sum_{k=5}^{n-2} f_{ik}\omega_k + f_{in} \omega + \omega t \(f_{i \ n-1} - f_{i \ n}\)
\ee

In terms of the new variables and the representation of the delta function the Mellin amplitude $\mathcal M_n$ can be written as,
\be
\begin{gathered}
\mathcal M_n  = \mathcal J \frac{4 z^8_{12}}{z_{12}z_{13}z_{23} z_{n-2 \ n-1}z_{n-2 \ n}} \\ \times  \prod_{i=5}^{n-2} \int_{0}^{\infty} d\omega_i \omega_i^{\D_i-1} \prod_{k=1}^4 \Theta(\omega^*_k) \int_{0}^{\infty}d\omega \ \omega^{\D_{n-1}+\D_n-3} \int_{-1}^{1}dt \ \Theta(t) \ t^{\D_{n-1}-2} \ (1-t)^{\D_{n}-2} \\ \times \frac{{\omega^*_1}^{\D_1+2} {\omega^*_2}^{\D_2+2}{\omega^*_3}^{\D_3-2}{\omega^*_4}^{\D_4 -1}}{ \omega_{n-2}} \frac{\text{det}( \Phi^{123}_{n-2 \ n-1 \ n })}{z_{n-1 \ n}} \bigg |_{\omega_k = \omega_k^*, k=1,2,3,4}
\end{gathered}
\ee

Here we have used the fact that $\mathcal J$ does not depend on $\omega_{i=1,..,n}$ and have also inserted the step function $\Theta(t)$. 

Since each $\omega_{i=5,6,\cdots,n}$ independently runs from $0$ to $\infty$, \eqref{theta} together with the theta function constraint $ \prod_{k=1}^4 \Theta(\omega^*_k)$, requires $\(f_{ik} > 0\)_{i=1,2,3,4; k=5,\cdots,n}$. We will assume that we are working in a region of parameter space where this condition is satisfied and will no longer write them explicitly in the rest of the formulas. 

Now we take the conformal soft limit using the formula,
\be\label{id2}
\alpha^{\D-2} \Theta(\alpha) \sim \frac{\delta(\alpha)}{\D-1} - \frac{\delta'(\alpha)}{\D} + \frac{1}{2} \frac{\delta''(\alpha)}{\D+1} + \cdots
\ee

\subsubsection{Leading conformal soft limit}

In the leading conformal soft limit $\D_{n-1}\rightarrow 1$ we get,
\be\label{ls}
\begin{split}
& \lim_{\D_{n-1} \rightarrow 1} \(\D_{n-1} -1\) \mathcal M_n \\  
&= \mathcal J \frac{4 z^8_{12}}{z_{12}z_{13}z_{23} z_{n-2 \ n-1}z_{n-2 \ n}}   \prod_{i=5}^{n-2} \int_{0}^{\infty} d\omega_i \omega_i^{\D_i-1} \int_{0}^{\infty}d\omega \ \omega^{\D_n-2} \int_{-1}^{1}dt \ \delta(t) \ (1-t)^{\D_{n}-2} \\ 
& \times \left( \frac{{\omega_1}^{\D_1+2} {\omega_2}^{\D_2+2}{\omega_3}^{\D_3-2}{\omega_4}^{\D_4 -1}}{ \omega_{n-2}} \ \frac{\text{det}( \Phi^{123}_{n-2 \ n-1 \ n })}{z_{n-1 \ n}} \right)\bigg |_{\omega_k = \omega_k^*, k=1,2,3,4} \\
&= \mathcal J \frac{4 z^8_{12}}{z_{12}z_{13}z_{23} z_{n-2 \ n-1}z_{n-2 \ n}} \times  \prod_{i=5}^{n-2} \int_{0}^{\infty} d\omega_i \omega_i^{\D_i-1} \int_{0}^{\infty}d\omega \ \omega^{\D_n-2} \\
&  \times \( \frac{{\omega_1}^{\D_1+2} {\omega_2}^{\D_2+2}{\omega_3}^{\D_3-2}{\omega_4}^{\D_4 -1}}{ \omega_{n-2}} \ \frac{\text{det}( \Phi^{123}_{n-2 \ n-1 \ n })}{z_{n-1 \ n}}\) \bigg |_{\omega_k = \omega_k^*\(t=0\), k=1,2,3,4}
\end{split}
\ee

where
\be
\omega^*_k(t=0) = \sum_{j=5}^{n-2} f_{kj}\omega_j + f_{kn} \omega, \quad k = 1,2,3,4
\ee 

which is obtained from \eqref{omega*}.

\subsubsection{Subleading conformal soft limit}

In the subleading conformal soft limit $\D_{n-1}\rightarrow 0$, we get
\be
\begin{split}
& \lim_{\D_{n-1} \rightarrow 0} \D_{n-1} \mathcal M_n \\ 
& = \mathcal J \frac{4 z^8_{12}}{z_{12}z_{13}z_{23} z_{n-2 \ n-1}z_{n-2 \ n}}   \prod_{i=5}^{n-2} \int_{0}^{\infty} d\omega_i \omega_i^{\D_i-1} \int_{0}^{\infty}d\omega \ \omega^{\D_n-3} \int_{-1}^{1}dt \ \[- \delta'(t)\] \ (1-t)^{\D_{n}-2} \\
& \times \left( \frac{{\omega_1}^{\D_1+2} {\omega_2}^{\D_2+2}{\omega_3}^{\D_3-2}{\omega_4}^{\D_4 -1}}{ \omega_{n-2}} \  \frac{\text{det}( \Phi^{123}_{n-2 \ n-1 \ n })}{z_{n-1 \ n}} \right)\bigg |_{\omega_k = \omega_k^*, k=1,2,3,4} \\
&= \mathcal J \frac{4 z^8_{12}}{z_{12}z_{13}z_{23} z_{n-2 \ n-1}z_{n-2 \ n}} \times  \prod_{i=5}^{n-2} \int_{0}^{\infty} d\omega_i \omega_i^{\D_i-1} \int_{0}^{\infty}d\omega \ \omega^{\D_n-3} \\ 
& \times \(\frac{d}{dt}\)_{t=0} \[\(1-t\)^{\D_n-2}  \left(\frac{{\omega^*_1}^{\D_1+2} {\omega^*_2}^{\D_2+2}{\omega^*_3}^{\D_3-2}{\omega^*_4}^{\D_4 -1}}{ \omega_{n-2}} \frac{\text{det}( \Phi^{123}_{n-2 \ n-1 \ n })}{z_{n-1 \ n}}\right) \bigg |_{\omega_k = \omega_k^*, k=1,2,3,4}\]
\end{split}
\ee

Now let us define the function,
\be\label{F}
F(\omega_1^*, \omega_2^*,\omega_3^*,\omega_4^*) = \frac{{\omega^*_1}^{\D_1+2} {\omega^*_2}^{\D_2+2}{\omega^*_3}^{\D_3-2}{\omega^*_4}^{\D_4 -1}}{ \omega_{n-2}} \frac{\text{det}( \Phi^{123}_{n-2 \ n-1 \ n })}{z_{n-1 \ n}} \bigg |_{\omega_k = \omega_k^*, k=1,2,3,4}
\ee

$F$ also depends on other variables which do not depend on $t$ implicitly. Also note that due to our choice of reference spinors, $\text{det}( \Phi^{123}_{n-2 \ n-1 \ n })$ does not depend on $\omega_{n-1}$ and $\omega_n$. So all the dependence of $F$ on $t$ is through its dependence on $\(\omega^*_{i=1,2,3,4}\)$. 

\vskip 4pt
Now
\be\label{ddt}
\begin{split}
& \(\frac{d}{dt}\)_{t=0} \[\(1-t\)^{\D_n-2} F(\omega_1^*, \omega_2^*,\omega_3^*,\omega_4^*)\] \\
&= -\(\D_{n}-2\) F(\omega_1^*, \omega_2^*,\omega_3^*,\omega_4^*)\bigg |_{\omega_k^*(t=0), k=1,2,3,4} \\
& + \sum_{i=1}^4 \frac{\partial}{\partial\omega_i^*} F(\omega_1^*, \omega_2^*,\omega_3^*,\omega_4^*)\bigg |_{\omega_k^*(t=0), k=1,2,3,4} \times \omega \(f_{i \ n-1} - f_{i \ n}\)
\end{split}
\ee

where we have used \eqref{omega*} to write
\be
\frac{d}{dt} \omega^*_i = \omega \(f_{i \ n-1} - f_{i \ n}\), \quad i = 1,2,3,4
\ee 

\subsubsection{OPE limit $(n-1)^+\rightarrow n^+$ after taking the leading soft limit} 

Now we want to take the OPE limit $(n-1)^+\rightarrow n^+$ of \eqref{ls}. This is of course completely determined by the leading soft theorem and should be given by \eqref{wsope}. In particular, we want to focus on the unique $\mathcal O(z_{n-1 \ n}^0 \bar z_{n-1 \ n}^0)$ term. This can be extracted by the following procedure,
\be\label{int1}
\begin{gathered}
\lim_{z_{n-1 \ n}\rightarrow 0} \ \lim_{\bar z_{n-1 \ n}\rightarrow 0} \ \lim_{\D_{n-1} \rightarrow 1} \(\D_{n-1} -1\) \mathcal M_n \\
= \lim_{z_{n-1 \ n}\rightarrow 0} \ \lim_{\bar z_{n-1 \ n}\rightarrow 0} \  \mathcal J \frac{4 z^8_{12}}{z_{12}z_{13}z_{23} z_{n-2 \ n-1}z_{n-2 \ n}} \times  \prod_{i=5}^{n-2} \int_{0}^{\infty} d\omega_i \omega_i^{\D_i-1} \int_{0}^{\infty}d\omega \ \omega^{\D_n-2} \\ \times \(\frac{{\omega_1}^{\D_1+2} {\omega_2}^{\D_2+2}{\omega_3}^{\D_3-2}{\omega_4}^{\D_4 -1}}{ \omega_{n-2}} \frac{\text{det}( \Phi^{123}_{n-2 \ n-1 \ n })}{z_{n-1 \ n}}\) \bigg |_{\omega_k = \omega_k^*\(t=0\), k=1,2,3,4} \\
= \mathcal P_{-2,0}(n) \langle{G^-_{\D_1}(1)G^-_{\D_2}(2)\prod_{i=3}^{n-2}G^+_{\D_i}(i) G^+_{\D_n}(n)}\rangle
\end{gathered}
\ee 

Here the order of limit should be strictly maintained. 

\subsubsection{OPE limit $(n-1)^+\rightarrow n^+$ after taking the subleading soft limit}
 
In this case also we want to focus on the unique $\mathcal O(z_{n-1 \ n}^0 \bar z_{n-1 \ n}^0)$ term which is also determined by the subleading soft theorem given in \eqref{se}. According to \eqref{se} we should have,
\be\label{int2}
\begin{gathered}
\lim_{z_{n-1 \ n}\rightarrow 0} \ \lim_{\bar z_{n-1 \ n}\rightarrow 0} \ \lim_{\D_{n-1} \rightarrow 0} \D_{n-1} \mathcal M_n \\
= \lim_{z_{n-1 \ n}\rightarrow 0} \ \lim_{\bar z_{n-1 \ n}\rightarrow 0} \  \mathcal J \frac{4 z^8_{12}}{z_{12}z_{13}z_{23} z_{n-2 \ n-1}z_{n-2 \ n}} \times  \prod_{i=5}^{n-2} \int_{0}^{\infty} d\omega_i \omega_i^{\D_i-1} \int_{0}^{\infty}d\omega \ \omega^{\D_n-3} \\ \times \(\frac{d}{dt}\)_{t=0} \[\(1-t\)^{\D_n-2} \ \frac{{\omega^*_1}^{\D_1+2} {\omega^*_2}^{\D_2+2}{\omega^*_3}^{\D_3-2}{\omega^*_4}^{\D_4 -1}}{ \omega_{n-2}} \frac{\text{det}( \Phi^{123}_{n-2 \ n-1 \ n })}{z_{n-1 \ n}} \bigg |_{\omega_k = \omega_k^*, k=1,2,3,4}\] \\
= - \mathcal J^1_{-1}(n) \langle{G^-_{\D_1}(1)G^-_{\D_2}(2)\prod_{i=3}^{n-2}G^+_{\D_i}(i) G^+_{\D_n}(n)}\rangle
\end{gathered}
\ee 

Let us now use \eqref{ddt} and \eqref{int1} to further simplify the second line of \eqref{int2}. In \eqref{ddt} there are two terms. The first one gives,
\be\label{int3}
\begin{gathered}
\lim_{z_{n-1 \ n}\rightarrow 0} \ \lim_{\bar z_{n-1 \ n}\rightarrow 0} \ \lim_{\D_{n-1} \rightarrow 0} \D_{n-1} \mathcal M_n \supset \\
 \lim_{z_{n-1 \ n}\rightarrow 0} \ \lim_{\bar z_{n-1 \ n}\rightarrow 0} \  \mathcal J \frac{4 z^8_{12}}{z_{12}z_{13}z_{23} z_{n-2 \ n-1}z_{n-2 \ n}} \times  \prod_{i=5}^{n-2} \int_{0}^{\infty} d\omega_i \omega_i^{\D_i-1} \int_{0}^{\infty}d\omega \ \omega^{\D_n-3} \\ \times
\[ -\(\D_{n}-2\) F(\omega_1^*, \omega_2^*,\omega_3^*,\omega_4^*)\bigg |_{\omega_k^*(t=0), k=1,2,3,4}\] \\
= - \(\D_n -2\) \lim_{z_{n-1 \ n}\rightarrow 0} \ \lim_{\bar z_{n-1 \ n}\rightarrow 0} \ \mathcal J \frac{4 z^8_{12}}{z_{12}z_{13}z_{23} z_{n-2 \ n-1}z_{n-2 \ n}} \times  \prod_{i=5}^{n-2} \int_{0}^{\infty} d\omega_i \omega_i^{\D_i-1} \int_{0}^{\infty}d\omega \ \omega^{\D_n-3} \\ \times \(\frac{{\omega_1}^{\D_1+2} {\omega_2}^{\D_2+2}{\omega_3}^{\D_3-2}{\omega_4}^{\D_4 -1}}{ \omega_{n-2}} \frac{\text{det}( \Phi^{123}_{n-2 \ n-1 \ n })}{z_{n-1 \ n}}\) \bigg |_{\omega_k = \omega_k^*\(t=0\), k=1,2,3,4} 
\end{gathered}
\ee 

where we have used the definition \eqref{F} of $F(\omega_1^*, \omega_2^*,\omega_3^*,\omega_4^*)$ in the last line. Now comparing \eqref{int3} and \eqref{int1} we can write,
\be\label{d-1}
\begin{gathered}
\lim_{z_{n-1 \ n}\rightarrow 0} \ \lim_{\bar z_{n-1 \ n}\rightarrow 0} \ \mathcal J \frac{4 z^8_{12}}{z_{12}z_{13}z_{23} z_{n-2 \ n-1}z_{n-2 \ n}} \times  \prod_{i=5}^{n-2} \int_{0}^{\infty} d\omega_i \omega_i^{\D_i-1} \int_{0}^{\infty}d\omega \ \omega^{\D_n-3} \\ \times \frac{{\omega^*_1}^{\D_1+2} {\omega^*_2}^{\D_2+2}{\omega^*_3}^{\D_3-2}{\omega^*_4}^{\D_4 -1}}{ \omega_{n-2}} \frac{\text{det}( \Phi^{123}_{n-2 \ n-1 \ n })}{z_{n-1 \ n}} \bigg |_{\omega_k = \omega_k^*\(t=0\), k=1,2,3,4}  \\
= \mathcal P_{-2,0}(n) \langle{G^-_{\D_1}(1)G^-_{\D_2}(2)\prod_{i=3}^{n-2}G^+_{\D_i}(i) G^+_{\D_n-1}(n)}\rangle
\end{gathered}
\ee

Note that the last line of \eqref{int1} and \eqref{d-1} differ by $\D_n \rightarrow \D_n -1$. This is due to the fact that the power of $\omega$ in \eqref{int1} and \eqref{d-1} are related by $\D_n \rightarrow \D_n -1$. So finally the contribution of the first term in \eqref{ddt} to the $\mathcal O(z_{n-1 \ n}^0\bar z_{n-1 \ n}^0)$ term in the subleading soft limit can be written as,
\be
\begin{gathered}
\lim_{z_{n-1 \ n}\rightarrow 0} \ \lim_{\bar z_{n-1 \ n}\rightarrow 0} \ \lim_{\D_{n-1} \rightarrow 0} \D_{n-1} \mathcal M_n \supset \\
- (\D_n -2) \mathcal P_{-2,0}(n) \langle{G^-_{\D_1}(1)G^-_{\D_2}(2)\prod_{i=3}^{n-2}G^+_{\D_i}(i) G^+_{\D_n-1}(n)}\rangle 
\end{gathered}
\ee

Now we have to consider the contribution of the second term in \eqref{ddt}. We will show that it does not contribute to the $\mathcal O(z_{n-1 \ n}^0\bar z_{n-1 \ n}^0)$ term in the subleading soft limit. 

\vskip 4pt
The contribution of the second term is given by,
\be\label{int23}
\begin{gathered}
\lim_{z_{n-1 \ n}\rightarrow 0} \ \lim_{\bar z_{n-1 \ n}\rightarrow 0} \ \lim_{\D_{n-1} \rightarrow 0} \D_{n-1} \mathcal M_n \supset \\
 \lim_{z_{n-1 \ n}\rightarrow 0} \ \lim_{\bar z_{n-1 \ n}\rightarrow 0} \  \mathcal J \frac{4 z^8_{12}}{z_{12}z_{13}z_{23} z_{n-2 \ n-1}z_{n-2 \ n}} \times  \prod_{i=5}^{n-2} \int_{0}^{\infty} d\omega_i \omega_i^{\D_i-1} \int_{0}^{\infty}d\omega \ \omega^{\D_n-3} \\ \times
\[\sum_{i=1}^4 \frac{\partial}{\partial\omega_i^*} F(\omega_1^*, \omega_2^*,\omega_3^*,\omega_4^*)\bigg |_{\omega_k^*(t=0), k=1,2,3,4} \times \omega \(f_{i \ n-1} - f_{i \ n}\) \] 
\end{gathered}
\ee 

where 
\be\label{F2}
F(\omega_1^*, \omega_2^*,\omega_3^*,\omega_4^*) = \frac{{\omega^*_1}^{\D_1+2} {\omega^*_2}^{\D_2+2}{\omega^*_3}^{\D_3-2}{\omega^*_4}^{\D_4 -1}}{ \omega_{n-2}} \frac{\text{det}( \Phi^{123}_{n-2 \ n-1 \ n })}{z_{n-1 \ n}} \bigg |_{\omega_k = \omega_k^*, k=1,2,3,4}
\ee

Let us now enumerate various properties of the integrand in \eqref{int23} : 
\begin{enumerate}
\item $\(f_{i \ n-1} - f_{i \ n}\)$ is a polynomial in $z_{n-1 \ n}$ and $\bar z_{n-1 \ n}$, which goes to \textit{zero} as $z_{n-1 \ n}$ and $\bar z_{n-1 \ n}$ go to zero.

\item $\text{det}(\Phi^{123}_{n-2 \ n-1 \ n })$ is non-singular as $z_{n-1 \ n}$ or $\bar z_{n-1 \ n}$ goes to zero and it does not depend on $\omega_{n-1}$ and $\omega_n$. 

\item Let us consider the last two rows of the matrix $\Phi^{123}_{n-2 \ n-1 \ n }$ given by
\be
\begin{pmatrix}
\Phi_{n-1 \ 1} & \Phi_{n-1 \ 2} & \cdots & \Phi_{n-1 \ n-3} \\
\Phi_{n1} & \Phi_{n2} & \cdots & \Phi_{n \ n-3}
\end{pmatrix}
\ee

Now the off-diagonal entries are given by \eqref{offd},
\be
\Phi_{ij} = \frac{\[ ij\]}{\langle{ij}\rangle} = - \epsilon_i \epsilon_j \frac{\bar z_{ij}}{z_{ij}}
\ee

So
\be
\Phi_{n-1 \ i} = - \epsilon_i \frac{\bar z_{n-1 \ i}}{z_{n-1 \ i}}, \quad \Phi_{ni} = - \epsilon_i \frac{\bar z_{ni}}{z_{ni}}, \quad i = 1,2,\cdots,n-3
\ee

where we have used $\epsilon_{n-1}= \epsilon_n =1$. Therefore if $z_{n-1 \ n}= \bar z_{n-1 \ n}=0$ then the two rows are equal and as a result $\text{det}(\Phi^{123}_{n-2 \ n-1 \ n})=0$. So we can expand $\text{det}(\Phi^{123}_{n-2 \ n-1 \ n})$ around $z_{n-1 \ n}=\bar z_{n-1 \ n}=0$ and write
\be
\text{det}(\Phi^{123}_{n-2 \ n-1 \ n}) = A z_{n-1 \ n} + B \bar z_{n-1 \ n} + C z_{n-1 \ n} \bar z_{n-1 \ n} + \cdots
\ee

where the coefficients are functions only of $(\omega_{i=1,2,\cdots,n-2} , z_{ij}, z_{in}, \bar z_{ij}, \bar z_{in})$. This tells us that 
\be
\frac{\text{det}(\Phi^{123}_{n-2 \ n-1 \ n })}{z_{n-1 \ n}} = A + B \frac{\bar z_{n-1 \ n}}{z_{n-1 \ n}} + C \bar z_{n-1 \ n} + \cdots
\ee

and 
\be
\lim_{z_{n-1 \ n}\rightarrow 0} \ \lim_{\bar z_{n-1 \ n}\rightarrow 0} \frac{\text{det}(\Phi^{123}_{n-2 \ n-1 \ n })}{z_{n-1 \ n}} = A < \infty
\ee

\item We have from \eqref{omega*},
\be
\omega^*_i(t=0) = \sum_{k=5}^{n-2} f_{ik}\omega_k + f_{in} \omega, \quad i = 1,2,3,4
\ee
By inspection one can see that $f_{ik}$ and $f_{in}$ do not depend on $z_{n-1 \ n}$ and $\bar z_{n-1 \ n}$ for $k=5,\cdots,n-2$ and $i=1,2,3,4$. 

\end{enumerate}

We can use these facts to conclude that,
\be
\begin{gathered}
\sum_{i=1}^4 \frac{\partial}{\partial\omega_i^*} F(\omega_1^*, \omega_2^*,\omega_3^*,\omega_4^*)\bigg |_{\omega_k^*(t=0), k=1,2,3,4} \times \omega \(f_{i \ n-1} - f_{i \ n}\) \\
= \( A '+ B' \frac{\bar z_{n-1 \ n}}{z_{n-1 \ n}} + C' \bar z_{n-1 \ n} + \cdots \) \( A'' z_{n-1 \ n} + B'' \bar z_{n-1 \ n} + C'' z_{n-1 \ n} \bar z_{n-1 \ n}+ \cdots\)
\end{gathered}
\ee 

for some coefficient functions independent of $z_{n-1 \ n}$ and $\bar z_{n-1 \ n}$. This leads to 
\be\label{int231}
\begin{gathered}
\lim_{z_{n-1 \ n}\rightarrow 0} \ \lim_{\bar z_{n-1 \ n}\rightarrow 0} \ \lim_{\D_{n-1} \rightarrow 0} \D_{n-1} \mathcal M_n \supset \\
 \lim_{z_{n-1 \ n}\rightarrow 0} \ \lim_{\bar z_{n-1 \ n}\rightarrow 0} \  \mathcal J \frac{4 z^8_{12}}{z_{12}z_{13}z_{23} z_{n-2 \ n-1}z_{n-2 \ n}} \times  \prod_{i=5}^{n-2} \int_{0}^{\infty} d\omega_i \omega_i^{\D_i-1} \int_{0}^{\infty}d\omega \ \omega^{\D_n-3} \\ \times
\[\sum_{i=1}^4 \frac{\partial}{\partial\omega_i^*} F(\omega_1^*, \omega_2^*,\omega_3^*,\omega_4^*)\bigg |_{\omega_k^*(t=0), k=1,2,3,4} \times \omega \(f_{i \ n-1} - f_{i \ n}\) \]  = 0
\end{gathered}
\ee 

So the contribution of the second term in \eqref{ddt} to the $\mathcal O(z_{n-1 \ n}^0 \bar z_{n-1 \ n}^0)$ term in the subleading soft limit is $0$ and it is solely given by
\be\label{order0}
\begin{gathered}
\lim_{z_{n-1 \ n}\rightarrow 0} \ \lim_{\bar z_{n-1 \ n}\rightarrow 0} \ \lim_{\D_{n-1} \rightarrow 0} \D_{n-1} \mathcal M_n \\ 
= - (\D_n -2) \mathcal P_{-2,0}(n) \langle{G^-_{\D_1}(1)G^-_{\D_2}(2)\prod_{i=3}^{n-2}G^+_{\D_i}(i) G^+_{\D_n-1}(n)}\rangle
\end{gathered}
\ee

Now comparing \eqref{int2} and \eqref{order0} we get
\be\label{nullho}
\begin{gathered}
(\D_n -2) \mathcal P_{-2,0}(n) \langle{G^-_{\D_1}(1)G^-_{\D_2}(2)\prod_{i=3}^{n-2}G^+_{\D_i}(i) G^+_{\D_n-1}(n)}\rangle \\
= \mathcal J^1_{-1}(n) \langle{G^-_{\D_1}(1)G^-_{\D_2}(2)\prod_{i=3}^{n-2}G^+_{\D_i}(i) G^+_{\D_n}(n)}\rangle
\end{gathered} 
\ee

Shifting the dimension $\D_n \rightarrow \D_n + 1$ and taking into account that fact that $n^+$ is outgoing we can write \eqref{nullho} as, 
\be\label{nullho1}
\[ \mathcal J^1_{-1}(n)\mathcal P_{-1,-1}(n) - (\D_n -1) \mathcal P_{-2,0}(n)\] \langle{G^-_{\D_1}(1)G^-_{\D_2}(2)\prod_{i=3}^{n-2}G^+_{\D_i}(i) G^+_{\D_n}(n)}\rangle =0
\ee

This is the null-state decoupling relation \eqref{de1} we wanted to prove. 

\vskip 4pt
Before we end the proof we would like to mention that the Mellin transformation of the graviton scattering amplitude in GR is generically UV divergent. One way of curing this is by modifying the Mellin transformation by the introduction of $e^{-i\sum_k \epsilon_k\omega_k u_k}$ which acts as a damping factor. This is particularly useful because $u$, thought of as a regulator, preserves all the symmetries of the problem. Since the null-states are completely determined by symmetries, as we have shown, \eqref{nullho1} still holds in the regulated theory as we have explicitly checked for $5$ graviton MHV amplitude in section \eqref{6mhvu}. With some more notation the above proof can be readily generalized to the regulated theory. In fact, it should not depend on what regulator one uses, as long as all the symmetries are preserved.

\end{document}